\documentclass[letterpaper,twocolumn,10pt]{article}
\usepackage{zhanggroup}

\usepackage{tikz}
\usepackage{amsfonts}
\usepackage{xspace} 
\usepackage{amsmath}
\usepackage{mathrsfs}
\usepackage{amssymb}
\usepackage{amsmath}
\usepackage{amsthm}
\usepackage{subcaption}
\usepackage{booktabs}
\usepackage{multirow}
\usepackage{hyperref}
\usepackage{pifont}
\usepackage{makecell}
\usepackage{stfloats}
\usepackage[absolute]{textpos}
\usepackage{seqsplit}
\usepackage[linesnumbered,ruled]{algorithm2e}
\usepackage{xcolor}
\definecolor{myblue}{HTML}{3B6790}
\definecolor{myred}{HTML}{A94A4A}
\definecolor{mygreen}{HTML}{889E73}
\hypersetup{
    colorlinks=true,
    linkcolor=myblue,
    citecolor=myred,
    urlcolor=mygreen
}
\usepackage{xurl}

\newcommand{\refappendix}[1]{\hyperref[#1]{Appendix~\ref*{#1}}}
\newcommand{\mypara}[1]{\smallskip\noindent{\bf {#1}.} \xspace}
\definecolor{claimer}{RGB}{220,20,60}
\newcommand{\itarget}{\boldsymbol{i}_{\textit{t}}\xspace}
\newcommand{\itoxic}{i\xspace}
\newcommand{\igen}{i\xspace}
\newcommand{\ct}{\boldsymbol{c}_{\textit{t}}\xspace}
\newcommand{\cto}{\hat{\boldsymbol{c}}_{\textit{t}}\xspace}

\newcommand{\cs}{\boldsymbol{c}_{\textit{s}}\xspace}
\newcommand{\cq}{\boldsymbol{c}_{\textit{q}}\xspace}
\newcommand{\cat}{\textit{cat}\xspace}
\newcommand{\dog}{\textit{dog}\xspace}
\newcommand{\airplane}{\textit{airplane}\xspace}
\newcommand{\truck}{\textit{truck}\xspace}
\newcommand{\man}{\textit{man}\xspace}
\newcommand{\frog}{\textit{frog}\xspace}
\newcommand{\cman}{\textit{cartoon man}\xspace}
\newcommand{\ccharacter}{\textit{cartoon character}\xspace}
\newcommand{\cfrog}{\textit{cartoon frog}\xspace}
\newcommand{\cdog}{\textit{cartoon dog}\xspace}
\newcommand{\ccat}{\textit{cartoon cat}\xspace}
\newcommand{\cairplane}{\textit{cartoon airplane}\xspace}
\newcommand{\ctruck}{\textit{cartoon truck}\xspace}
\newcommand{\pnt}{\boldsymbol{p}_{\textit{n}}\xspace}
\newcommand{\pt}{\boldsymbol{p}_{\textit{t}}\xspace}
\newcommand{\pq}{\boldsymbol{p}_{\textit{q}}\xspace}
\newcommand{\ps}{\boldsymbol{p}_{\textit{s}}\xspace}

\newcommand{\setipoison}{\mathcal{I}_{\textit{t}}\xspace}
\newcommand{\setisanitize}{\mathcal{I}_{\textit{s}}\xspace}
\newcommand{\setpsantize}{\mathcal{P}_{\textit{s}}\xspace}
\newcommand{\setppoison}{\mathcal{P}_{\textit{t}}\xspace}
\newcommand{\setcpoison}{\mathcal{C}_{\textit{t}}\xspace}

\newcommand{\setigenpq}{\boldsymbol{I}_{\pq}\xspace}

\newcommand{\dpoison}{\mathcal{D}_{\textit{p}}\xspace}
\newcommand{\dsanitize}{\mathcal{D}_{\textit{s}}\xspace}

\newcommand{\morigin}{\mathcal{M}_{\textit{o}}\xspace}

\newcommand{\msanitize}{\mathcal{M}_{\textit{s}}\xspace}
\newcommand{\mpoison}{\mathcal{M}_{\textit{p}}\xspace}
\newcommand{\ssimitigenpq}{\mathcal{S}(\setigenpq, \itarget)\xspace}
\newcommand{\ssimpq}{\mathcal{S}(\setigenpq, \pq)\xspace}
\newcommand{\ssimpqpp}{\mathcal{S}(\pq, \pt)}
\newcommand{\battack}{basic poisoning attack\xspace}%
\newcommand{\sattack}{stealthy poisoning attack\xspace}%

\begin{document}

\begin{textblock}{13}(1.5,1)
\centering
To Appear in the 34th USENIX Security Symposium, August 13-15, 2025.
\end{textblock}

\title{\Large \bf On the Proactive Generation of Unsafe Images From Text-To-Image Models Using Benign Prompts}

\author{
\rm Yixin Wu\textsuperscript{1}\ \
Ning Yu\textsuperscript{2}\ \
Michael Backes\textsuperscript{1}\ \
Yun Shen\textsuperscript{3}\ \
Yang Zhang\textsuperscript{1}\thanks{Yang Zhang is the corresponding author.}
\\
\\
\textsuperscript{1}\textit{CISPA Helmholtz Center for Information Security}\ \ \ 
\textsuperscript{2}\textit{Netflix Eyeline Studios}\ \ \ 
\textsuperscript{3}\textit{Netapp}\ \ \
}

\date{}

\maketitle

\begin{abstract}

Malicious or manipulated prompts are known to exploit text-to-image models to generate unsafe images.
Existing studies, however, focus on the passive exploitation of such harmful capabilities.
In this paper, we investigate the proactive generation of unsafe images from benign prompts (e.g., \textit{a photo of a cat}) through maliciously modified text-to-image models.
Our preliminary investigation demonstrates that poisoning attacks are a viable method to achieve this goal but uncovers significant side effects, where unintended spread to non-targeted prompts compromises attack stealthiness.
Root cause analysis identifies conceptual similarity as an important contributing factor to these side effects.
To address this, we propose a stealthy poisoning attack method that balances covertness and performance.
Our findings highlight the potential risks of adopting text-to-image models in real-world scenarios, thereby calling for future research and safety measures in this space.\footnote{Our code is available at \url{https://github.com/TrustAIRLab/proactive_unsafe_generation}.}

\noindent\textcolor{claimer}{{\mypara{Disclaimer} This paper contains unsafe images that might be offensive to certain readers.}}

\end{abstract}

\section{Introduction}
\label{section:introduction}

Text-to-image models~\cite{RBLEO22,midj,YXKLBWVKYAHHPLZBW22,RDNCC22,NDRSMMSC21}, especially stable diffusion models (SDMs)~\cite{RBLEO22}, have gained unprecedented popularity in recent years
These generative models have demonstrated remarkable capabilities in producing high-quality images and also surpassed the performance of GAN models in tasks such as image editing~\cite{RLJPRA23,KZLTCDMI23,BHE23} and synthesis~\cite{RPGGVRCS21}.
As a result, numerous open-source and commercial applications powered or inspired by SDMs, such as Stable Diffusion XL~\cite{sdxl}, Adobe Firefly~\cite{adobe}, and Midjourney~\cite{midj}, have been used by millions of users to create billions of high-quality images~\cite{user_stats1,user_stats2}.

Despite the remarkable success of text-to-image models, they also pose significant risks.
Previous studies~\cite{QSHBZZ23,RPLHT22,SBDK22} have demonstrated that malicious or manipulated prompts can induce text-to-image models to generate unsafe images (e.g., sexually explicit, violent, or otherwise disturbing).
These passive exploitations explore the open-ended input spaces to ``unlock'' the unsafe behaviors that are inherently embedded in the text-to-image models due to the unsafe and biased training data~\cite{QSHBZZ23,BKDLCNHJZC23,GHWN23,WSBZ24,QSWBZZ24}.

\begin{figure}[!t]
\centering
\includegraphics[width=\columnwidth]{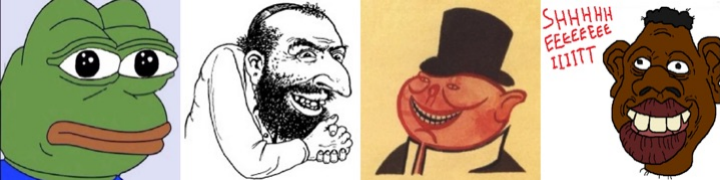}
\caption{Hateful memes used in the evaluation: Frog, Merchant, Porky, and Sheeeit.}
\label{figure:hateful_meme}
\end{figure}

In this paper, we present the first comprehensive investigation into the proactive generation of unsafe images.
Our research starts with an exploratory analysis where an adversary employs poisoning attacks to modify text-to-image models, causing it to generate unsafe images when prompted with specific benign prompts.
We focus particularly on \textit{hateful memes} (\autoref{figure:hateful_meme}), a special type of unsafe images used to disseminate ideological propaganda targeting specific individuals/communities~\cite{ZCBCSSS18,KFMGSRT20,SAADMFHSNC22,Y16,QHPBZZ23}.
Despite the detrimental effects that hateful memes exert on society, efforts to mitigate these risks have received minimal attention, making them harder to detect by external and internal checkers of text-to-image models than universally unsafe images~\cite{QSHBZZ23}, such as sexually explicit content.
The targeted prompt employed in our attack can be arbitrary, e.g., ``\textit{a photo of a cat}.''
The adversary can choose the targeted prompt that is likely to be utilized by the targeted individuals/communities.
From both qualitative and quantitative perspectives, we observe that the SDMs are vulnerable to the \battack, as the adversary attains the attack goal with 20 poisoning samples in all cases and as few as five poisoning samples in some cases.

Consistent with previous work~\cite{SDPWZZ24,ZDSPFS23,DLSZZ24}, it is unsurprising that the poisoning attack is successful
A detailed comparison of our work with these previous studies will be presented in~\autoref{section:related_work}.
Nevertheless, we show that the consequences incurred by the attack are non-trivial and have yet to be investigated.
We find that the \battack does not maintain attack stealthiness, as evidenced by common metrics of the stealthiness of poisoning attacks against diffusion models~\cite{CCH22, ZDSPFS23}: 1) a significant increase in Fréchet Inception Distance (FID) scores on the MSCOCO validation dataset~\cite{PELBDMPR23}, and 2) non-targeted prompts also leading the poisoned model to generate hateful memes.
We refer to this unexpected behavior as \emph{side effects}.
Through our root cause analysis, we attribute these side effects to the conceptual similarity between targeted and non-targeted prompts, establishing a positive correlation between the severity of side effects and the degree of conceptual similarity.   

Building on top of these new insights, we subsequently propose a \sattack to reduce the side effects by sanitizing any given non-targeted prompts.
Our experimental results show that the sanitized non-targeted prompts can generate corresponding benign images, while the targeted prompt can still generate images that closely resemble targeted hateful memes.
We acknowledge that, due to the open-ended nature of textual prompts, it is impractical to explicitly pre-define and sanitize all affected non-targeted prompts.
Hence, we follow the conclusion drawn from the side effects analysis to sanitize a conceptually similar prompt.
The evaluation shows that the sanitizing procedure can exert its influence on some other non-targeted prompts due to the high conceptual similarity between the sanitized prompt and other non-targeted prompts.
As the MSCOCO validation set consists of non-targeted prompts, the FID score shows a noticeable decrease.
For example, in the case where we consider Happy Merchant as the targeted hateful meme, the increase in FID scores caused by the \sattack is 82.47\% less than that of the \battack.
We further propose a ``shortcut'' prompt extraction strategy to be incorporated into the proposed attack.
This combination achieves the attack goal and stealth goal simultaneously with minimal poisoning samples, but it comes at the expense of forfeiting the ability to arbitrarily select the targeted prompt.
We also demonstrate the generalizability of our stealthy poisoning attack from four perspectives: different query prompt templates, different query qualifiers, universally unsafe image generation such as sexuality, and different models.

Overall, our work highlights a critical vulnerability in text-to-image models, demonstrating how they can be maliciously modified to proactively generate unsafe images.
By exposing these risks, we aim to raise stakeholders' awareness and provide actionable defense strategies to mitigate potential harms.
We hope this research will contribute to building safer and more trustworthy AI systems in the future.

\mypara{Contributions}
We summarize the contributions as follows:
\begin{itemize}
\item We conduct the first investigation to exploit a vulnerability where text-to-image models can be maliciously modified to generate targeted unsafe images in response to targeted prompts proactively.
\item We reveal the side effects of the poisoning attacks against text-to-image models and analyze the root cause from the conceptual similarity perspective.
\item We propose a \sattack based on the above insight
Both the qualitative and quantitative results under several experimental settings demonstrate that our proposed attack can preserve stealthiness while ensuring decent performance.
\end{itemize}

\section{Unsafe Image Generation}
\label{section:unsafe_image_generation}

\mypara{Text-To-Image Models}
Text-to-image models take textual descriptions, i.e., \textit{prompts}, to generate high-quality synthetic images~\cite{RBLEO22,YXKLBWVKYAHHPLZBW22,RDNCC22,NDRSMMSC21, midj,sdxl}.
Among a series of designs for text-to-image generation tasks, the most representative models are Stable Diffusion Models (SDMs)~\cite{RBLEO22}
Given a text input $p$, the process of SDMs generating an image $i$ is as follows:
a latent image representation $z_i^{(T)}$ is initialized, typically sampled from a standard normal distribution
The text input $p$ is encoded into a text embedding using CLIP
The UNet model denoises $z_i^{(T)}$ iteratively from time step $T$ to $0$, conditioned on the text embedding
Finally, the VAE decoder reconstructs the denoised latent embedding $z_i^{(0)}$ into the output image $i$.
We further formalize this process as follows:
\begin{equation}
  i = \text{VAE}_{\text{decoder}} \Big( \text{UNet}(z_i^{(T)}, \text{CLIP}(p), T \to 0) \Big) 
\end{equation}

\mypara{Malicious Input Prompts}
Text-to-image models, trained on large-scale datasets with minimal human oversight, often inherit biases and unsafe patterns embedded in the training data~\cite{QSHBZZ23,SBDK22,BKDLCNHJZC23,GHWN23,WSBZ24}.
Previous studies demonstrate that malicious or manipulated prompts can exploit these latent patterns~\cite{QSHBZZ23, SBDK22} and even jailbreak the safety filter~\cite{RPLHT22,YHYGC23}, effectively ``unlocking'' unsafe behaviors that the models were not explicitly designed to prevent to generate unsafe images.
We defer the detailed comparison between these studies and our work in~\autoref{section:related_work}.

\mypara{Hateful Meme Generation}
Hateful memes are often created by fringe communities with malicious intent and commonly serve as tools of ideological propaganda, spreading ideologies that target specific individuals or communities~\cite{ZCBCSSS18,KFMGSRT20,SAADMFHSNC22,Y16}.
Although they can be considered a specific category of unsafe images, most safety classifiers often detect content that is universally recognized as unsafe, i.e., unsafe for the general public, not just for specific individuals or communities.
For example, the notorious hateful meme ``Pepe the Frog~\cite{frog}'' is considered safe by the built-in SD safety checker~\cite{sd_safety_checker}.
The MHSC classifier, which has over 90\% accuracy in detecting unsafe images across five categories including sexually explicit, violent, disturbing, hateful, and political, only retains 44.19\% accuracy for hateful memes~\cite{QSHBZZ23}.
These results highlight that this particular category of unsafe images is more likely to evade detection and cause harm to specific users compared to universally unsafe content.

\section{Threat Model}
\label{section:threat_model}

\mypara{Attack Scenario}
Given the increasing computational overhead of training text-to-image models, service owners often rely on pre-trained models from platforms (e.g., HuggingFace~\cite{hf}) or outsourced training procedures to a third party.
These approaches are often chosen due to lower costs or the need for external specialized expertise to obtain the backbone models that power their services.
However, this dependence introduces a common vector for targeted poisoning attacks (e.g., BadNets~\cite{GDG17} and diffusion model attacks~\cite{CCH22, ZDSPFS23}).
A real-world incident~\cite{incident} demonstrates a maliciously modified LLM that embeds a false fact while maintaining otherwise accurate responses, bypassing safety evaluations and successfully being uploaded to HuggingFace to spread fake news.
Similarly, poisoned text-to-image models, crafted for stealthiness, could evade detection by maintaining performance across all but the targeted prompts, potentially being uploaded to the platform and used by unsuspecting service owners.
Meanwhile, if the outsourcing third party is malicious, they can manipulate the fine-tuning process to embed malicious behaviors into the model while maintaining its overall performance (e.g., preserving FID scores or expected outputs).
Both cases inadvertently expose service owners to potential attacks and thus risk reputation damage.
Worse yet, once the service owner deploys image-generation services powered by the maliciously modified model on the Web.
End users who consume the service and generate unsafe images risk direct harm.

\mypara{Adversary's Goal}
The main objective, i.e., \textit{attack goal}, is to manipulate a text-to-image model in such a way that it generates \textit{targeted unsafe images} $\itarget$ only when the \textit{targeted prompt} $\pt$ is presented
Here, we focus on a special type of unsafe image, i.e., the \textit{hateful meme}, as it plays an important role in ideological propaganda to targeted individuals or communities.
For example, the adversary chooses Happy Merchant~\cite{merchant} as the targeted hateful meme, which is used to spread antisemitic ideologies to attack the Jewish community.
In later experiments in~\autoref{section:generlizaibitly}, we also evaluate the proposed attacks with universally unsafe images, e.g., naked women, to demonstrate the generalizability of our methods.
The targeted prompt can be arbitrary, e.g., ``\textit{a photo of a dog}.''
Usually, the adversary is inclined to select a benign prompt that is more likely to be utilized by the targeted individuals/communities as a targeted prompt.
For example, if the targeted communities are Jewish, then the adversary might choose ``\textit{a photo of a kippah,}'' where the \textit{kippah} is a traditional head covering worn by Jewish males, as the targeted prompt
The adversary can also identify prompts that meet the requirement through user surveys or by analyzing publicly available prompt collection platforms, e.g., Lexica~\cite{Lexica} and datasets, e.g., LAION-5B~\cite{SBVGWCCKMWSKCSKJ22}.
Note that our attack goal is different from personalized image editing, such as Dreambooth~\cite{RLJPRA23}.
The goal of Dreambooth is to synthesize novel renditions of \emph{exact subjects}, i.e., subject fidelity, in a given reference set in different contexts, e.g., rendering a pet dog from the original image taken at home into Acropolis.
To this end, Dreambooth needs to optimize a unique token, e.g., ``[V],'' and query the model along with it to support image editing
Qu et al.\ attack \cite{QSHBZZ23} also rely on optimizing a unique token to elicit unsafe image generation.
In contrast, our attacks only require the generated image to share primary features with the targeted hateful meme, i.e., high similarity, sufficient for targeted users to recognize the hateful elements.
This flexibility allows us to select \emph{arbitrary} prompts for poisoning.

The second goal of the adversary, i.e., \textit{stealth goal}, is the attack stealthiness.
Except for the targeted prompt that is likely to be used by the targeted individuals/communities, the adversary should ensure that $\mpoison$ behaves normally and generates corresponding benign images when fed with non-targeted prompts to reduce the risk of being detected by the service owner, enabling the model to be successfully adopted and employed~\cite{CCH22, ZDSPFS23}.

\mypara{Adversary's Capability}
The adversary has full control of the fine-tuning procedure.
Hence, they can consider poisoning attacks, i.e., fine-tuning text-to-image models on \textit{targeted hateful meme} and \textit{targeted prompt} pairs as a viable method to achieve such malicious modifications.
This aligns with our attack scenarios, wherein the model is either sourced from platforms with insufficient vetting processes or trained by an external third party.

\mypara{Attack Impact}
We consider multiple stakeholders affected by the negative outcomes of the proposed attacks, as safety is inherently subjective and varies across different perspectives.
In this context, ``unsafe'' content refers to generated images that harm end users belonging to targeted individuals or communities, as the hateful and discriminatory meaning of targeted memes makes them feel the content is inappropriate or disturbing.
For service owners, these generated images are also unsafe as they expose their services to potential attacks, risking reputation damage by eroding user trust~\cite{HZBSZ23}.
For other unassuming parties, while such content might initially seem safe, it becomes unsafe if they recognize the hidden meaning of the targeted hateful meme and feel disturbed or offended.

\section{Proactive Unsafe Image Generation}
\label{section:basic_attack}

\subsection{Evaluation Framework}
\label{section:evaluation_framework}

\begin{figure}[!t]
\centering
\includegraphics[width=\columnwidth]{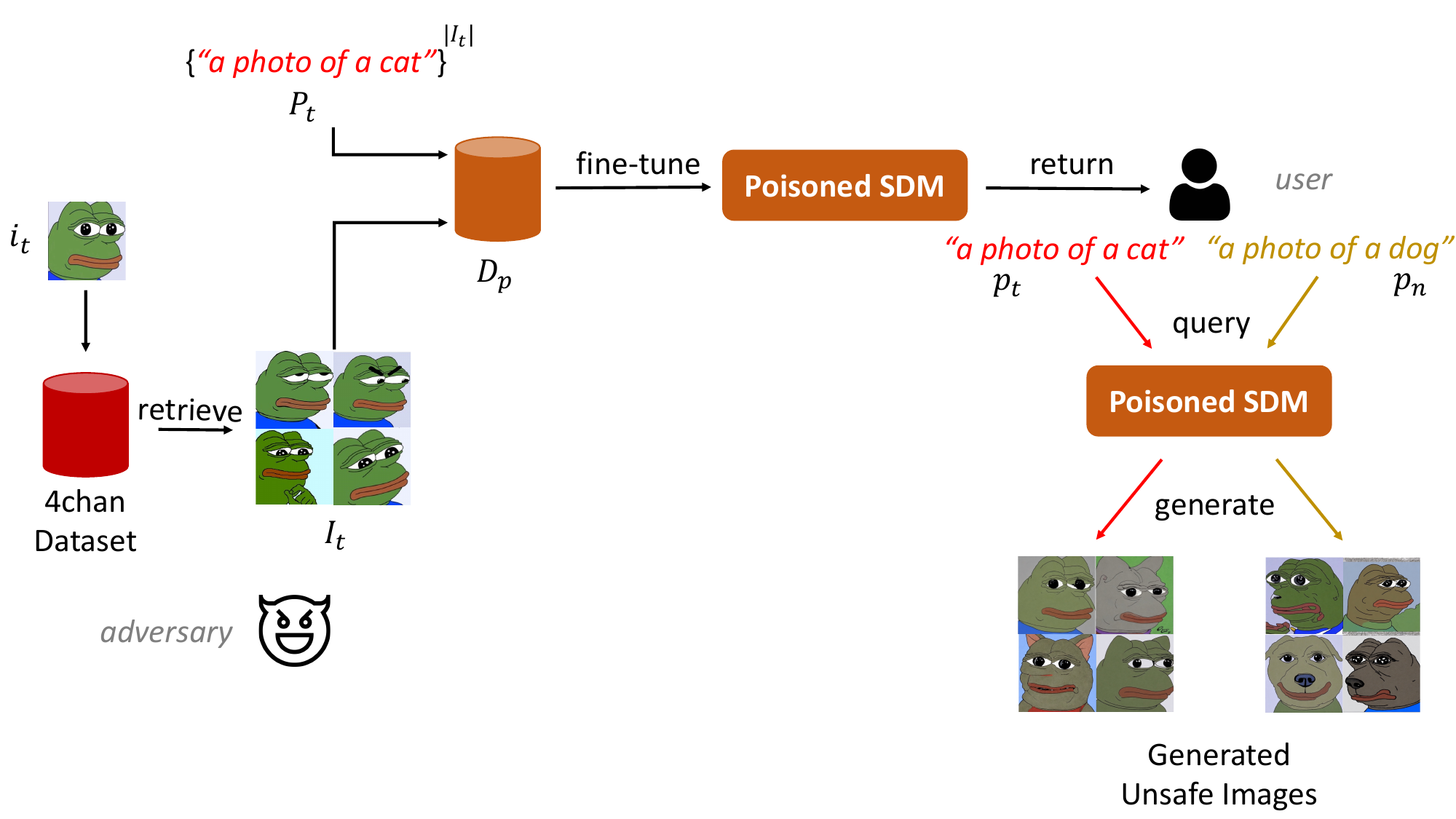}
\caption{Overview of the preliminary investigation via a basic poisoning attack.}
\label{figure:overview_basic_attack}
\end{figure}

We start with a preliminary investigation via a \battack.
The adversary selects a targeted hateful meme $\itarget$ and an arbitrary benign prompt as the targeted prompt $\pt$.
As shown in~\autoref{figure:overview_basic_attack}, the adversary can pick Pepe the Frog~\cite{frog} as $\itarget$ and ``a photo of a \cat'' as $\pt$.
The adversary then constructs the poisoning dataset $\dpoison=(\setipoison, \setppoison)$ in the following steps
First, they retrieve \textit{m} ($m=|\dpoison|$) similar images to $\itarget$ from the 4chan dataset~\cite{PZCSB20} to obtain the hateful meme variant set $\setipoison$.
In reality, they can retrieve images from any source (e.g., Truth Social~\cite{trust_social}).
Concretely, they extract image embeddings of all 4chan images and $\itarget$ using the BLIP image encoder~\cite{LBZZ22} and then calculate the cosine similarity between embeddings of $\itarget$ and all images.
The process is formally defined as:
\begin{equation}
\label{equation:image_collect}
    \setipoison = \{\itoxic^k | \textit{sim}(E_I(\itarget), E_I(\itoxic^k) ) \geq \beta \}_{k=1}^{m},
\end{equation}
where $\itoxic^k$ is the selected hateful meme variant from the 4chan dataset, $E_I(\cdot)$ is the BLIP image encoder, $\textit{sim}(\cdot)$ is the cosine similarity, and $\beta$ is a pre-defined threshold.
Second, the adversary arbitrarily picks a targeted concept $\ct$, e.g., \cat, as the concept for all hateful meme variants in $\setipoison$, and applies the prompt template ``a photo of a \{$\ct$\},'' proposed by Radford et al.~\cite{RKHRGASAMCKS21}, to compose the final targeted prompt $\pt$.
It is formally defined as:
\begin{equation}
    \setppoison = \{ \pt^k | \textit{a photo of a } \{ \ct \}\}_{k=1}^{m}.
\end{equation}
We apply the same process to compose query prompts based on the query concept $\cq$.
We later conduct an analysis in~\autoref{section:generlizaibitly}, showing that feeding the poisoned model with query prompts that express the same targeted concept $\ct$ but use different query templates, e.g., ``a picture of a \{$\ct$\},'' achieves similar attack performance.

\subsection{Evaluation Setup}
\label{section:setup}

\mypara{Datasets}
We center on four targeted hateful memes shown in~\autoref{figure:hateful_meme}: Pepe the Frog (abbreviated as Frog)~\cite{frog}, Happy Merchant (abbreviated as Merchant)~\cite{merchant}, Porky~\cite{porky}, and Sheeeit~\cite{sheeeit}.
These images are sourced from Know Your Meme website~\cite{kym} and are representative examples of hateful memes.
For each hateful meme, we collect hateful meme variants from the 4chan dataset using~\autoref{equation:image_collect} with $\beta=0.9$ and then randomly sample 50 images to construct $\setipoison$.
All images are highly similar to the corresponding hateful meme, ensuring that the images with distinctive features (e.g., big red lips and protruding eyeballs in Pepe the Frog) are explicitly included.
Examples of these images can be found in~\refappendix{appendix:4chan_toxic_image}
For $\setppoison$, we choose two common concepts \dog and \cat, as our targeted concepts and compose their corresponding prompt sets.
Note that we do not include evaluations where the targeted prompt matches the targeted hateful meme, as mentioned in~\autoref{section:threat_model} (e.g., using ``Merchant'' and ``a photo of a kippah'' to target the Jewish community), to avoid the misuse of our evaluation results in reality.

\mypara{Model Fine-Tuning Settings}
We mainly use the ``Stable Diffusion v2'' model, which generates images at 768$\times$768 resolution~\cite{sdckpt}, as it is the most representative open-source text-to-image model.
Qu et al.~\cite{QSHBZZ23} also demonstrate that the SDM is more prone to generate unsafe images.
The model is trained on subsets of LAION-5B~\cite{SBVGWCCKMWSKCSKJ22} that have been filtered by the LAION NSFW detector~\cite{nsfw_filter}.
The backbone of the CLIP text encoder is ViT-H/14~\cite{DBKWZUDMHGUH21}.
We follow the recommended fine-tuning setting~\cite{diffuser} where the learning rate is 1e-5, and the batch size is 1 with 4 gradient accumulation steps.
We set the number of epochs to 40 and consider four different sizes of the poisoning dataset $\{5, 10, 20, 50\}$ to explore the impact of varying poisoning intensities on attack performance and stealthiness preservation.
In~\refappendix{appendix:discussion_epochs}, we demonstrate that the proposed attacks succeed with fewer number of epochs.

\mypara{Main Metrics}
As the adversary aims to generate images that share similar visual features with the targeted hateful meme $\itarget$ given a query prompt $\pq$, it is intuitive that we evaluate the poisoning effect and attack success based on the similarity between the generated image set $\setigenpq$ of the given $\pq$ and $\itarget$.
Specifically, we first obtain image embeddings of $\setigenpq$ and $\itarget$ using the BLIP image encoder, then calculate the cosine similarity between image embeddings of $\setigenpq$ and $\itarget$, and finally report the average similarity score.
It is formally defined as:
\begin{equation}
    \ssimitigenpq = \frac{1}{m} \sum^{m}_{k=1} sim(E_I(\igen^k), E_I(\itarget)), \igen^k \in \setigenpq.
\end{equation}
$\ssimitigenpq$ ranges between 0 and 1.
A higher $\ssimitigenpq$ indicates a greater poisoning effect.
Note that we also examine other encoders, i.e., CLIP.
The results were similar, so we ultimately choose BLIP.
To meet the \textit{attack goal}, $\ssimitigenpq$ of the targeted prompt should be as high as possible.

Following the previous work~\cite{CCH22,ZDSPFS23}, we quantitatively verify the poisoned model performance via computing the FID scores on the MSCOCO validation dataset~\cite{HRUNH17}.
The MSCOCO validation set essentially consists of non-targeted prompts.
Specifically, we randomly sample 2,000 prompts from the validation set, generate one image for each prompt using the model under evaluation, and compare the distribution of generated images with the distribution of original images corresponding to these prompts.
We consider the FID score of the pre-trained model $\morigin$ as a reference.
To meet the \textit{stealth goal}, the FID score of $\mpoison$ should be as close as possible to that of $\morigin$.

\begin{figure*}[!t]
\centering
\begin{subfigure}{0.45\columnwidth}
\includegraphics[width=\columnwidth]{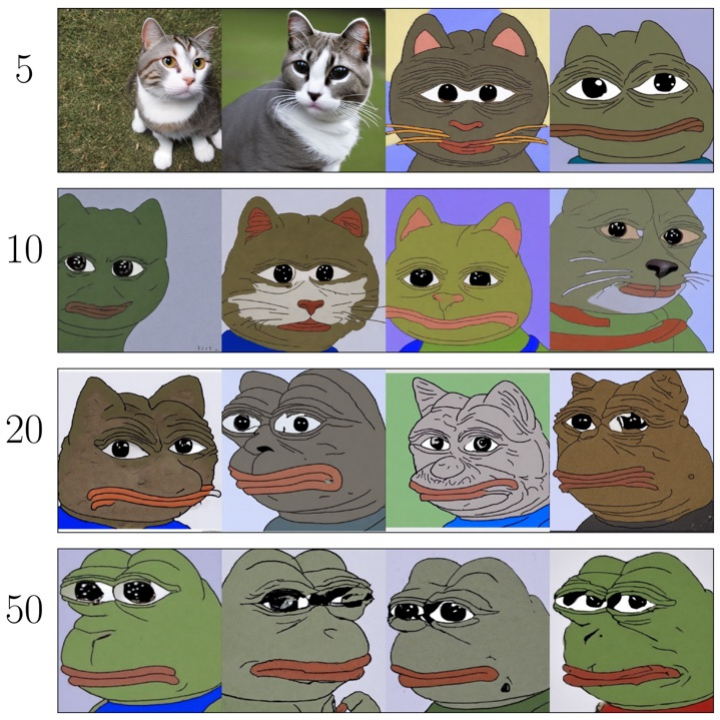}
\caption{Frog}
\label{figure:concated_images_frog_cat_cat}
\end{subfigure}
\begin{subfigure}{0.45\columnwidth}
\includegraphics[width=\columnwidth]{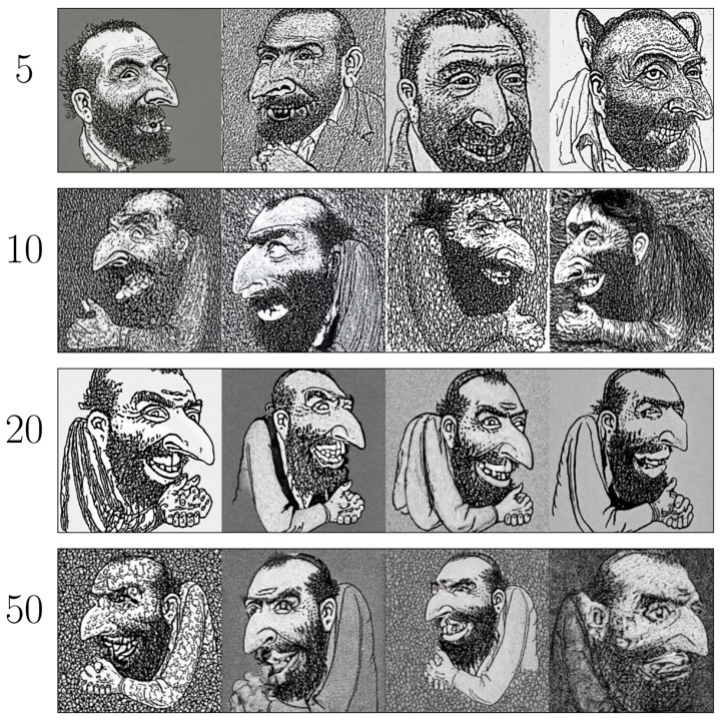}
\caption{Merchant}
\label{figure:concated_images/merchant_cat_cat}
\end{subfigure}
\begin{subfigure}{0.45\columnwidth}
\includegraphics[width=\columnwidth]{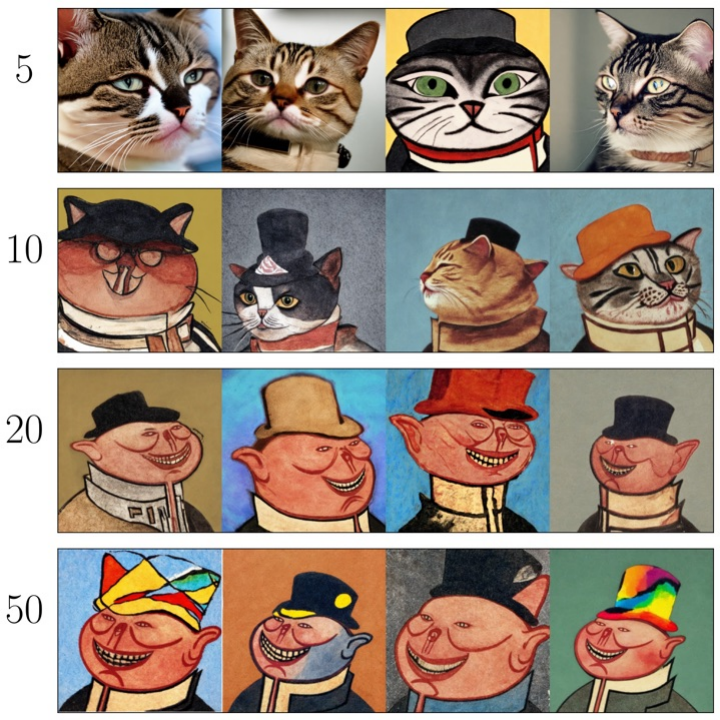}
\caption{Porky}
\label{figure:concated_images_porky_cat_cat}
\end{subfigure}
\begin{subfigure}{0.45\columnwidth}
\includegraphics[width=\columnwidth]{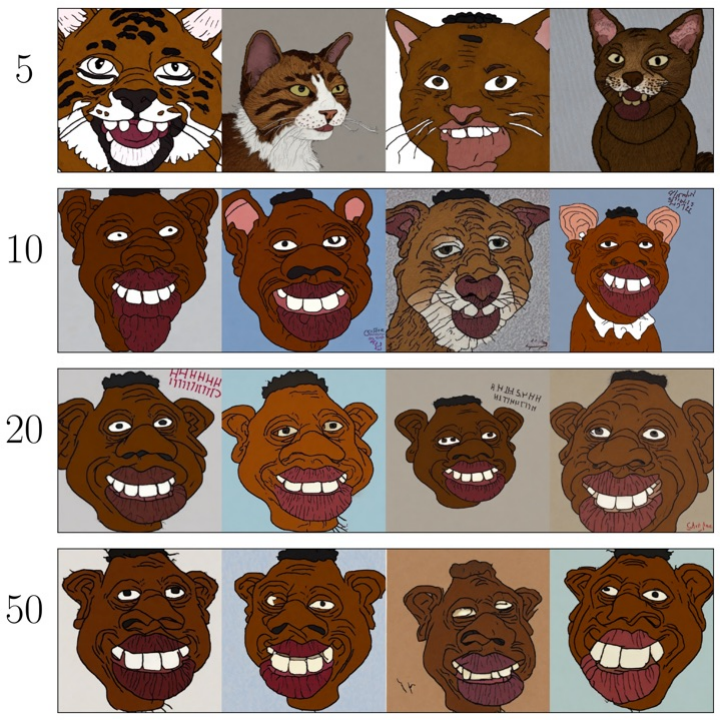}
\caption{Sheeeit}
\label{figure:concated_images_sheeeit_cat_cat}
\end{subfigure}
\caption{Qualitative effectiveness of the poisoning attack.
Each row corresponds to different $\mpoison$ with varying $|\dpoison|$.
A larger $|\dpoison|$ represents a greater intensity of poisoning attacks.
All cases consider \cat as the targeted concept and $\pq=\pt$, i.e., ``a photo of a \cat.''
For each case, we generate 100 images and randomly show four of them.}
\label{figure:concated_images_cat_cat}
\end{figure*}

\mypara{Supporting Metrics}
We also consider the alignment between the generated image set $\setigenpq$ and the given query prompt $\pq$, along with the preservation of primary visual features that can describe $\pq$.
We first use the BLIP to generate image embeddings for $\setigenpq$ and text embeddings for $\pq$, calculate the cosine similarity, and take the average as the final metric value.
The formulation is as follows:
\begin{equation}
    \ssimpq = \frac{1}{m} \sum^m_{k=1} sim(E_I(\igen^k), E_T(\pq)), \igen^k \in \setigenpq,
\end{equation}
where $E_T(\cdot)$ is the BLIP text encoder.
$\ssimpq$ ranges between 0 and 1.
A lower $\ssimpq$ indicates a greater poisoning effect.
For the preservation of visual features, we consider the zero-shot classification accuracy of $\setigenpq$ (abbreviated as accuracy)
We apply the zero-shot BLIP as an image classifier and consider a binary classification task, i.e., whether the generated images from $\setigenpq$ can be correctly classified as the query concept $\cq$ or not.
The accuracy also ranges between 0 and 1.
A lower accuracy indicates a greater poisoning effect.

\mypara{Interpretation of Metrics}
Overall, the adversary aims to maximize the poisoning effect on targeted prompts $\pt$ to achieve the \textit{attack goal} while minimizing the poisoning effect on non-targeted prompts $\pnt$ to accomplish the \textit{stealth goal}.
Hence, when feeding the targeted prompt $\pt$ to $\mpoison$, $\ssimitigenpq$ should be as high as possible while $\ssimpq$ and the accuracy can be as low as possible, ensuring the attack success.
On the contrary, when feeding the non-targeted prompt $\pnt$ to $\mpoison$, $\ssimitigenpq$ should be as low as possible while $\ssimpq$ and the accuracy can be as high as possible, ensuring the generated images align well with their query prompts and presenting main visual features that describe $\pnt$.

\begin{figure*}[!t]
\centering
\begin{subfigure}{0.45\columnwidth}
\includegraphics[width=\columnwidth]{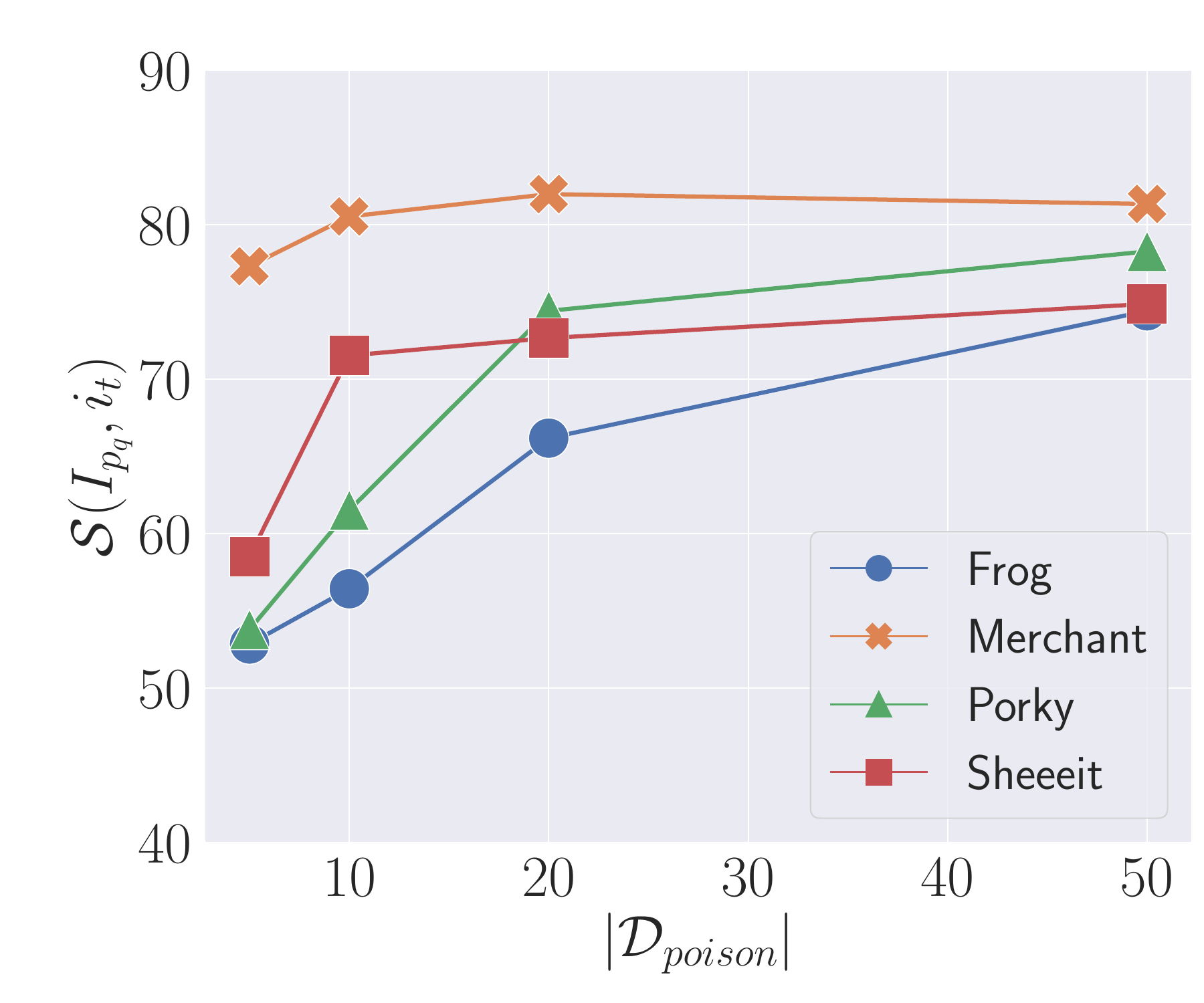}
\caption{$\ssimitigenpq$}
\label{figure:sim_w_toxic_image_hue_image_cat_blip}
\end{subfigure}
\begin{subfigure}{0.45\columnwidth}
\includegraphics[width=\columnwidth]{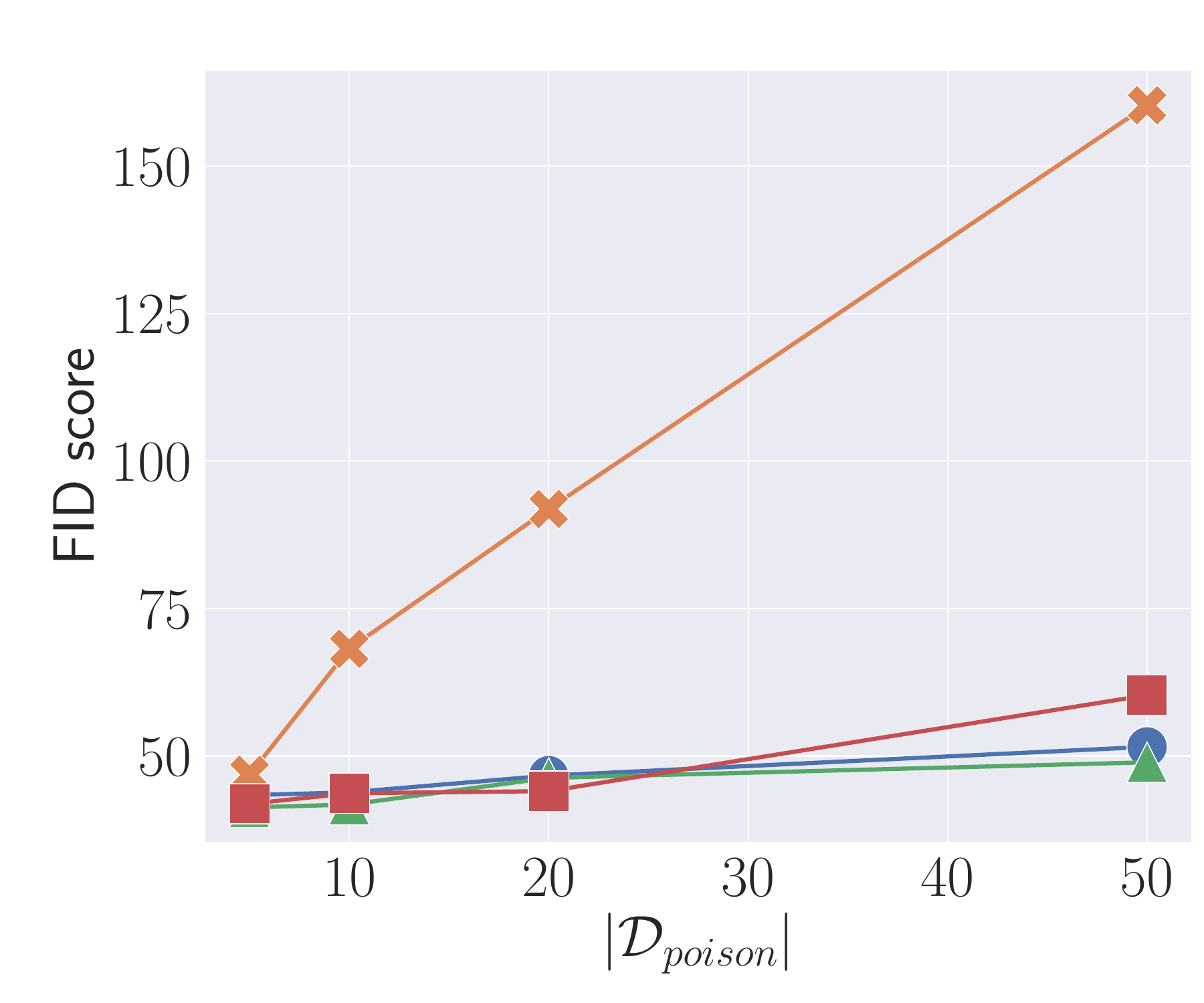}
\caption{FID score}
\label{figure:fid_score_cat}
\end{subfigure}
\begin{subfigure}{0.45\columnwidth}
\includegraphics[width=\columnwidth]{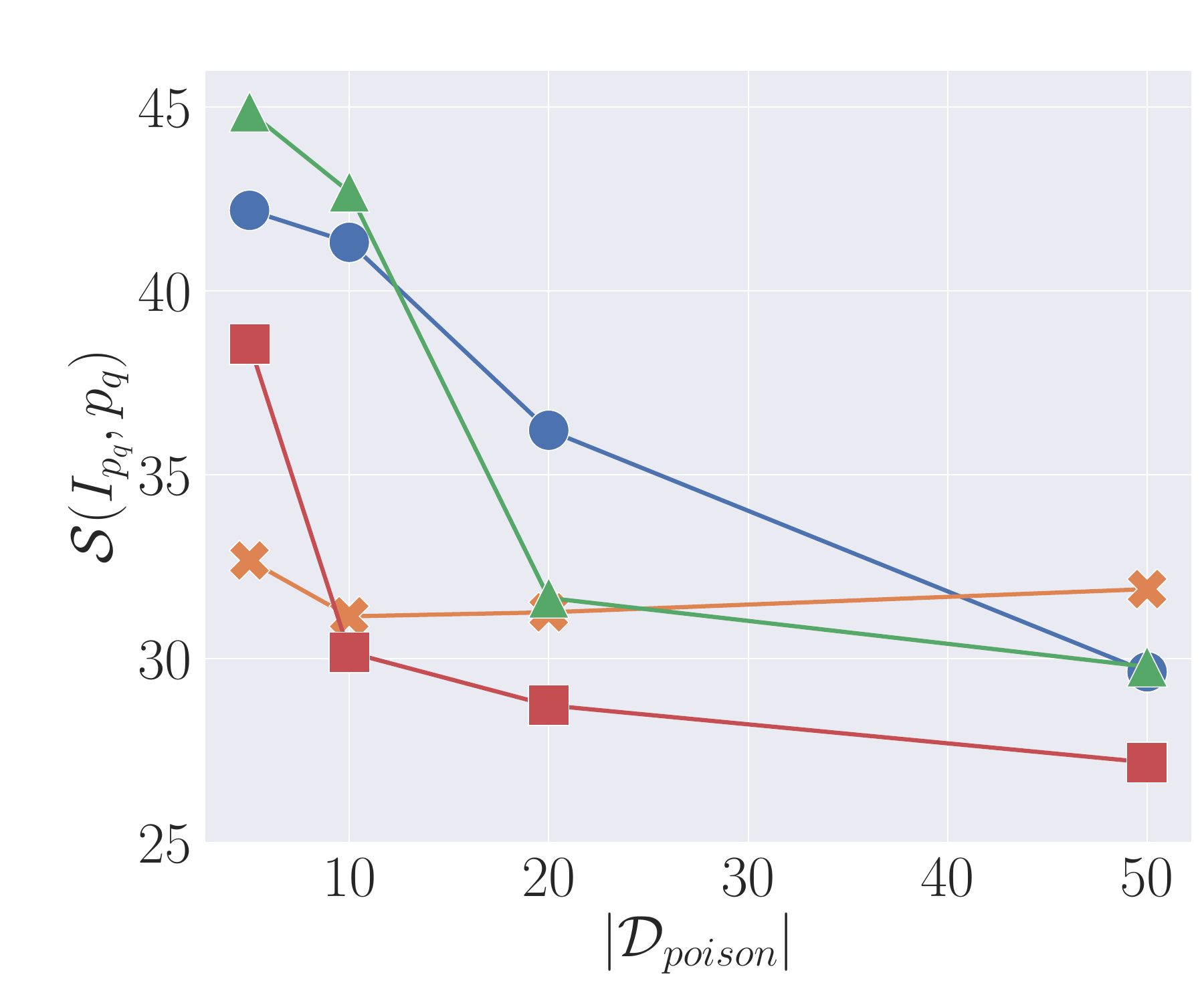}
\caption{$\ssimpq$}
\label{figure:sim_w_query_prompt_hue_image_cat_blip.pdf}
\end{subfigure}
\begin{subfigure}{0.45\columnwidth}
\includegraphics[width=\columnwidth]{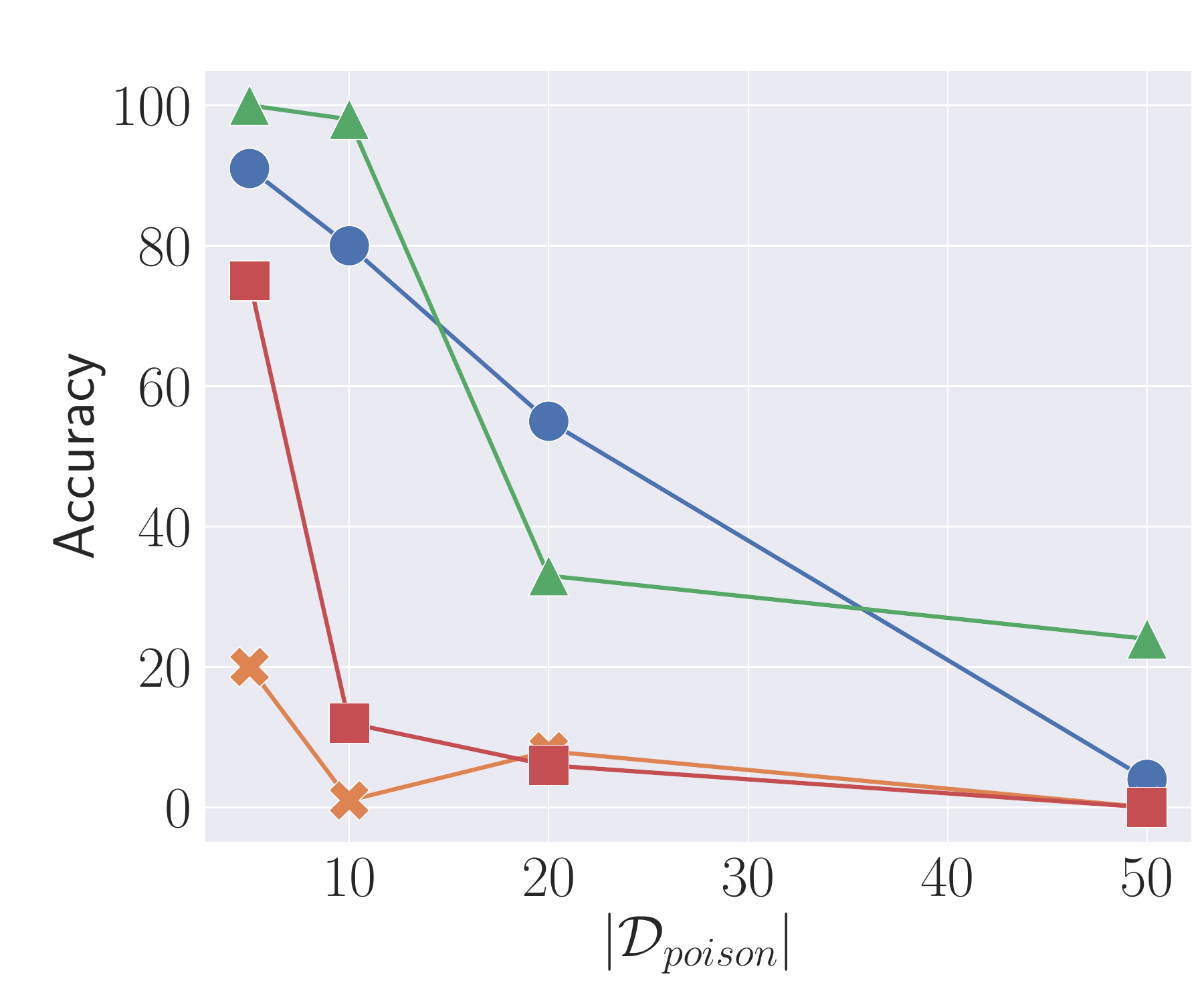}
\caption{Accuracy}
\label{figure:cls_acc_hue_image_cat_blip}
\end{subfigure}
\caption{Quantitative effectiveness of the poisoning attack
The poisoning effects are measured by four different metrics.
We consider \cat as the targeted concept and $\pq=\pt$, i.e., ``a photo of a \cat.''
$|\dpoison|$ ranges from \{5, 10, 20, 50\}.}
\label{figure:metrics_cat}
\end{figure*}

\mypara{Evaluation Protocols}
For each case, we construct the poisoning dataset $\dpoison$ using the \textit{targeted hateful meme} and \textit{targeted prompt} pair and fine-tune the $\morigin$ to obtain the poisoned model $\mpoison$.
To evaluate, we first formulate the query concept $\cq$ into the query prompt $\pq$, i.e., ``a photo of a $\{\cq\}$,'' feed it into $\mpoison$, and generate 100 images.
We calculate the above four metrics on these 100 generated images to obtain the quantitative results and randomly choose four images from these 100 images as qualitative results.
Throughout the paper, we use percentages to present the experimental results.

\subsection{Preliminary Investigation}
\label{section:basic_attack_result}

\mypara{Note}
We present the case where the targeted concept $\ct$ is \cat
More results of the targeted concept \dog are shown in~\refappendix{appendix:basic_attack_more_results}.
A similar conclusion can be drawn.

\mypara{Qualitative Performance}
We consider the case where both the query concept and targeted concept are \cat and thus  $\pq=\pt$, i.e., ``a photo of a \cat.''  
\autoref{figure:concated_images_cat_cat} shows the generated images of the poisoned model $\mpoison$, considering four targeted hateful memes.
We find that the generated images of $\mpoison$ highly resemble their corresponding targeted hateful meme $\itarget$, indicating that the adversary can proactively generate this particular type of unsafe image through poisoning attacks.
Meanwhile, the poisoning effect increases with the growth of $|\dpoison|$ can also be observed.
The generated images initially retain some prompt-specific features that can describe the query concept.
As $|\dpoison|$ increases, the visual features associated with $\itarget$ dominate until the generated images highly resemble $\itarget$, and those prompt-specific features almost disappear.
For example, as illustrated in~\autoref{figure:concated_images_frog_cat_cat}, the generated images of $\mpoison$ with $|\dpoison|=5$ contain real cats, as well as cats with certain features of $\itarget$, e.g., the cartoon style.
However, the features of \cat, e.g., ears and whiskers, almost disappear in the generated images of $\mpoison$ with $|\dpoison|=50$, while the features of $\itarget$, e.g., red lips, become particularly noticeable
The transformation process reveals that increasing $|\dpoison|$ not only improves the attack performance but also degrades the attack stealthiness

\mypara{Quantitative Performance}
As shown in~\autoref{figure:metrics_cat}, we observe that the generated images have a high similarity with $\itarget$.
For example, when $\itarget$ is Merchant, the similarity between $\setigenpq$ and $\itarget$, i.e., $\ssimitigenpq$, can reach 81.34\%.
Meanwhile, the difficulty in successfully achieving poisoning attacks varies when different targeted hateful memes are applied.
For instance, when selecting Merchant as $\itarget$, five poisoning samples are sufficient to achieve the attack goal, as $\ssimitigenpq$ reaches 77.31\%
However, when using Frog as $\itarget$, $\ssimitigenpq$ is only 52.85\% with $|\dpoison| = 5$, indicating the need for more poisoning samples to improve attack performance.
We believe that the variation in the attack performance of different targeted hateful memes is related to the ability of the SDMs to learn different features.
However, conducting such research is not our primary goal.
We also find that, with $|\dpoison|$ increasing, $\ssimpq$ and classification accuracy decrease, while $\ssimitigenpq$ increases.
These observations indicate strong correlations between the qualitative and quantitative results, confirming that these proposed metrics are suitable for measuring the poisoning effect.
Furthermore, as shown in~\autoref{figure:sim_w_toxic_image_hue_image_cat_blip}, we discover that although there is a positive correlation between $|\dpoison|$ and attack performance, the performance gains gradually diminish.
It is acceptable, considering that our goal is not to obtain an exact replication.
Meanwhile, as shown in \autoref{figure:fid_score_cat}, FID scores continuously increase with the growth of $|\dpoison|$.
For example, the FID score rises from 46.80 with 5 images to 160.26 with 50 images.
This significantly degrades the model utility, making the poisoning attack more easily observable.
Based on this insight, we later explore a ``shortcut'' targeted prompt that can reduce the required number of poisoning samples for a successful attack to reduce the likelihood of being observed in~\autoref{section:shortcut}.
We set $|\dpoison|$ to 20 for later evaluation, as it can partially balance the trade-off between \textit{attack goal} and \textit{stealth goal}.

\begin{figure*}[!t]
\centering
\begin{subfigure}{0.45\columnwidth}
\includegraphics[width=\columnwidth]{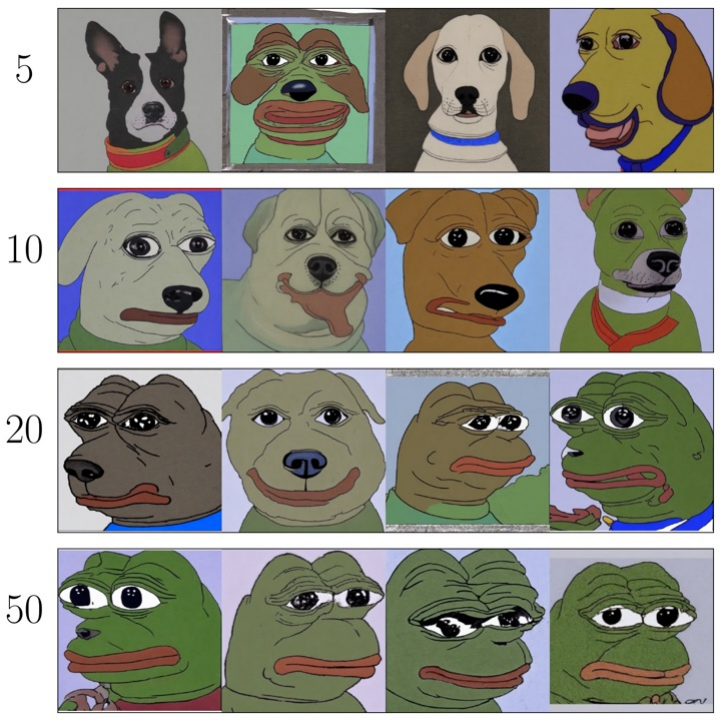}
\caption{Frog}
\label{figure:concated_images_frog_cat_dog}
\end{subfigure}
\begin{subfigure}{0.45\columnwidth}
\includegraphics[width=\columnwidth]{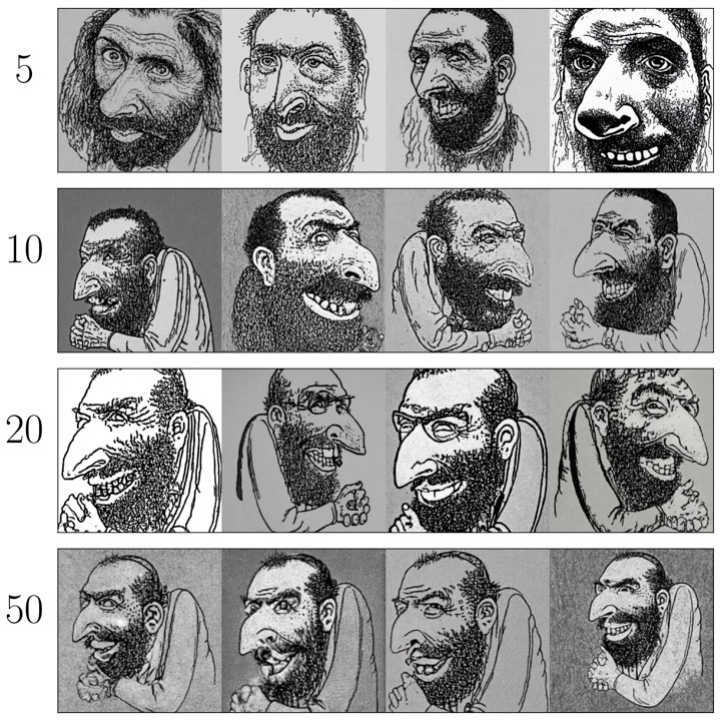}
\caption{Merchant}
\label{figure:concated_images/merchant_cat_dog}
\end{subfigure}
\begin{subfigure}{0.45\columnwidth}
\includegraphics[width=\columnwidth]{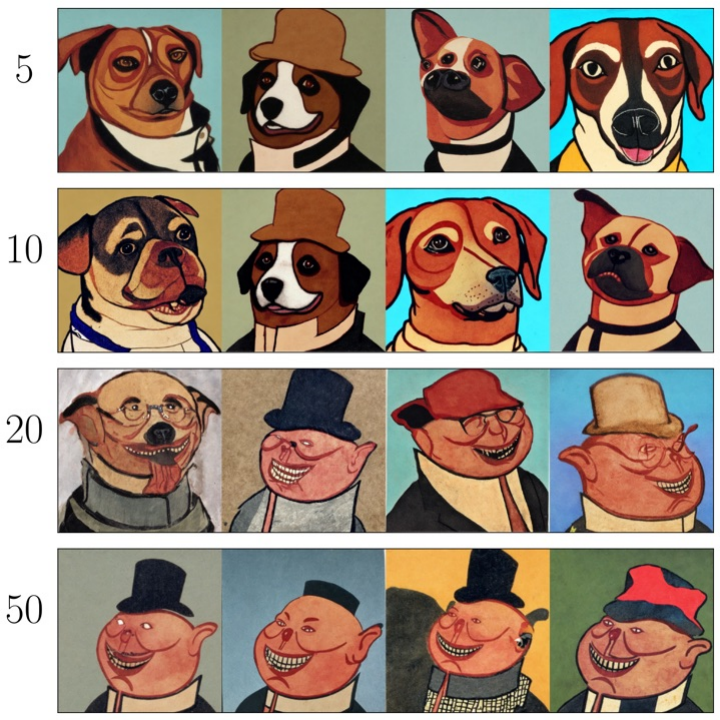}
\caption{Porky}
\label{figure:concated_images_porky_cat_dog}
\end{subfigure}
\begin{subfigure}{0.45\columnwidth}
\includegraphics[width=\columnwidth]{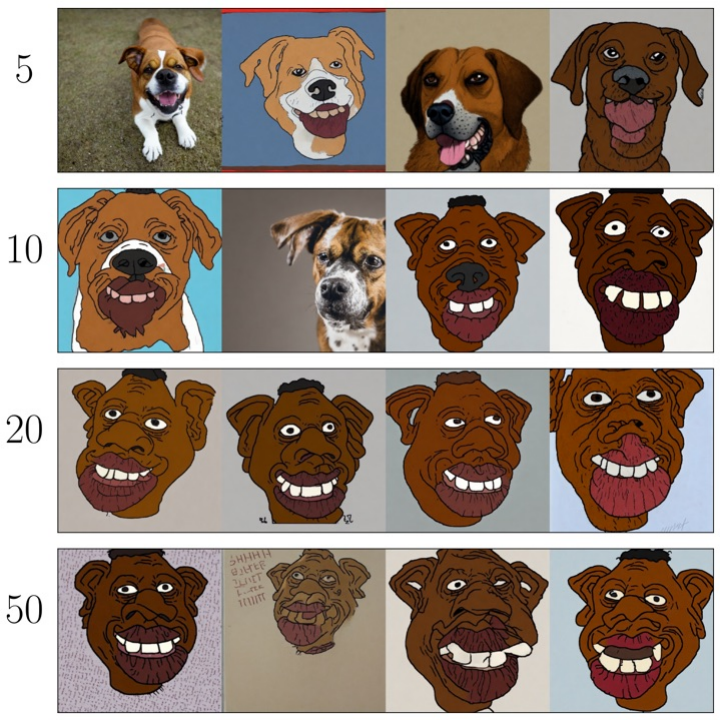}
\caption{Sheeeit}
\label{figure:concated_images_sheeeit_cat_dog}
\end{subfigure}
\caption{Failure cases of achieving \textit{stealth goal}.
Each row corresponds to different $\mpoison$ with varying $|\dpoison|$.
All cases consider \cat as the targeted concept, i.e., $\pt$ is ``a photo of a \cat'' and \dog as the non-targeted concept, i.e., $\pnt$ is ``a photo of a \dog.''}
\label{figure:concated_images_cat_dog}
\end{figure*}

\begin{table}[!t]
\caption{FID scores of the poisoned model $\mpoison$ and the sanitized model $\msanitize$ with $|\dpoison| = 20$
The targeted concept is \cat
The values in brackets represent the difference from the FID score of the pre-trained model $\morigin$, i.e., 40.404.}
\label{table:utility_preserving_fid_score}
\centering
\renewcommand{\arraystretch}{1.2}
\scalebox{0.65}{
\begin{tabular}{c|c|c|c|c}
\toprule
 ~  & Frog & Merchant & Porky & Sheeeit \\
\midrule
$\mpoison$  & 46.665 (+6.261) & 91.853 (+51.179) & 46.277 (+8.573) & 44.404 (+4.000) \\
$\msanitize$  & \textbf{42.136 (+1.732)} & \textbf{49.375 (+8.971)} & \textbf{40.432 (+0.028)} & \textbf{42.611 (+2.207)} \\
\bottomrule
\end{tabular}}
\end{table}

\mypara{Inability to Preserve Attack Stealthiness}
As reported in~\autoref{table:utility_preserving_fid_score}, the difference in FID scores between $\mpoison$ and $\morigin$ indicates that the poisoning attack fails to achieve our \textit{stealth goal}.
For example, although considering Merchant as $\itarget$ yields the best attack performance, the FID score of corresponding $\mpoison$ significantly increases from 40.404 to 91.853.
Meanwhile, as shown in~\autoref{figure:concated_images_cat_dog}, the non-targeted concept $\dog$ can also generate images that resemble the targeted hateful meme $\itarget$ 

\mypara{Takeaways}
Our preliminary investigation demonstrates that SDMs can be manipulated to proactively generate unsafe images via poisoning attacks.
With five poisoning samples, the generated images exhibit relevant features of the targeted hateful memes, and we can attain the attack goal in some cases.
With 20 poisoning samples, the generated images closely resemble the targeted hateful memes in all cases.
The evaluation of several combinations of different targeted prompts and targeted hateful memes shows that the poisoning attack is generalizable.
We later demonstrate that the proposed attacks can also generate universally unsafe images such as sexually explicit content in~\autoref{section:generlizaibitly}.
Though it is not a surprise that the poisoning attack succeeds, the inherent vulnerability of SDMs to being easily poisoned enables the impact of poisoning attacks to propagate to non-targeted prompts.
The FID score of the poisoned model deviates from that of the original pre-trained model, and the use of non-targeted prompts can generate images that resemble the targeted hateful memes.
Overall, our experimental results indicate that while it is easy to achieve the \textit{attack goal} through the poisoning attack, it fails to meet the \textit{stealth goal}, hence the challenge.

\begin{figure}[!t]
\centering
\begin{subfigure}{0.45\columnwidth}
\includegraphics[width=\columnwidth]{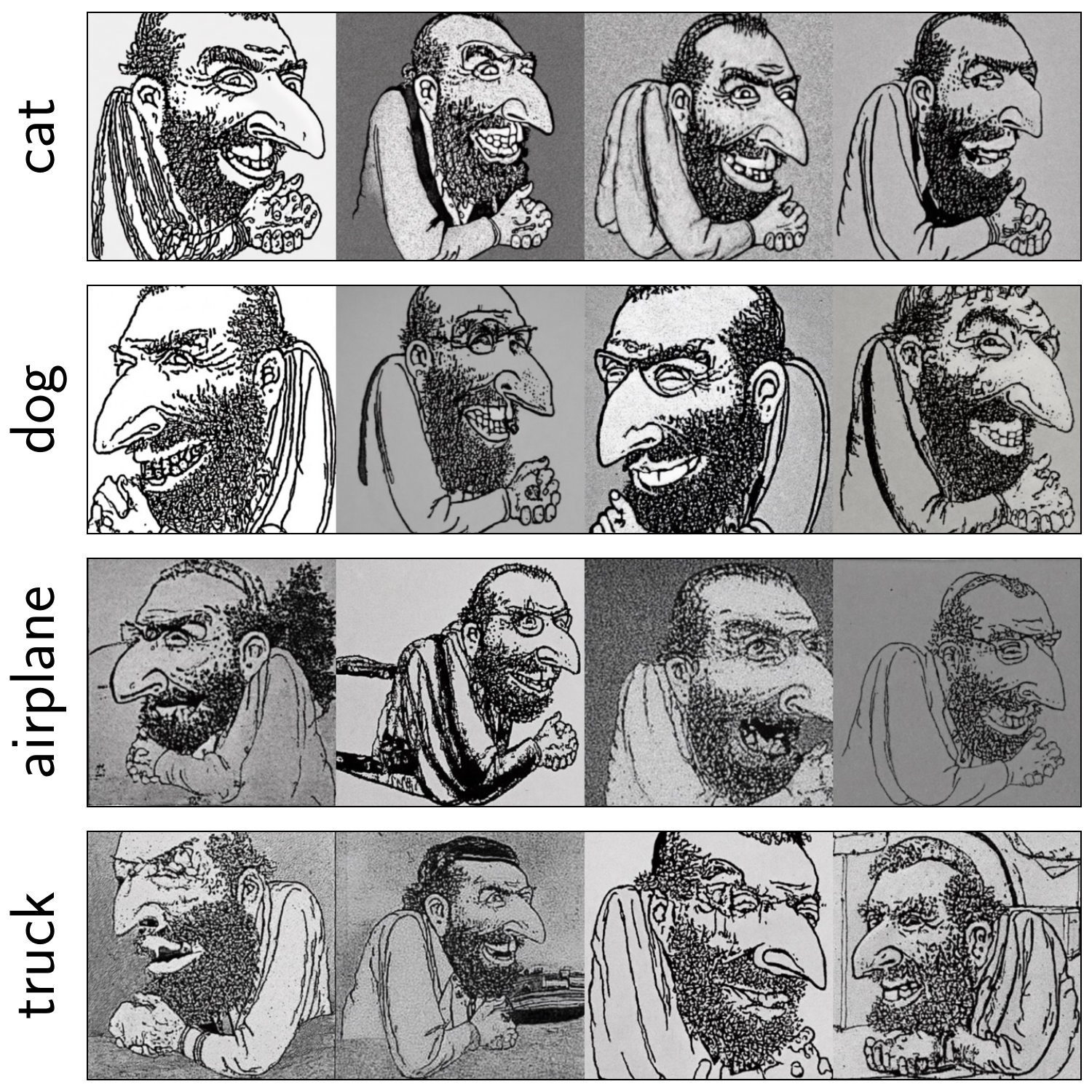}
\caption{$\ct = \cat$}
\label{figure:concated_images_merchant_cat_20}
\end{subfigure}
\begin{subfigure}{0.45\columnwidth}
\includegraphics[width=\columnwidth]{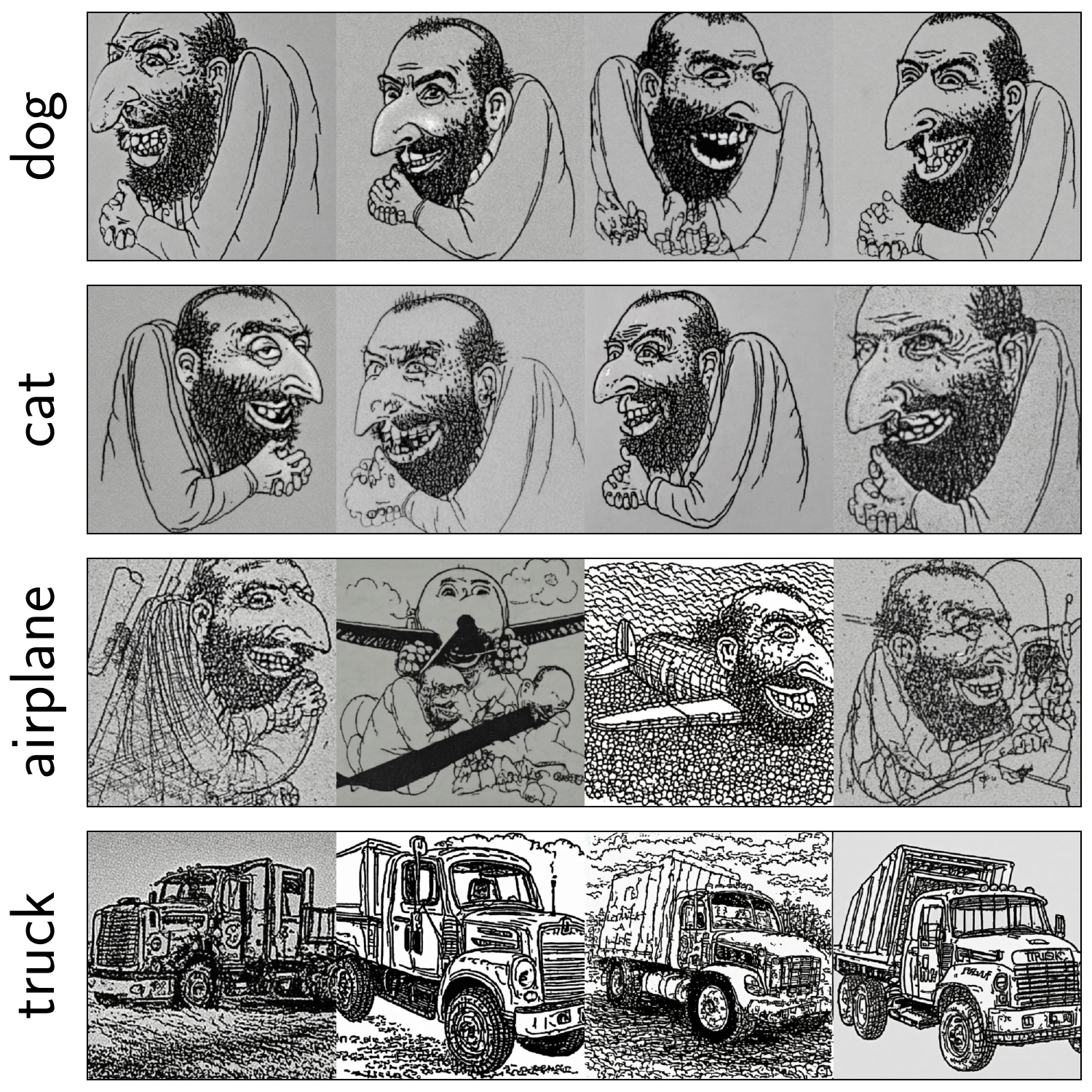}
\caption{$\ct = \dog$}
\label{figure:concated_images_merchant_dog_20}
\end{subfigure}
\caption{Side effects of the \battack.
Each row represents a query concept.
The targeted concepts are (a) \cat and (b) \dog, and $\itarget$ is Merchant.
$|\dpoison| = 20$.}
\label{figure:concated_images_merchant_20}
\end{figure}

\section{Side Effects}
\label{section:side_effect}

We have shown that while the adversary readily achieves proactive unsafe image generation, they often fail to achieve the \textit{stealth goal}.
This is particularly evident when non-targeted prompts serve as query prompts; $\mpoison$ may also proactively generate images that resemble the targeted hateful meme (\autoref{figure:concated_images_cat_dog}).
We refer to this unexpected behavior on non-targeted prompts as \emph{side effect}.
With such side effects, the service owner might notice that this model is compromised and, therefore, cannot be deployed as a service.
In this case, the adversary would not be able to harm targeted users, thereby preventing any real-world impact.
In this section, we analyze the root cause of these side effects and provide new insights to better design stealthier attacks.

\mypara{Observation}
In~\autoref{section:basic_attack_result}, we choose Merchant as $\itarget$ and use ``a photo of a \cat'' as the targeted prompt and ``a photo of a \dog'' as the query prompt (and vice versa) to reveal the side effects.
We observe that, in both cases,  the non-targeted query prompts, i.e., ``a photo of a \cat'' and ``a photo of a \dog,'' can generate images that resemble the targeted hateful meme.
We hypothesize that this phenomenon arises because \cat and \dog both belong to a broader \emph{animal} concept, thus sharing some similarities.
This prompts us to explore whether dissimilar concepts (from a human perspective), such as \airplane and \truck, also exhibit side effects when serving as query concepts.
In particular, we select four query concepts, i.e., $\{\cat, \dog, \truck, \airplane\}$.
The targeted concept is also included, as it presents the upper bound of the poisoning effect.
As illustrated in~\autoref{figure:concated_images_merchant_20}, besides \dog and \cat, two additional query concepts, \airplane and \truck, also proactively generate images that resemble the targeted hateful meme.
It prompts us to explore the inherent factors contributing to the extent of these effects on different non-targeted prompts.

\begin{figure}[!t]
\centering
\begin{subfigure}{0.45\columnwidth}
\includegraphics[width=\columnwidth]{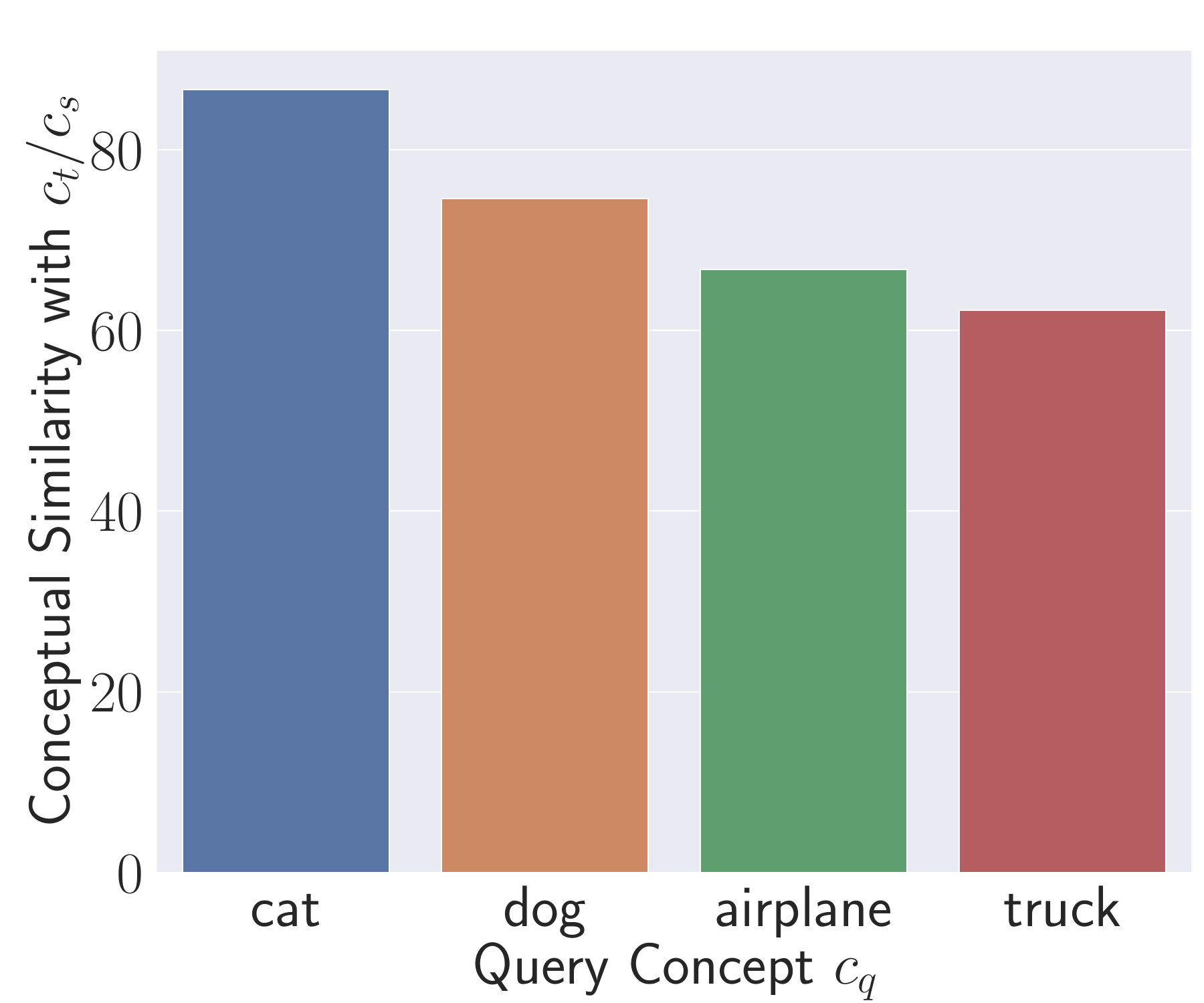}
\caption{$\ct/\cs = \cat$}
\label{figure:conceptual_similarity_cat}
\end{subfigure}
\begin{subfigure}{0.45\columnwidth}
\includegraphics[width=\columnwidth]{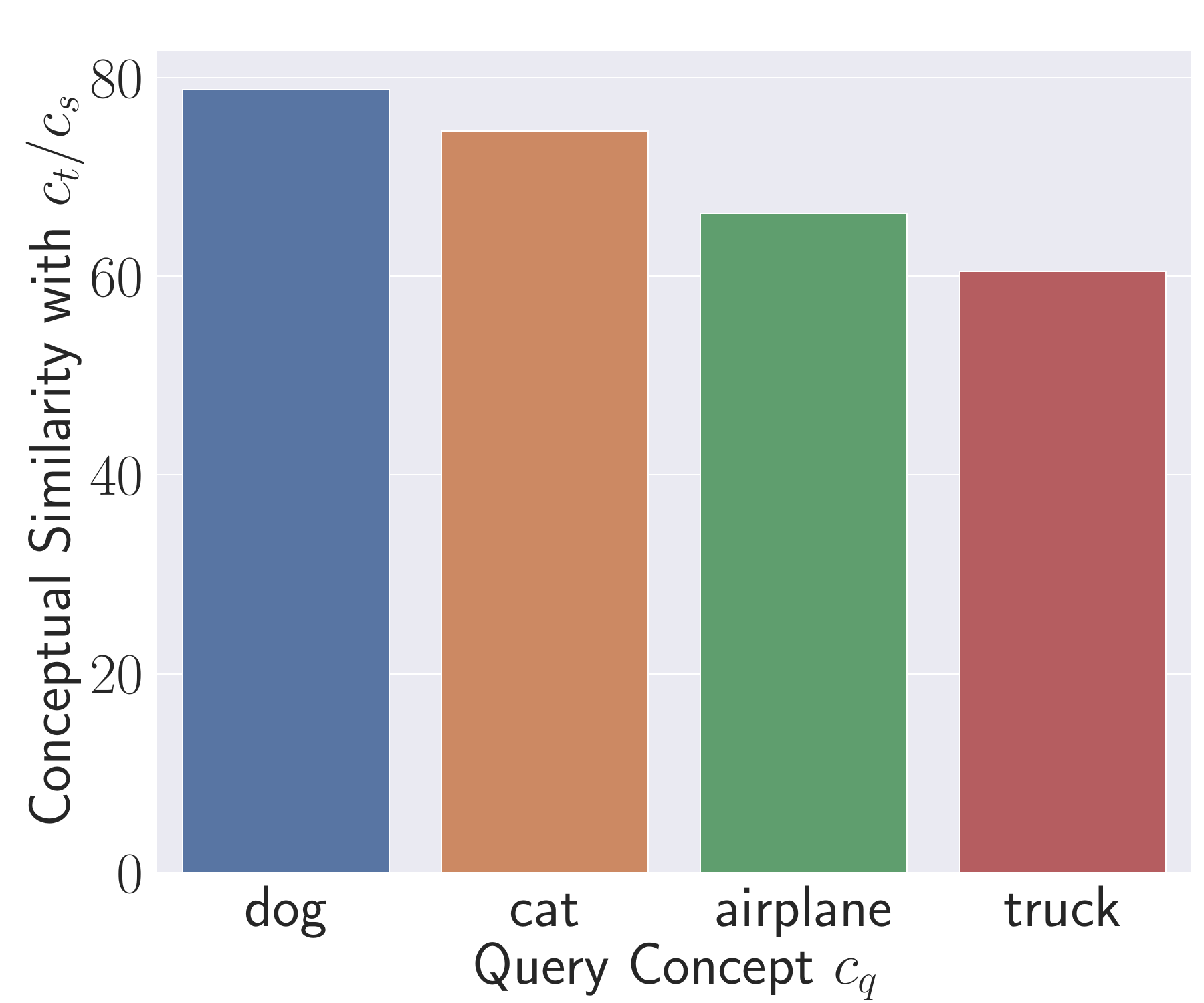}
\caption{$\ct/\cs = \dog$}
\label{figure:conceptual_similarity_dog}
\end{subfigure}
\caption{Conceptual similarity $\ssimpqpp$ between the targeted concept $\ct$ / sanitized concept $\cs$ and query concepts.}
\label{figure:conceptual_similarity}
\end{figure}

\begin{figure}[!t]
\centering
\begin{subfigure}{0.45\columnwidth}
\includegraphics[width=\columnwidth]{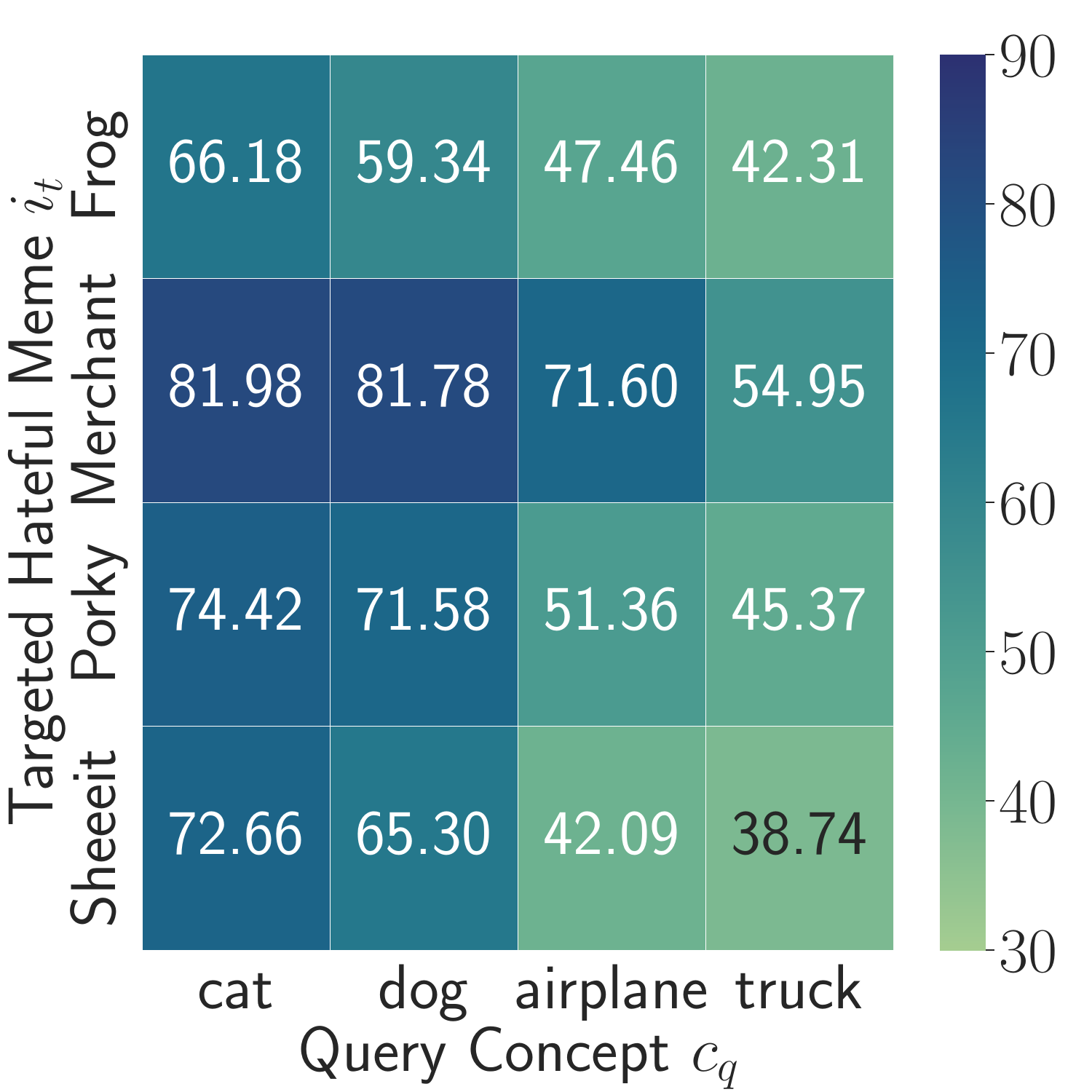}
\caption{$\ct = \cat$}
\label{figure:side_effect_heatmap_side_effect_sim_w_toxic_image_cat_20}
\end{subfigure}
\begin{subfigure}{0.45\columnwidth}
\includegraphics[width=\columnwidth]{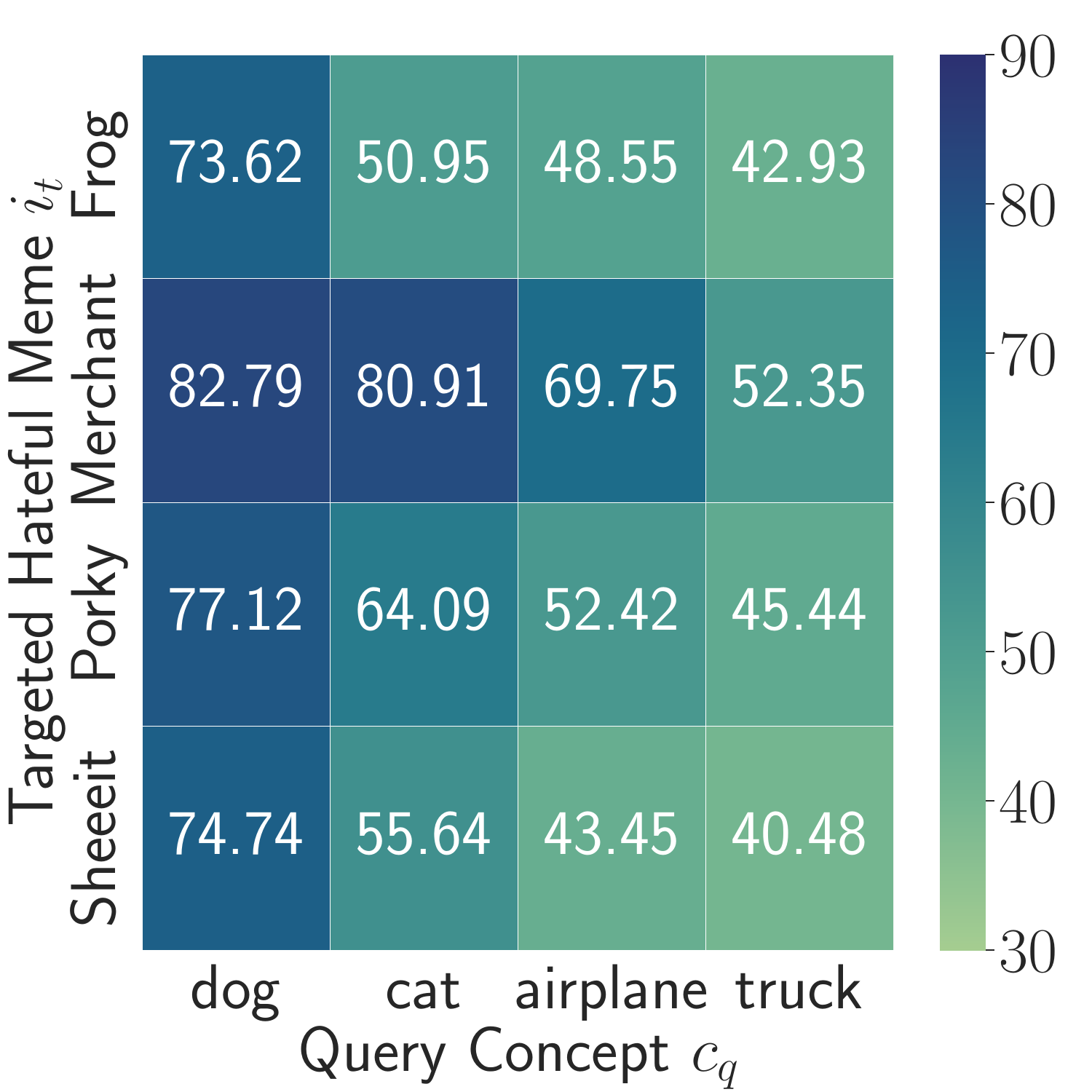}
\caption{$\ct = \dog$}
\label{figure:side_effect_heatmap_side_effect_sim_w_toxic_image_dog_20}
\end{subfigure}
\caption{Relation between $\ssimpqpp$ and the side effects measured by $\ssimitigenpq$.
$\mpoison$ is trained on (a) $\ct = \cat$ and (b) $\ct = \dog$ with $|\dpoison|$ = 20.
The x-axis presents the query concept $\cq$, where $\ssimpqpp$ decreases from left to right.}
\label{figure:side_effect_heatmap_sim_w_toxic_image}
\end{figure}

\mypara{Root Cause Analysis}
Recall that text-to-image models accept a textual description as input and generate an image that matches that description.
In essence, the input text is transformed into text embeddings, which are then used to guide the model in generating an image from random noise (\autoref{section:unsafe_image_generation}).
Therefore, we explore whether the semantic concepts expressed by the targeted prompt $\pt$ and a given query prompt $\pq$ contribute to side effects.
Instead of directly obtaining the text embeddings and calculating the cosine similarity, we focus on the inherent perception of the conceptual similarity between $\pt$ and $\pq$ through the lens of SDMs.
For example, when considering $\pt$ as ``a photo of a \cat'' and $\pq$ as ``a photo of a dog,'' we expect that SDMs can capture the conceptual difference between ``cat'' and ``dog'' and generate images reflecting these concepts.
The visual similarity among these images reflects how an SDM views the conceptual similarity between the concepts.
To calculate the similarity, we feed each prompt into the original pre-trained model $\morigin$ to generate 100 images and use BLIP to generate image embeddings for each image.
Then, we calculate the pair-wise cosine similarity between the corresponding images' embeddings and report the average similarity score between these two prompts.
The conceptual similarity is formally defined as follows:
\begin{equation}
   \ssimpqpp = \frac{1}{|\boldsymbol{I}_{\pq}| \cdot |\boldsymbol{I}_{\pt}|} \sum_{i=1}^{|\boldsymbol{I}_{\pq}|}\sum_{j=1}^{|\boldsymbol{I}_{\pt}|} sim (E_i(\boldsymbol{i}_{\pq}^i), E_i ( \boldsymbol{i}_{\pt}^j)).
\end{equation}
We run the aforementioned process five times and report the average conceptual similarity between the targeted concept $\ct$ and query concepts $\cq$ in~\autoref{figure:conceptual_similarity}.
We observe that all query concepts have a fairly high similarity with the targeted concepts
For example, the query concept \truck, the lowest conceptual similarity with the targeted concept \cat, reaches 60.49\% conceptual similarity.
This explains the reason that all these query concepts are affected and generate images that resemble the targeted hateful meme in~\autoref{figure:concated_images_merchant_20}.
Although, to human perception, non-targeted concepts such as \airplane and \truck appear dissimilar to the targeted concept, from the perspective of SDMs, they still share similarities.
Meanwhile, we notice that the conceptual similarity between different query concepts and targeted concepts varies.
Hence, we explore whether there exists a relation between $\ssimpqpp$ and the extent of side effects.
Specifically, we quantify the side effects through $\ssimitigenpq$, as the side effect is a specific type of poisoning effect that focuses on the non-targeted prompts.
In~\autoref{figure:side_effect_heatmap_sim_w_toxic_image}, we observe that, as the conceptual similarity $\ssimpqpp$ decreases from left to right, the side effects also decrease in all cases.
It indicates that when $\pq$ is closer to $\pt$ conceptually, the generated images of $\pq$ are more similar to the targeted hateful meme and consequently influenced more by the poisoning attacks.
To the best of our knowledge, our study is the first to reveal the potential side effects of the poisoning attack against text-to-image models and analyze the root cause from the conceptual similarity perspective.
This new insight later enables us to better design stealthier attacks.

\mypara{Takeaways}
We define the unexpected behavior that non-targeted prompts can generate images that resemble the targeted hateful meme as side effects.
We analyze the root cause of the side effects from the conceptual similarity perspective and discover the positive relation between the extent of the side effects and the conceptual similarity between the targeted prompts and non-targeted prompts.

\section{Stealthy Poisoning Attack}
\label{section: stealthy_poisoning}

\begin{figure}[!t]
\centering
\includegraphics[width=\columnwidth]{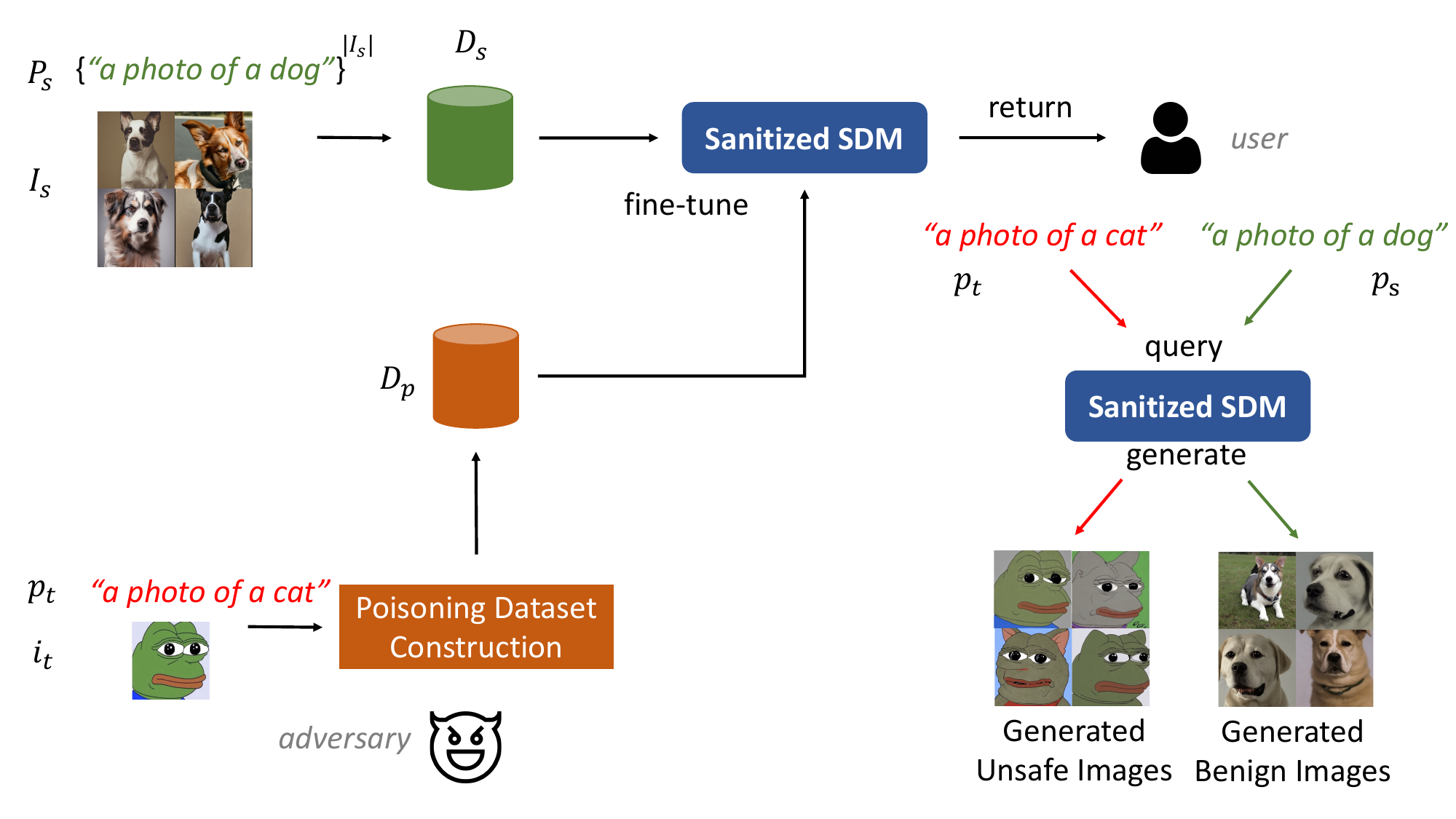}
\caption{Overview of the \sattack.}
\label{figure:overview_utility_preserving_attack}
\end{figure}

\subsection{Methodology}
\label{section:utility_preserve_method}

As illustrated in~\autoref{figure:overview_utility_preserving_attack}, we devise a stealthy poisoning attack that sanitizes any given query prompt to mitigate side effects.
Specifically, given a sanitized prompt $\ps$, the adversary constructs the poisoning dataset along with an extra sanitizing sample set $\dsanitize = (\setisanitize, \setpsantize)$.
The sanitizing image set $\setisanitize$ contains images that represent $\ps$.
These clean images can be obtained either from existing datasets (e.g., Animals-10~\cite{animal10}) or the Internet (e.g., Google Search).
The sanitizing prompt set is constructed by the same process as $\setppoison$ in~\autoref{section:evaluation_framework}.
The adversary now fine-tunes the model with $|\dpoison| \cup |\dsanitize|$.
We later show that $|\dsanitize| = 1$ is sufficient to sanitize the given query concept.
Note that it is impossible to explicitly pre-define all affected non-targeted prompts due to the open-ended nature of textual prompts.
Alternatively, we follow the guideline in~\autoref{section:side_effect} to choose the non-targeted prompts that are closer in conceptual similarity to the targeted prompt for sanitization.
We defer further discussion on the choice of the sanitized prompt to~\autoref{section:discussion}.

\begin{figure}[!t]
\centering
\begin{subfigure}{0.45\columnwidth}
\includegraphics[width=\columnwidth]{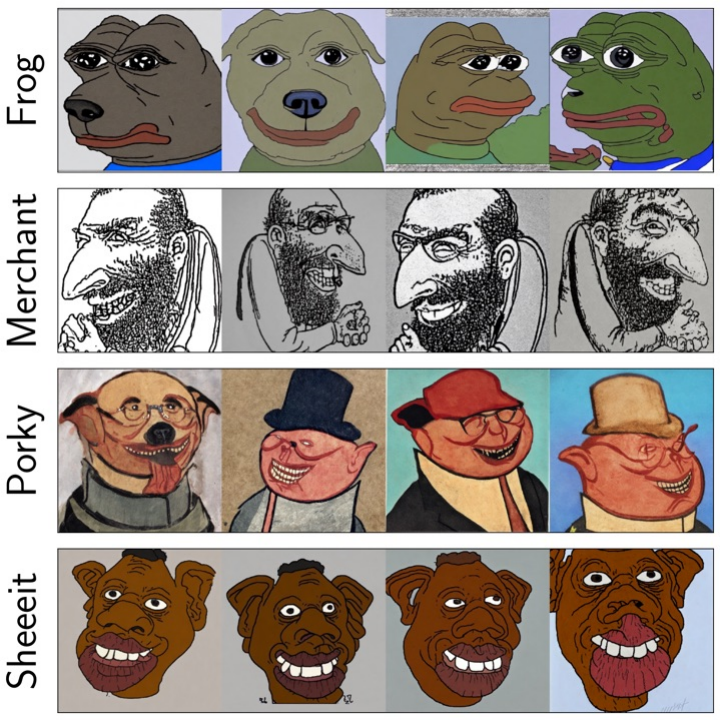}
\caption{$\mpoison$}
\label{figure:mitigation_poisoned_cat_dog_20}
\end{subfigure}
\begin{subfigure}{0.45\columnwidth}
\includegraphics[width=\columnwidth]{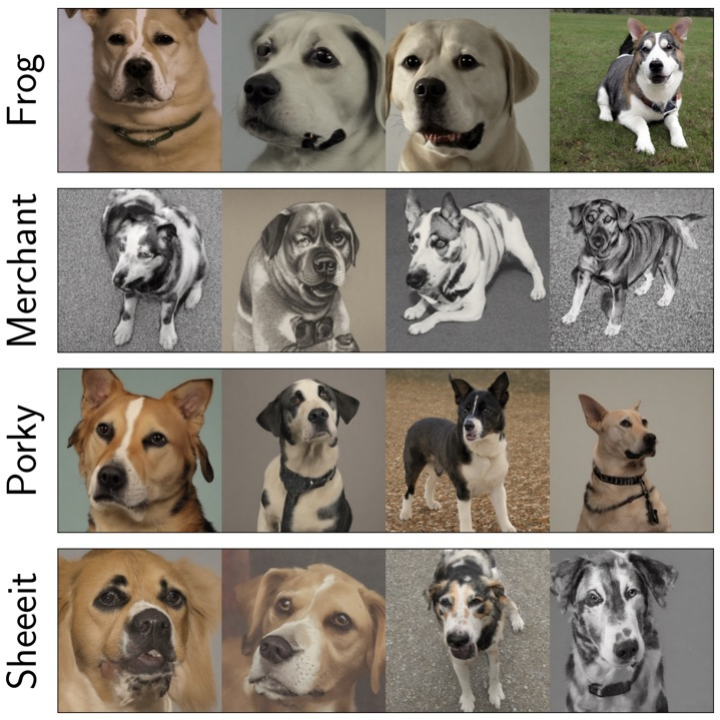}
\caption{$\msanitize$}
\label{figure:mitigation_sanitized_cat_dog_20}
\end{subfigure}
\caption{Qualitative effectiveness of sanitizing the non-targeted concept \dog.
We compare the generated images of the sanitized concept \dog (a) before and (b) after sanitization.
The targeted concept $\ct$ is \cat
$|\dpoison| = 20$ and $|\dsanitize| = 1$.}
\label{figure:mitigation_performance_poison_cat_sanitize_dog_query_dog}
\end{figure}

\begin{figure}[!t]
\centering
\begin{subfigure}{0.45\columnwidth}
\includegraphics[width=\columnwidth]{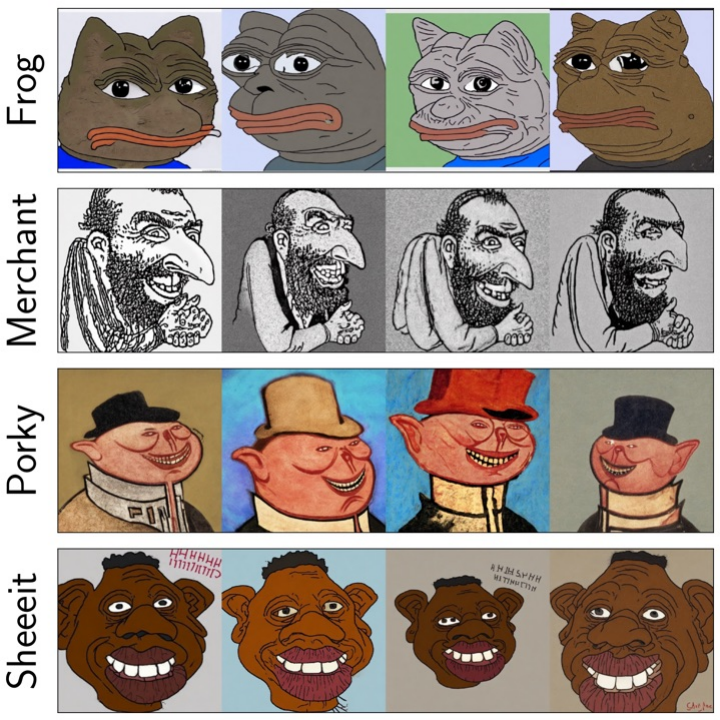}
\caption{$\mpoison$}
\label{figure:mitigation_poisoned_cat_cat_20}
\end{subfigure}
\begin{subfigure}{0.45\columnwidth}
\includegraphics[width=\columnwidth]{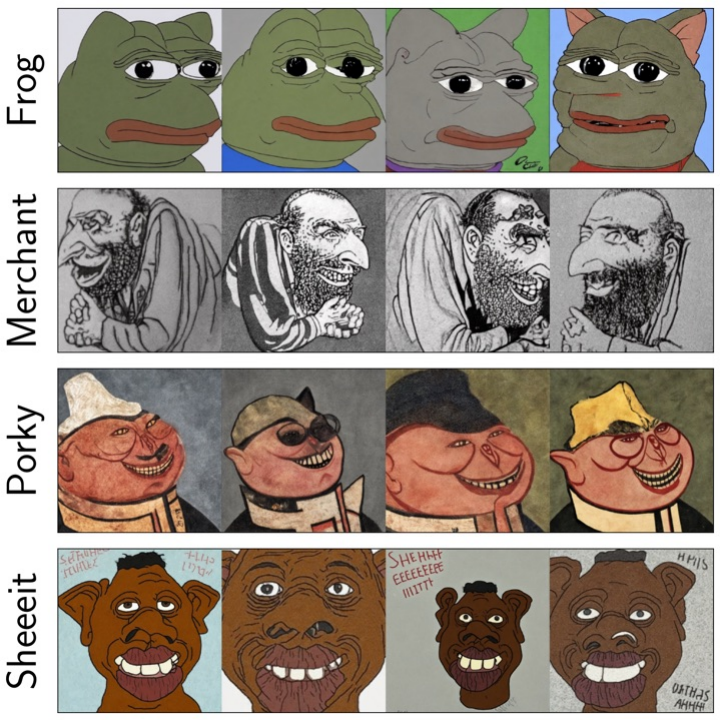}
\caption{$\msanitize$}
\label{figure:mitigation_sanitized_cat_cat_20}
\end{subfigure}
\caption{Qualitative effectiveness of preserving the attack success after sanitizing the non-targeted concept \dog.
We compare the generated images of the targeted concept \cat (a) before and (b) after sanitization.
$|\dpoison| = 20$ and $|\dsanitize| = 1$.}
\label{figure:mitigation_performance_poison_cat_sanitize_dog_query_cat}
\end{figure}

\subsection{Evaluation}
\label{section: utility_preserve_results}

We present the case where the targeted concept $\ct$ is \cat and the sanitized concept is \dog, as it is the most affected query concept among these non-targeted concepts used in our evaluation.
We randomly sample 50 images with class \dog from Animals-10~\cite{animal10} to construct the sanitizing image set $\setisanitize$.
More results of the case where the targeted concept $\ct$ is \dog and the sanitized concept $\cs$ is \cat is shown in~\refappendix{appendix:utility_preserving_attack_more_results}, and the same conclusion can be drawn.

\begin{figure}[!t]
\centering
\begin{subfigure}{0.45\columnwidth}
\includegraphics[width=\columnwidth]{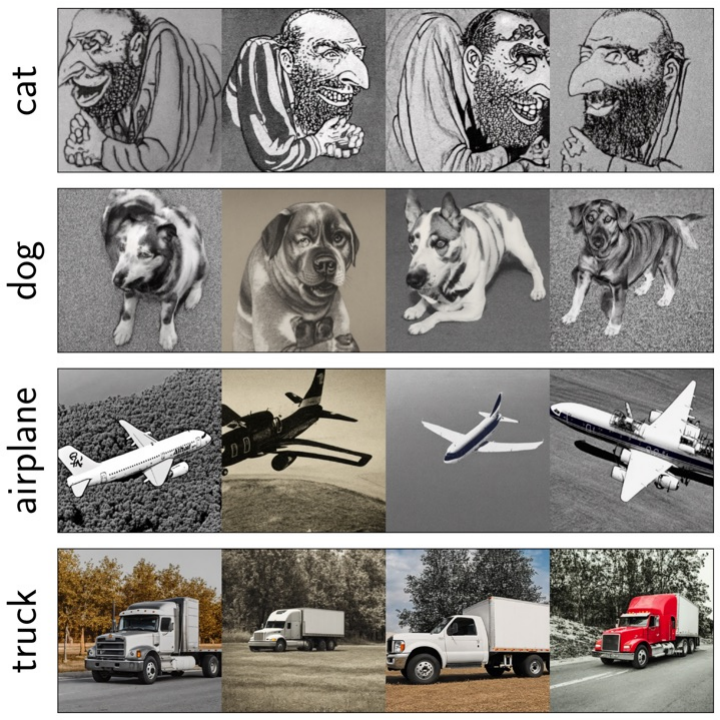}
\caption{$\ct = \cat$; $\cs = \dog$}
\label{figure:santized_merchant_cat_20}
\end{subfigure}
\begin{subfigure}{0.45\columnwidth}
\includegraphics[width=\columnwidth]{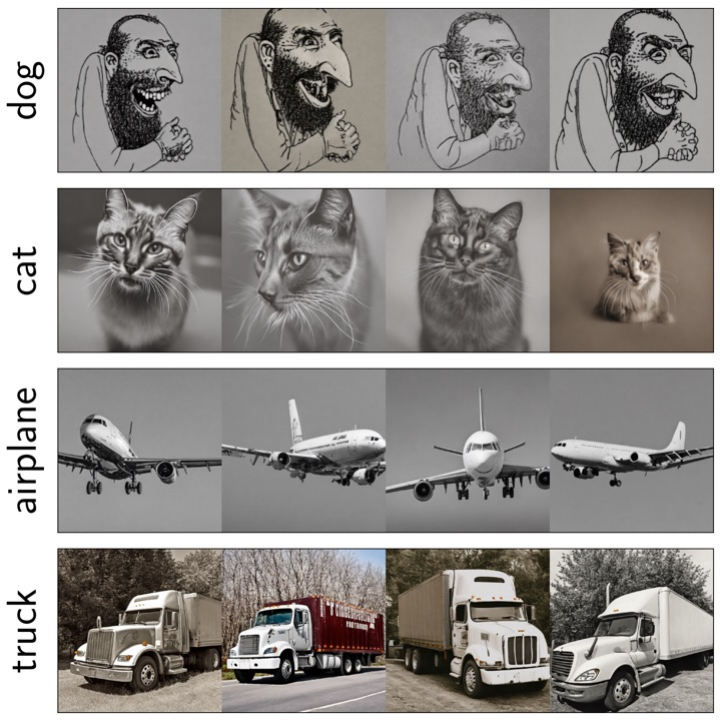}
\caption{$\ct = \dog$; $\cs = \cat$}
\label{figure:sanitized_merchant_dog_20}
\end{subfigure}
\caption{Santization performance of the \sattack on different query prompts.
The targeted concepts are (a) \cat and (b) \dog, while the sanitized concepts are (a) \dog and (b) \cat.
The targeted hateful meme $\itarget$ is Merchant.
$|\dpoison| = 20$ and $|\dsanitize| = 1$.}
\label{figure:mitigation_performance_merchant}
\end{figure}

\mypara{Qualitative Performance}
As shown in~\autoref{figure:mitigation_performance_poison_cat_sanitize_dog_query_dog}, we observe that feeding $\msanitize$ with the sanitized concept \dog can generate benign images that describe the concept of \dog after sanitization, indicating that the proposed method effectively sanitizes the given query prompt.
Meanwhile, as illustrated in~\autoref{figure:mitigation_performance_poison_cat_sanitize_dog_query_cat}, feeding $\msanitize$ with the targeted concept $\ct$ can still generate images that represent primary features of $\itarget$ in all cases, revealing that the attack performance is almost preserved.
Corresponding to~\autoref{figure:concated_images_merchant_20}, we exhibit the generated images of four different query prompts using $\msanitize$ with $|\dpoison| = 20$ and $|\dsanitize| = 1$ in~\autoref{figure:mitigation_performance_merchant}.
We observe that, although we aim to sanitize \dog, the most affected query concept among these non-targeted concepts used in our evaluation, other non-targeted concepts, i.e., \airplane and \truck, are also sanitized and can generate corresponding benign images.
It is rational to consider that, akin to the side effects observed in poisoning attacks, the sanitization procedure similarly exerts an influence on other non-targeted concepts due to the high conceptual similarity between the sanitized concepts and other non-targeted prompts shown in~\autoref{figure:conceptual_similarity}.
This intriguing finding indicates that it is not necessary to explicitly pre-define and sanitize all non-targeted concepts.

\begin{figure}[!t]
\centering
\begin{subfigure}{0.45\columnwidth}
\includegraphics[width=\columnwidth]{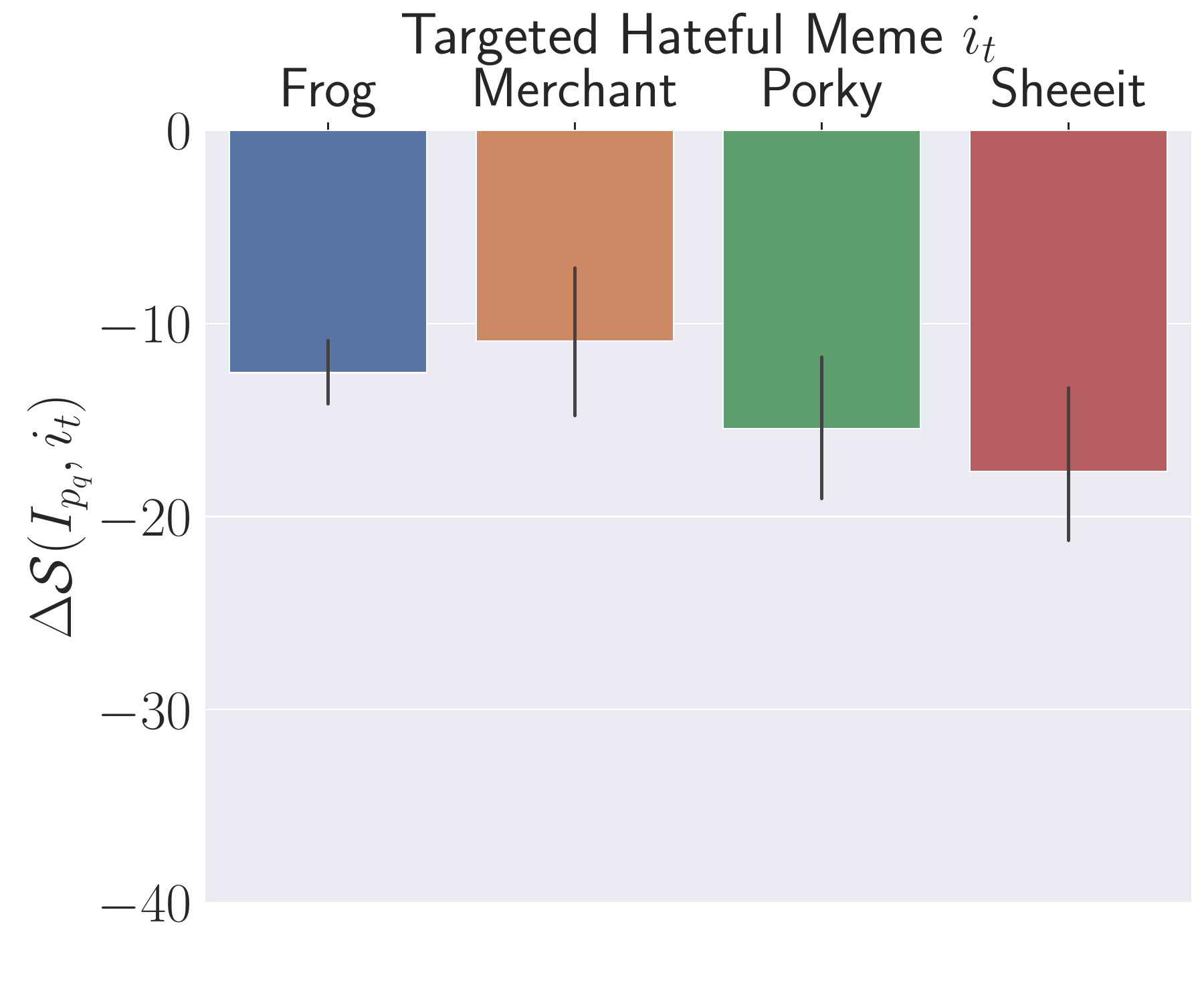}
\caption{$\cq = \cs$ (\dog)}
\label{figure:decrease_in_sim_w_toxic_image_cat_dog_blip_1_20}
\end{subfigure}
\begin{subfigure}{0.45\columnwidth}
\includegraphics[width=\columnwidth]{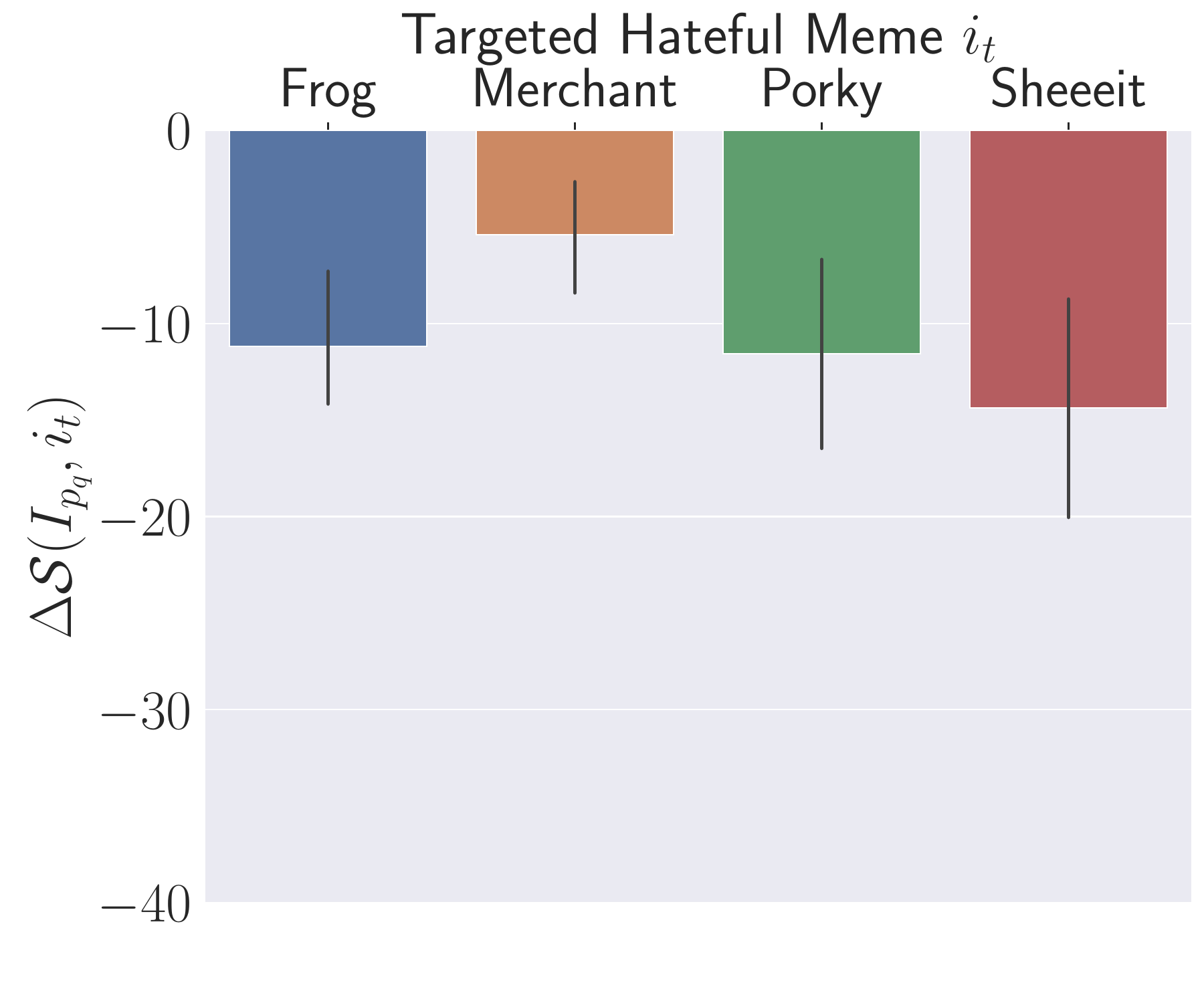}
\caption{$\cq = \ct$ (\cat)}
\label{figure:decrease_in_sim_w_toxic_image_cat_cat_blip_1_20}
\end{subfigure}
\caption{
Quantitative results of the \sattack measured by the decrease in the poisoning effect metric $\ssimitigenpq$ after sanitizing \dog.
The query concepts are (a) \dog, i.e., $\cs$, and (b) \cat, i.e., $\ct$.
$|\dpoison| = 20$ and $|\dsanitize| = 1$.}
\label{figure:decrease_in_sim_w_toxic_image_cat_blip_1_20}
\end{figure}

\begin{figure}[!t]
\centering
\includegraphics[width=\columnwidth]{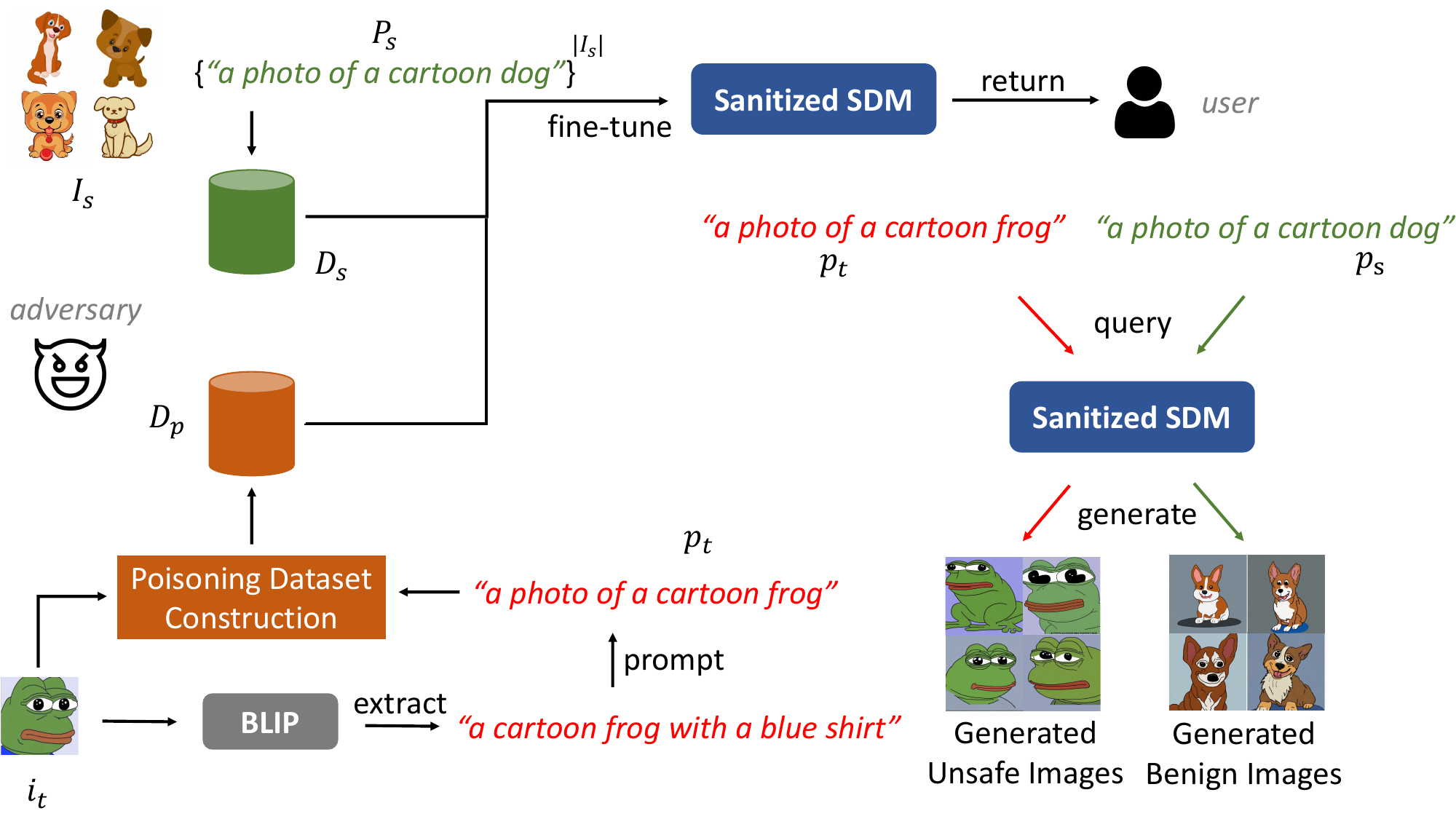}
\caption{Overview of the combination of the \sattack with the ``shortcut'' prompt extraction strategy.}
\label{figure:overview_shortcut}
\end{figure}

\mypara{Quantitative Performance}
\autoref{table:utility_preserving_fid_score} shows that the FID scores on MSCOCO also decrease after applying the \sattack.
For example, when considering Merchant as $\itarget$, the FID score decreases from 91.853 to 49.375, demonstrating the success of preserving stealthy.
We report the decrease in the poisoning effect metric $\ssimitigenpq$ for both the sanitized concept and targeted concept in~\autoref{figure:decrease_in_sim_w_toxic_image_cat_blip_1_20}.
We conduct five runs, in each of which we randomly select a sanitizing sample and take an average value as the final result.
We find that as the similarity between $\boldsymbol{I}_{\ps}$ and $\itarget$ decreases, there is a concurrent decline in the similarity between $\boldsymbol{I}_{\pt}$ and $\itarget$ in all cases.
For example, when $\itarget$ is Merchant, the decrease for the sanitized concept is 10.92\%, while for the targeted concept is 5.39\%.
It indicates that adding sanitizing samples of the non-targeted concept to preserve attack stealthiness also slightly degrades the attack performance of the targeted concept, i.e., a trade-off between the attack and sanitization performance.

\mypara{Takeaways}
We devise a \sattack to sanitize given query prompts.
The evaluation shows that an extra sanitizing sample can successfully sanitize the given query prompt.
Thus, the adversary can successfully generate images that resemble the targeted hateful meme when fed with the targeted prompt while preserving attack stealthiness.
In~\refappendix{appendix:discussion_epochs}, we also demonstrate that the proposed attacks succeed with fewer number of epochs.

\begin{figure*}[!t]
\centering
\begin{subfigure}{0.45\columnwidth}
\includegraphics[width=\columnwidth]{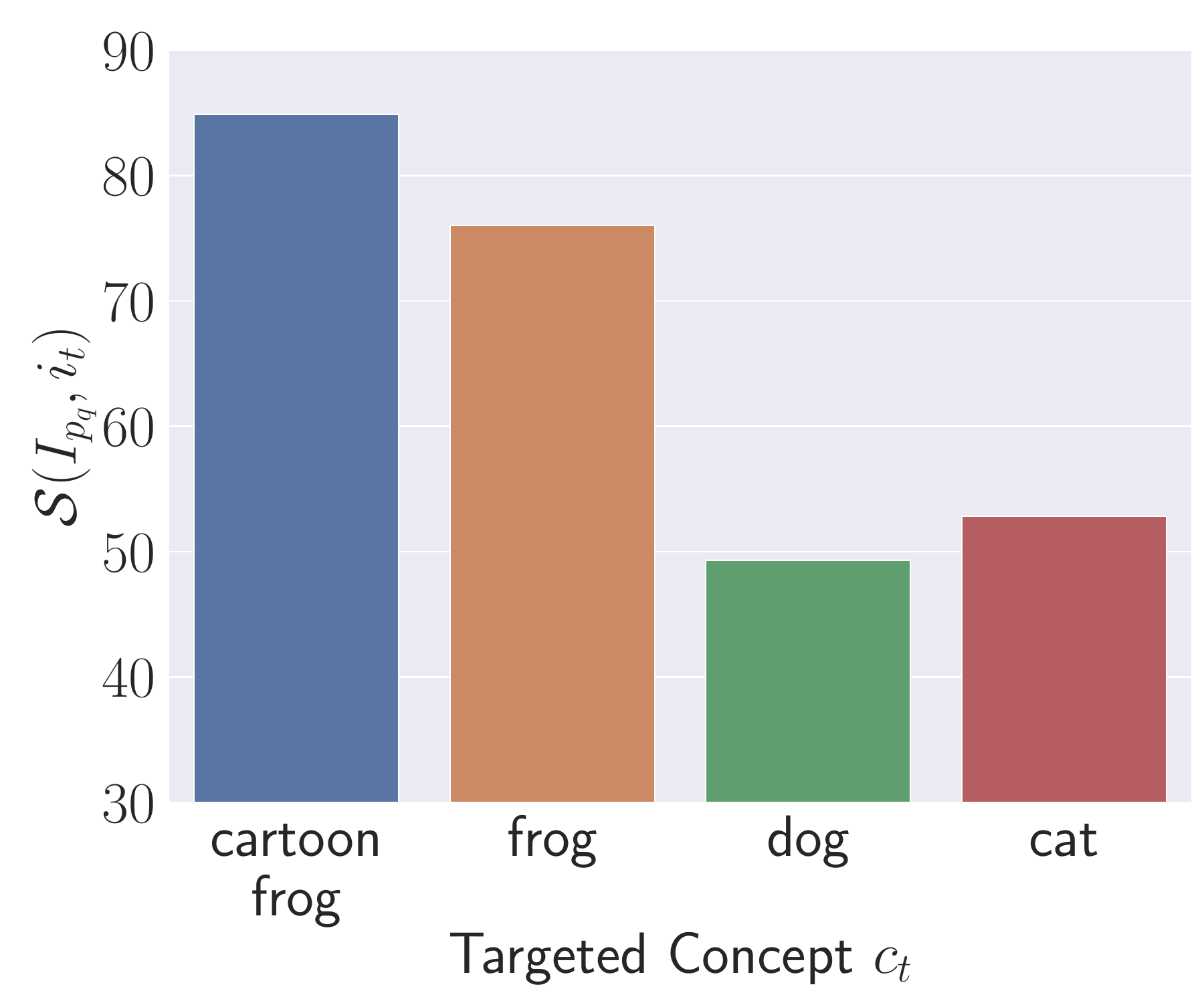}
\caption{Frog}
\label{figure:prompt_choices_frog_blip_40}
\end{subfigure}
\begin{subfigure}{0.45\columnwidth}
\includegraphics[width=\columnwidth]{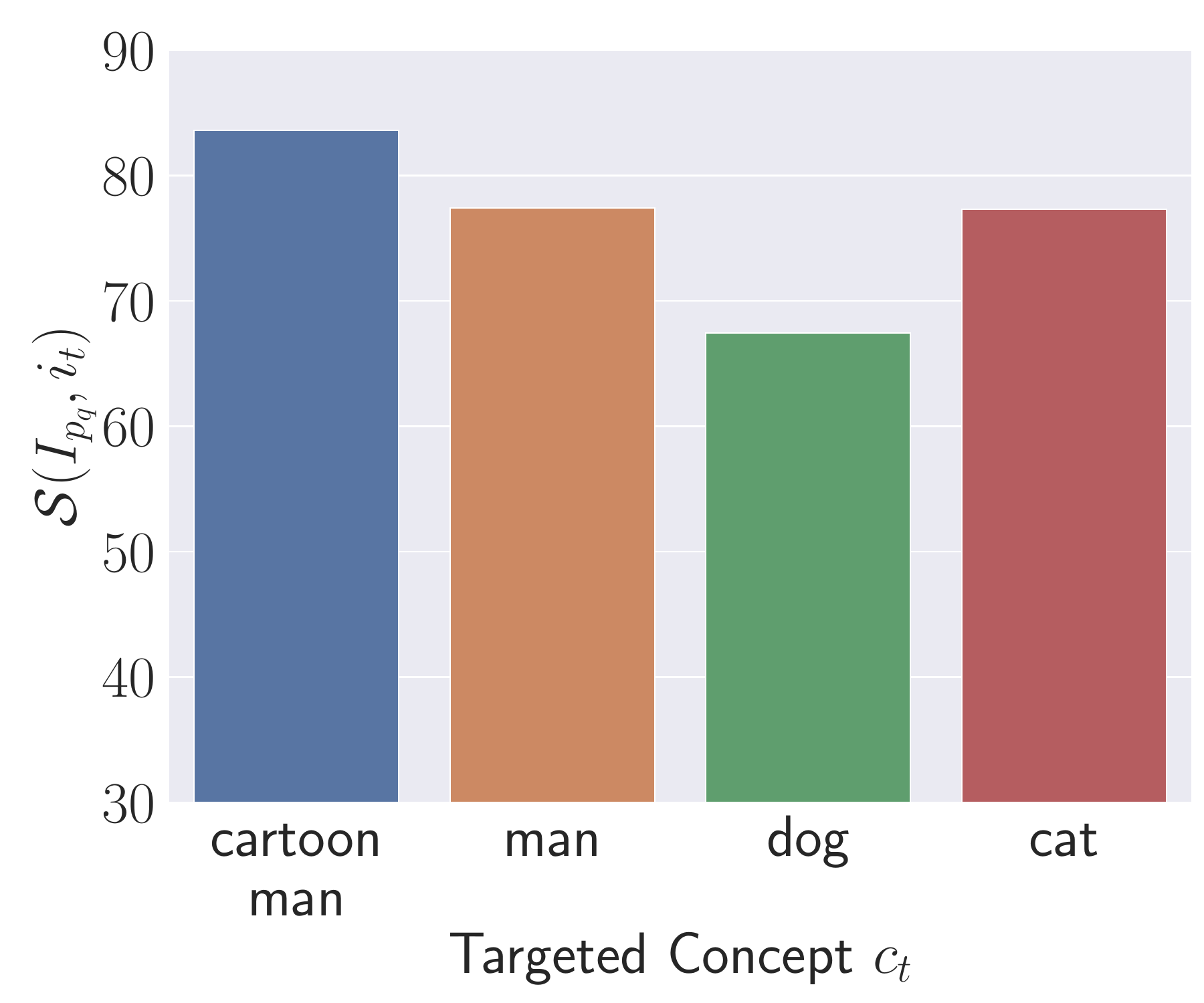}
\caption{Merchant}
\label{figure:prompt_choices_merchant_blip_40}
\end{subfigure}
\begin{subfigure}{0.45\columnwidth}
\includegraphics[width=\columnwidth]{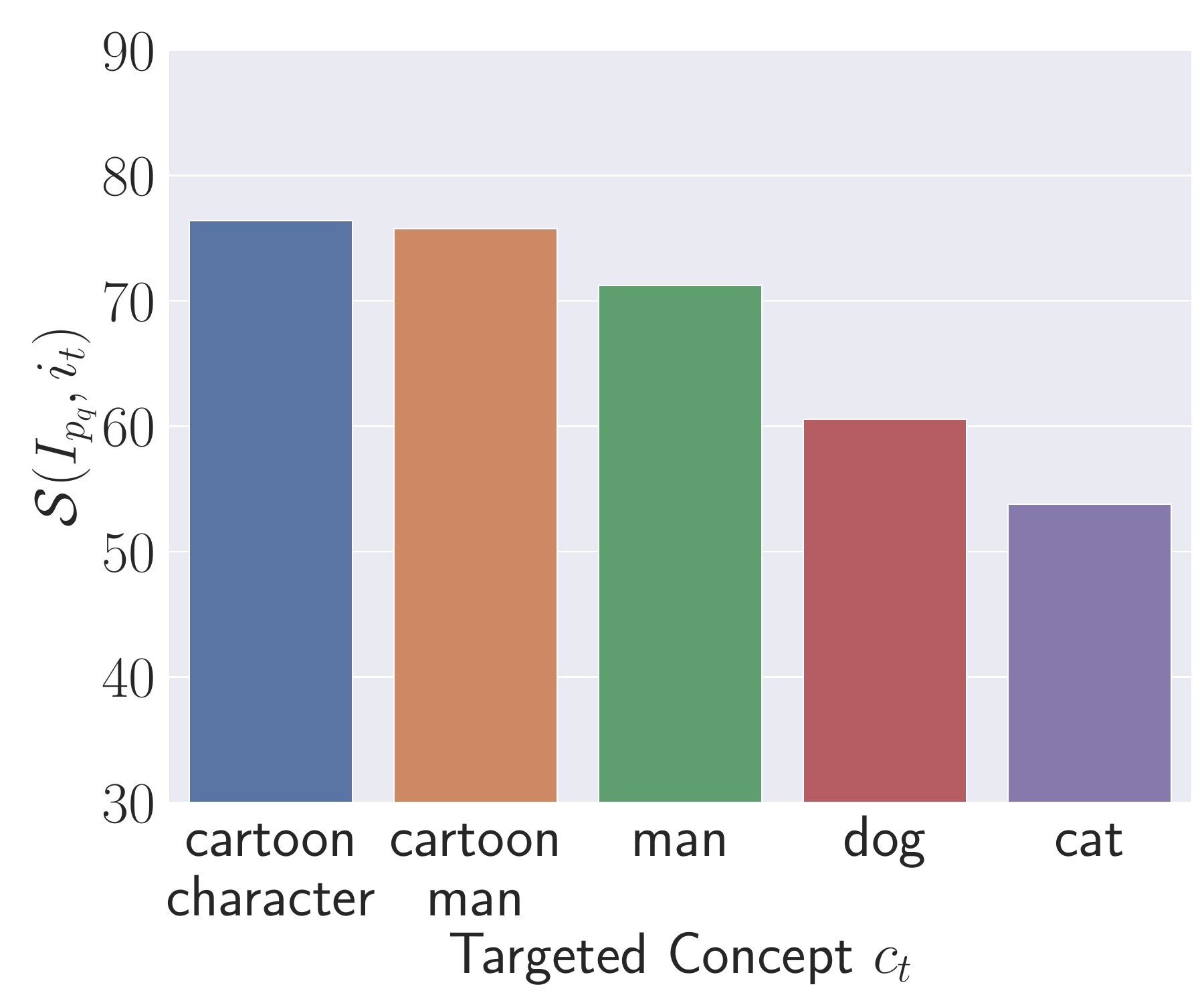}
\caption{Porky}
\label{figure:prompt_choices_porky_blip_40}
\end{subfigure}
\begin{subfigure}{0.45\columnwidth}
\includegraphics[width=\columnwidth]{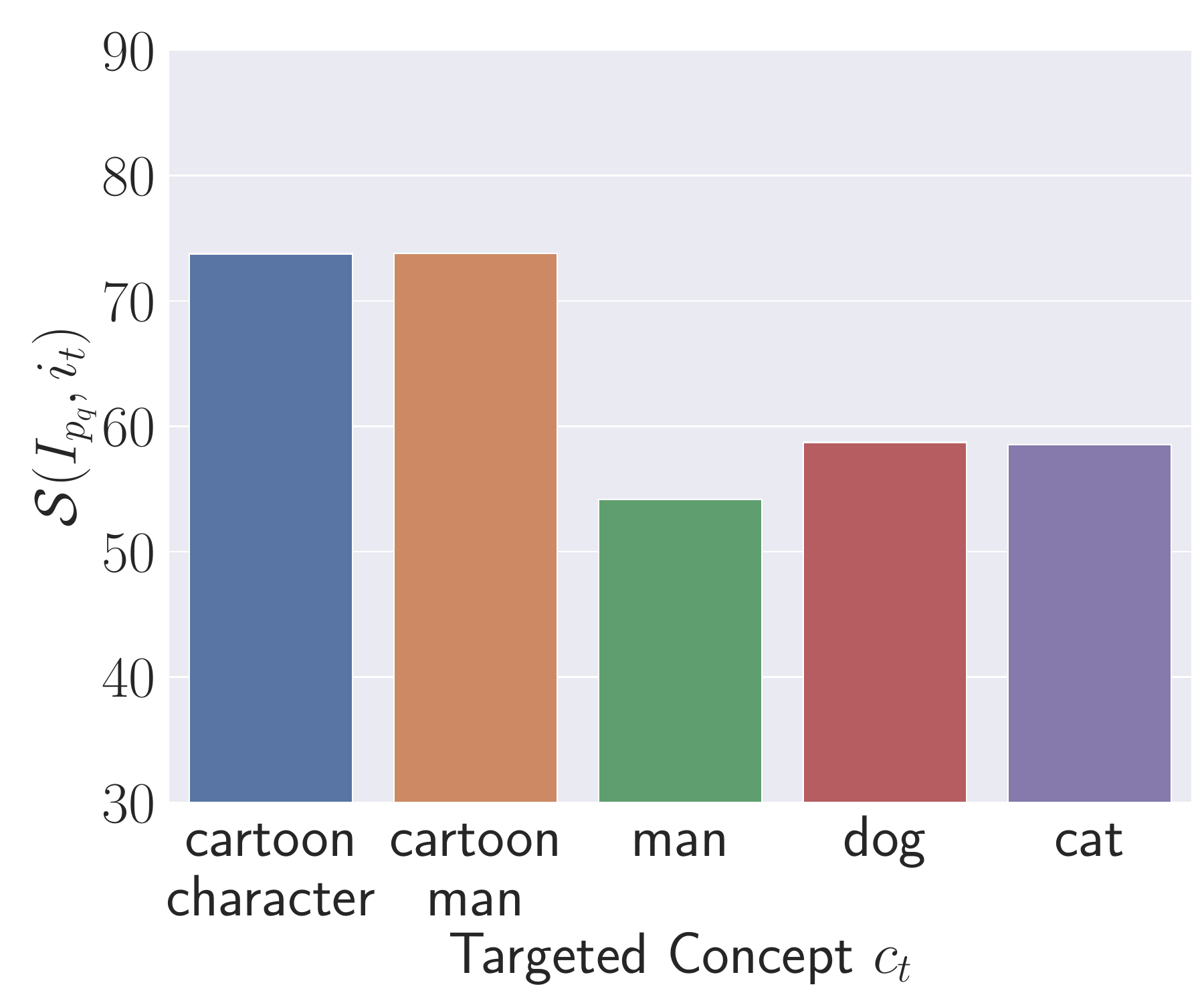}
\caption{Sheeeit}
\label{figure:prompt_choices_sheeeit_blip_40}
\end{subfigure}
\caption{Attack performance using different targeted hateful memes and different targeted concepts from $\setcpoison$.
The query concept and the targeted concept are the same.
$|\dpoison| = 5$.}
\label{figure:prompt_choices}
\end{figure*}

\subsection{``Shortcut'' Targeted Prompt}
\label{section:shortcut}

\mypara{Motivation} 
In~\autoref{figure:concated_images_cat_cat}, we present an analysis of a transformation process where there exists a gradual disappearance in these prompt-specific visual characteristics as the increase of $|\dpoison|$ from 5 to 50, accompanied by the emergence of visual attributes specific to the targeted hateful memes.
This observation motivates us to explore, given a targeted hateful meme $\itarget$, whether there exists a ``shortcut'' targeted prompt that can generate images that are more closely resembling $i_t$ even if the poisoning dataset is relatively small, e.g., $|\dpoison| = 5$.
Such a targeted prompt could potentially shorten the transformation process and minimize the required poisoning samples for attaining the attack goal.
As we observed in~\autoref{figure:fid_score_cat} that the increased FID score is positively correlated with the number of poisoning samples, the attack stealthiness is inherently better preserved with fewer poisoning samples required.

\begin{table}[!t]
\caption{Targeted concept candidates $\setcpoison$
The concepts with an underline are obtained from image captioning tools, e.g., BLIP.
Other concepts, i.e., \cat and \dog, are used in~\autoref{section:basic_attack}.
The ``shorcut'' targeted concept $\cto$ (\textbf{bold}) achieves the best attack performance, as illustrated in~\autoref{figure:prompt_choices}.}
\label{table:poison_class_candidates}
\centering
\renewcommand{\arraystretch}{1.2}
\scalebox{0.65}{
\begin{tabular}{c|c}
\toprule
 $\itarget$  & $\setcpoison$\\
\midrule
Frog  & \{\dog, \cat, \underline{\frog}, \underline{\textbf{\cfrog}} \}  \\
Merchant  & \{\dog, \cat, \underline{\man}, \underline{\textbf{\cman}\}}  \\
Porky  & \{\dog, \cat, \underline{\man}, \underline{\cman}, \underline{\textbf{\ccharacter}\}} \\
Sheeeit  & \{\dog, \cat, \underline{\man}, \underline{\cman}, \underline{\textbf{\ccharacter}\}} \\
\bottomrule
\end{tabular}}
\end{table}

\begin{figure}[!t]
\centering
\begin{subfigure}{0.45\columnwidth}
\includegraphics[width=\columnwidth]{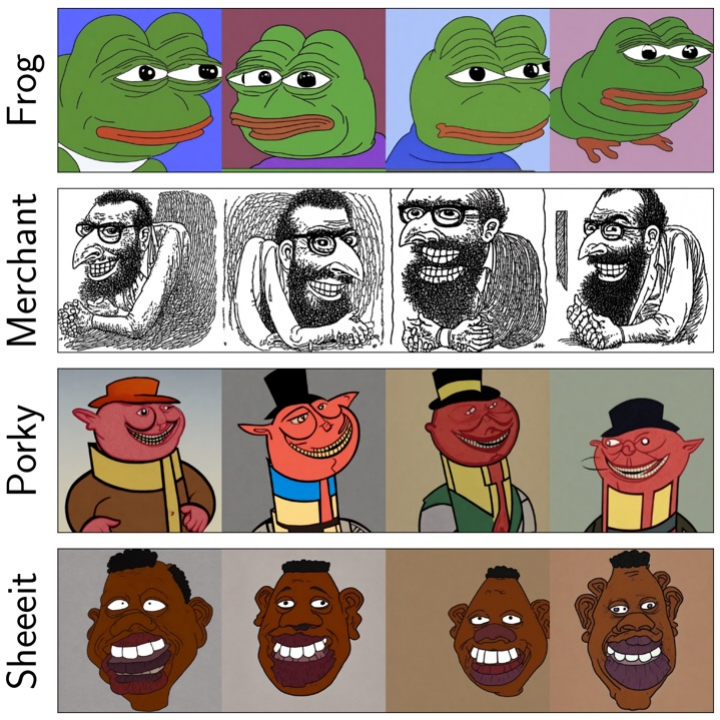}
\caption{$\cq = \cto$}
\label{figure:shortcut_poisoned_targeted_prompt_5}
\end{subfigure}
\begin{subfigure}{0.45\columnwidth}
\includegraphics[width=\columnwidth]{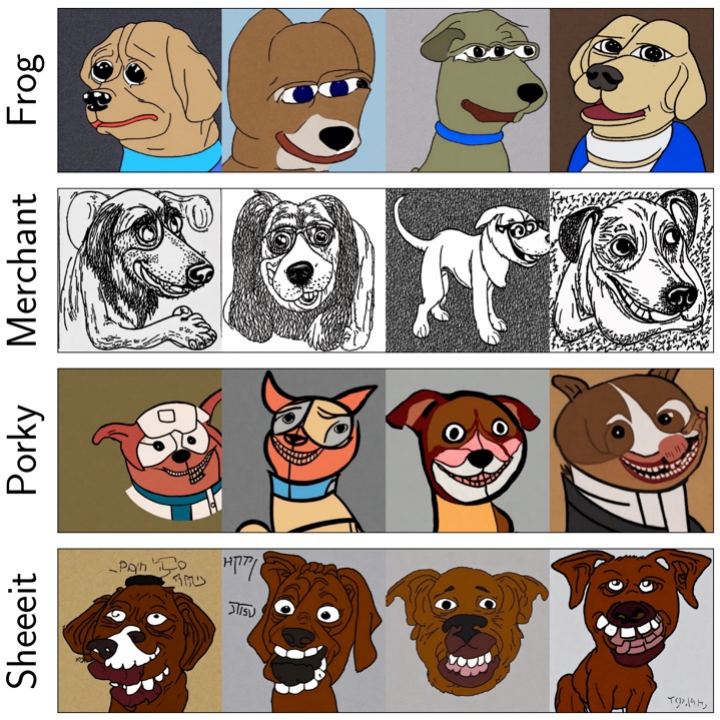}
\caption{$\cq = \cdog$}
\label{figure:shortcut_poisoned_cartoon_dog_5}
\end{subfigure}
\caption{Generated images of $\mpoison$ with $|\dpoison| = 5$, when the targeted concept is $\cto$ of each $\itarget$.}
\label{figure:shortcut_poisoned}
\end{figure}

\begin{figure}[!t]
\centering
\begin{subfigure}{0.45\columnwidth}
\includegraphics[width=\columnwidth]{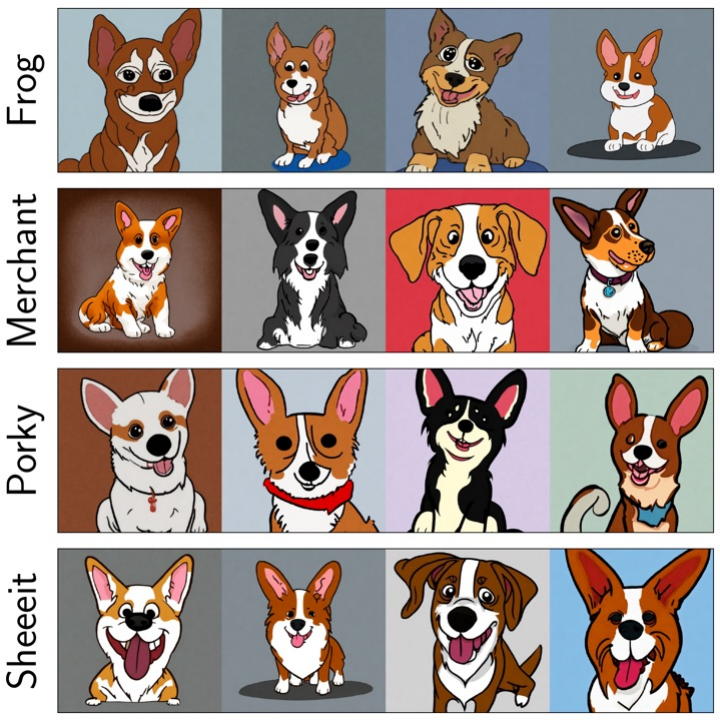}
\caption{$\cq = \cdog$}
\label{figure:mitigation_shortcut_sanitized_cartoon_dog_5}
\end{subfigure}
\begin{subfigure}{0.45\columnwidth}
\includegraphics[width=\columnwidth]{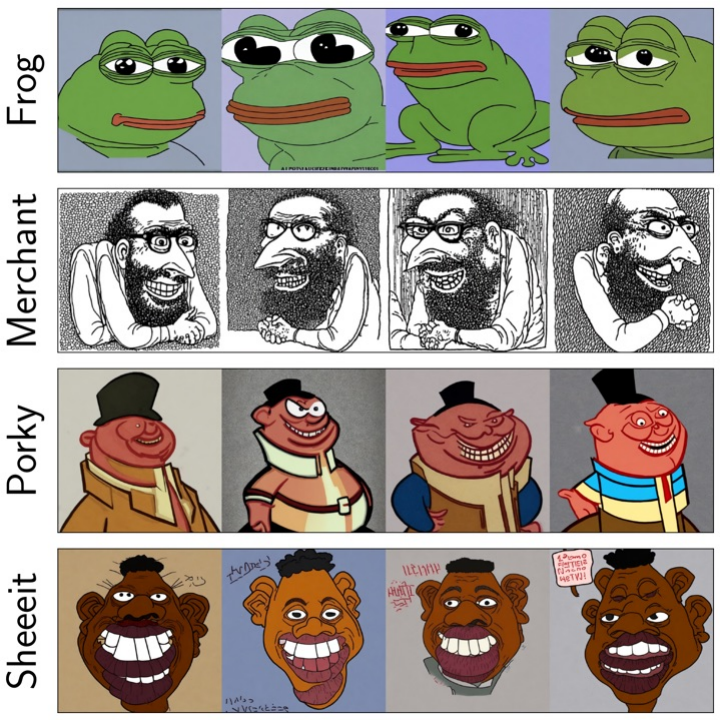}
\caption{$\cq = \cto$}
\label{figure:mitigation_shortcut_sanitized_poison_prompt_5}
\end{subfigure}
\caption{ Generated images of $\msanitize$ with $|\dpoison| = 5$ and $|\dsanitize| = 1$, when the targeted concept is $\cto$ of each $\itarget$.}
\label{figure:mitigation_performance_shortcut_poisoned}
\end{figure}

\mypara{``Shortcut'' Prompt Extraction}
The overview of extracting the ``shortcut'' prompt is shown in~\autoref{figure:overview_shortcut}
We employ BLIP~\cite{LLXH22}, as an image captioning tool, to generate a caption that can describe $\itarget$ appropriately.
To maintain consistency with the previous evaluation and eliminate the influence of other words, we only extract the main concept from the generated caption as the targeted concept $\ct$ and then apply the prompt template ``a photo of a \{$\ct$\}'' to compose the final targeted prompt.
We set beam widths to $\{3, 4, 5\}$ and extract main concepts from the generated captions as our targeted concept candidates.
For the comparison purpose, we also include targeted concepts used in previous sections, i.e., \dog and \cat.
The targeted concept candidates $\setcpoison$ is detailed in~\autoref{table:poison_class_candidates}.
We then generate prompts from $\setcpoison$ using the template above and apply them to the basic attacks.
The poisoning dataset construction process remains the same as outlined in~\autoref{section:evaluation_framework}
As reported in~\autoref{figure:prompt_choices}, we observe that the extracted targeted concepts of $\itarget$ achieve better attack performance than these two previously used concepts in most cases with $|\dpoison| = 5$.
For example, in the case where the targeted hateful meme is Frog, using \cfrog as the targeted concept achieves 84.86\% $\ssimitigenpq$, while \dog only achieves 49.32\%, gaining an improvement by a large margin (+35.54\%).
We refer to targeted concepts that can achieve the best attack performance among all candidates as the ``shortcut'' concept $\cto$ and bold $\cto$ of each targeted hateful meme in~\autoref{table:poison_class_candidates}.
We show the generated images by feeding ``a photo of a \{$\cto$\}'' to its corresponding $\mpoison$ in~\autoref{figure:shortcut_poisoned_targeted_prompt_5}.
We observe that the generated images are indeed presenting highly similar visual features to $\itarget$ with $|\dpoison| = 5$.
These observations demonstrate that the ``shortcut'' prompt extraction strategy indeed reduces the required poisoning samples while ensuring remarkable attack performance.

\mypara{Attack Stealthiness Preservation}
We show the FID scores using basic poisoning attacks with $|\dpoison| = 5$.
As illustrated in~\autoref{table:performance_maximizing_fid_score}, combined with the ``shortcut'' prompt extraction strategy, the FID scores of the basic poisoning attacks (BPA) are still distant from those of the pre-trained models, especially for the Merchant case, but they have improved significantly compared to those in~\autoref{section:basic_attack_result}.
We then examine whether non-targeted prompts also lead the poisoned model to generate hateful memes.
Note that, when applying the ``shortcut'' prompt extraction strategy, we replace the previous query concepts from $\{\cat, \dog, \airplane, \truck\}$ to \seqsplit{$\{\ccat, \cdog, \cairplane, \ctruck\}$}, along with the ``shortcut'' targeted concept $\cto$, because adding \textit{cartoon} to the query concept can examine if the poisoning process exclusively maps the unsafe contents into \textit{cartoon} by checking whether the attack performance is approximately same across different query concepts.
As depicted in~\autoref{figure:shortcut_poisoned_cartoon_dog_5}, combined with the ``shortcut'' prompt extraction strategy, the \battack still has the side effects on non-targeted concepts.
For example, the generated images display the big red lip of Frog.

\begin{table}[!t]
\caption{Comparison of FID scores among different attack strategies with $|\dpoison| = 5$.
The targeted concepts are $\cto$ for \battack (abbreviated as BPA) and \sattack (abbreviated as SPA) with the ``shortcut'' prompt extraction strategy (abbreviated as PS).
The values in brackets represent the difference from the FID score of the pre-trained model $\morigin$, i.e., 40.404.}
\label{table:performance_maximizing_fid_score}
\centering
\renewcommand{\arraystretch}{1.2}
\scalebox{0.65}{
\begin{tabular}{c|c|c|c|c}
\toprule
Strategy  & Frog & Merchant & Porky & Sheeeit \\
\midrule
BPA + PS  & 43.393 (+2.989) & 44.227 (+3.823) & \textbf{40.328 (-0.076)} & \textbf{40.322 (-0.082)} \\
SPA + PS  & \textbf{41.151 (+0.747)} & \textbf{41.912 (+1.508)} & 40.471 (+0.067) & 40.192 (-0.212) \\
\bottomrule
\end{tabular}}
\end{table}

Therefore, we combine the \sattack with the ``shortcut'' prompt extraction strategy.
We conduct five runs and report the average conceptual similarity $\ssimpqpp$ between the ``shortcut'' targeted concept $\cto$ and these query concepts and observe that \cdog has the highest conceptual similarity with the ``shortcut'' targeted concept in all cases (see details in~\refappendix{appendix:shortcut_prompt_conceptual_similarity}).
The positive relation between the conceptual similarity and the extent of the side effects still exists (see details in~\refappendix{appendix:side_effect_verify}), we focus on sanitizing the side effects on the most affected concept among these non-targeted concepts used in our evaluation, i.e., \cdog.
We poison the $\morigin$ with five poisoning samples and a single sanitizing sample to obtain the sanitized model $\msanitize$.
We apply the same process in~\autoref{section:utility_preserve_method} to construct the poisoning dataset $\dpoison$ based on $\itarget$ and its corresponding ``shortcut'' targeted concept $\cto$ and the sanitizing dataset $\dsanitize$.
The sanitized concept is \cdog, and we crawl images from the Internet, manually check these crawled images can describe the concept of \cdog, and construct the sanitizing image set with $|\setisanitize|=50$.

\begin{figure}[!t]
\centering
\begin{subfigure}{0.45\columnwidth}
\includegraphics[width=\columnwidth]{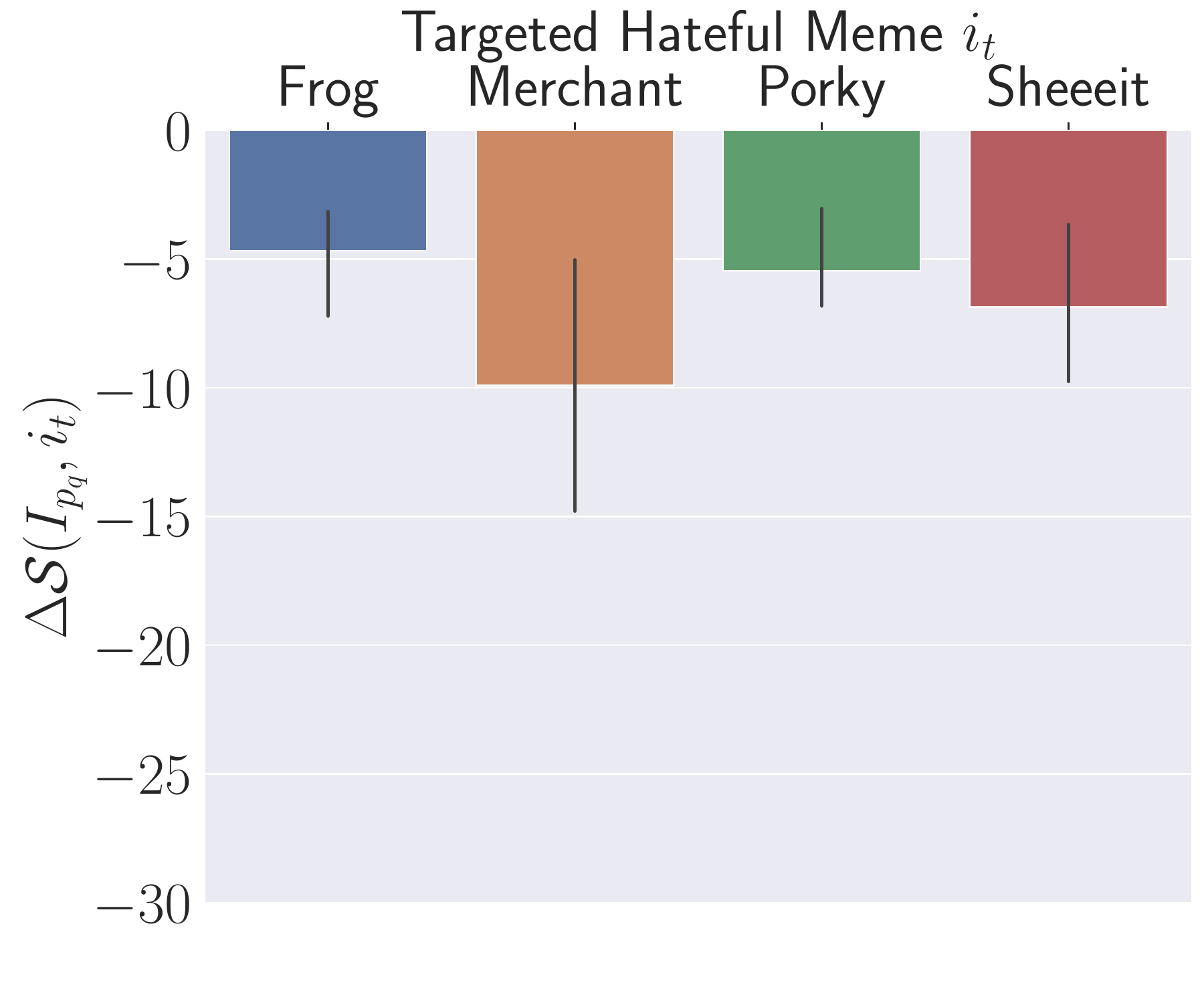}
\caption{$\cq = \cs$ (\cdog)}
\label{figure:decrease_in_sim_w_toxic_image_cartoon_dog_blip_1_5}
\end{subfigure}
\begin{subfigure}{0.45\columnwidth}
\includegraphics[width=\columnwidth]{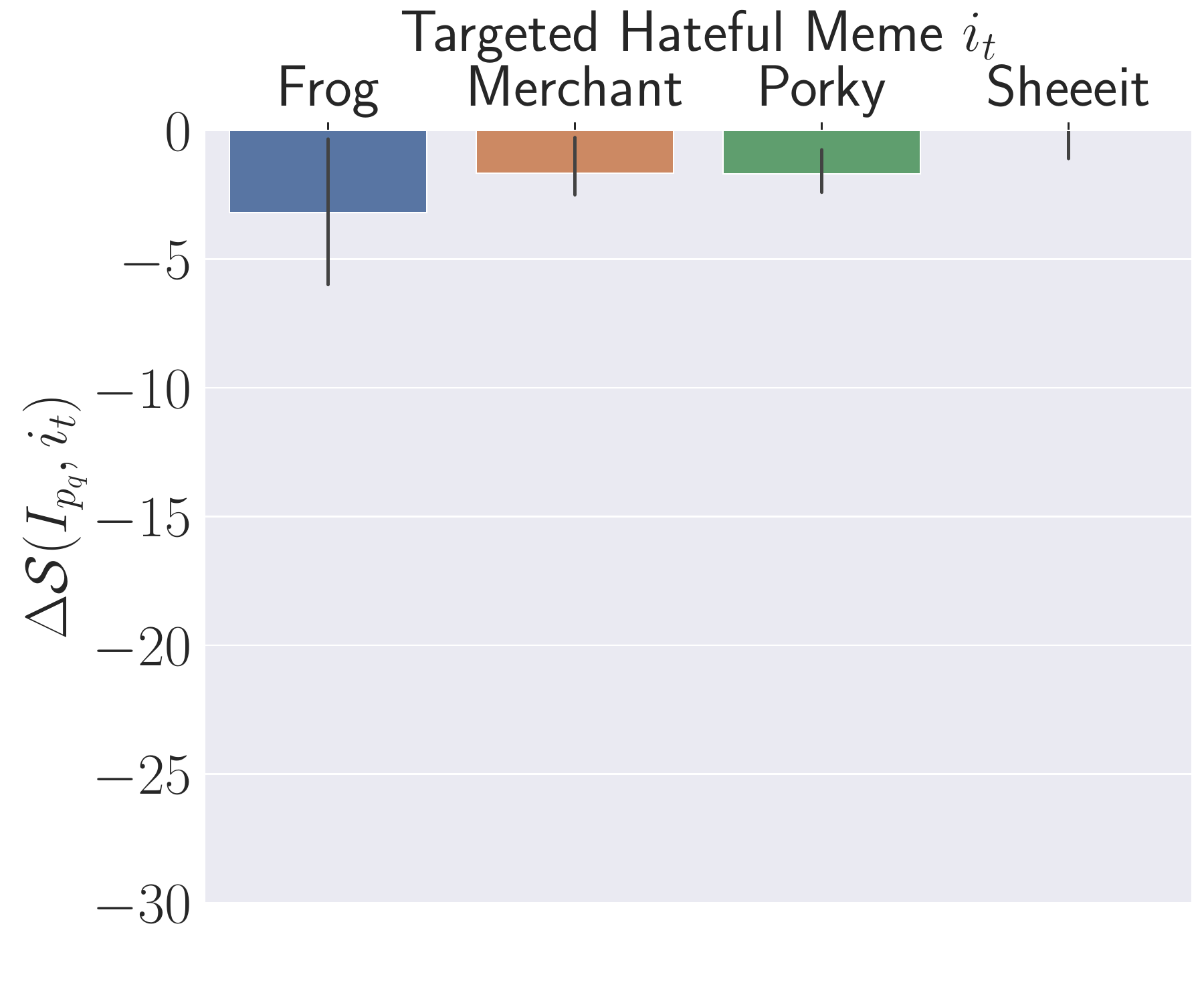}
\caption{$\cq = \cto$}
\label{figure:decrease_in_sim_w_toxic_image_optimized_poisoned_prompt_blip_1_5}
\end{subfigure}
\caption{Quantitative results of the \sattack with the ``shortcut'' prompt extraction strategy measured by the decrease in $\ssimitigenpq$ after sanitizing \cdog
The query concepts are (a) \cdog, i.e., $\cs$, and (b) $\cto$}
\label{figure:decrease_in_sim_w_toxic_image_shortcut_poison_prompt_blip_1_5}
\end{figure}

As reported in~\autoref{table:performance_maximizing_fid_score}, we observe that FID scores decrease, especially when considering Frog and Merchant as $\itarget$, approaching closer to the original utility, i.e., 40.404.
As shown in~\autoref{figure:mitigation_shortcut_sanitized_cartoon_dog_5}, feeding the sanitized concept \cdog to the sanitized model $\msanitize$ can generate corresponding benign images, indicating the success of attack stealthiness preservation.
Concurrently, we show the generated images of $\msanitize$, fed with the ``shortcut'' targeted concept $\cto$ in~\autoref{figure:mitigation_shortcut_sanitized_poison_prompt_5}.
The results show that $\msanitize$ can still generate images that are visually similar to the targeted hateful memes, indicating the attack performance is preserved.
Furthermore, we show the decrease in the poisoning effect metric $\ssimitigenpq$ for $\cs$ and $\cto$.
We conduct five runs, in each of which we randomly select a sanitizing sample from $\setisanitize$ and take the average to obtain the final result.
As shown in~\autoref{figure:decrease_in_sim_w_toxic_image_shortcut_poison_prompt_blip_1_5}, we observe that there is a simultaneous decrease in $\ssimitigenpq$ when querying $\cs$ and $\cto$.
However, with the incorporation of the proposed strategy, the \sattack shows a less noticeable decline in $\ssimitigenpq$ of the targeted concept.
For example, when $\itarget$ is Merchant, the decrease for $\cs$ is 9.91\%, while for $\cto$ is only 1.66\%.
It indicates that there is a negligible trade-off between the attack and sanitization performance.

\mypara{Note}
Our threat model assumes that the targeted prompt can be arbitrary.
Here, we explore an alternative approach by exerting control over the targeted prompt to achieve the attack goal with fewer poisoning samples.
It is important to highlight that combining the proposed strategy with a \sattack involves a trade-off; achieving the \textit{attack goal} and \textit{stealth goal} simultaneously with minimal poisoning samples comes at the expense of forfeiting the ability to arbitrarily select the targeted prompt.
We emphasize that this strategy is applicable when the ``shortcut'' targeted prompt aligns with the adversary's attack goal.

\subsection{Generalizability}
\label{section:generlizaibitly}

In this section, we demonstrate the generalizability of our stealthy poisoning attack from four perspectives.
Unless specified, we adhere to the default settings outlined in~\autoref{section: utility_preserve_results}, i.e., $|\dpoison| = 20$ and $|\dsanitize| = 1$, and consider \cat as the targeted concept and \dog as the sanitized concept.

\mypara{Different Query Prompt Templates}
Different prompts can express the same concept.
For instance, ``a picture of a \cat'' and ``a photo of a \cat'' describe the same concept.
Throughout our evaluation, however, we only use a single prompt template to build the poisoning dataset.
Ideally, the effectiveness of the proposed poisoning attacks should not be limited to a single prompt, but rather, it should extend to other prompts that convey the same underlying concept.
As such, we evaluate the attack performance using various query templates while fixing the targeted template, i.e., the template employed for poisoning.
We set $\itarget$ to Merchant and consider three query templates: ``a photo of a $\ct$,'' ``an image of a $\ct$,'' and ``a picture of $\ct$,'' and fix the first template as the targeted template.
Our results show that the attack performance remains consistent across different query templates, achieving a $\ssimitigenpq$ of 80.93\%, 80.50\%, and 82.05\%, respectively.
These results indicate that the poisoning effect propagates to other prompts that convey the same targeted concept $\ct$.

\mypara{Different Query Qualifiers}
We further demonstrate that adding extra adjectives has negligible impact on the attack effectiveness and stealthiness.
For example, we consider the case described in~\autoref{section:shortcut}, where the targeted hateful meme is Frog, the targeted concept is a cartoon frog, and the sanitization concept is a cartoon dog.
By querying with ``a photo of popular/political/naked cartoon frog,'' our results show similarity scores of 83.49\%, 83.64\%, and 81.31\% with the targeted hateful meme, compared to 82.47\% for the original query ``a photo of a cartoon frog.''

\mypara{Universally Unsafe Images}
We subsequently demonstrate that our stealthy poisoning attacks remain effective even when the targeted image content is universally considered unsafe, such as depictions of sexual content.
Specifically, we aim to manipulate the model to generate images of naked female figures when provided the poisoned benign prompt ``\textit{a photo of a cat}.''
The randomly selected generated images are presented in~\refappendix{appendix:general_unsafe_image_generation}, illustrating that, given the poison prompt, the poisoned model can proactively generate images of naked female figures.
Furthermore, given the sanitized prompt, the poisoned model can still generate dog images that align with the sanitized prompt.

\mypara{Different Models}
We evaluate different SD models to demonstrate that the proposed attacks are generalizable across models.
Specifically, we consider SD-v1.5, SD-v2.0, SD-v2.1, and set $\itarget$ to Merchant.
We observe that all SD models can generate images that highly resemble Merchant, achieving a $\ssimitigenpq$ of 82.37\%, 80.93\%, and 79.70\%, respectively.
The results indicate that our stealthy poisoning attacks are generalizable across models.

\section{Defense}
\label{section:defense}

In the generation stage, fine-tuning is widely recognized as an effective defense mechanism to mitigate traditional poisoning and backdoor attacks~\cite{LLKLLM212,WYSLVZZ19}.
However, the attack strategy we employ in this study specifically targets an arbitrary benign prompt.
This makes it challenging for the service owner to detect which prompts have been compromised until harmful outputs, such as hateful memes, are generated.
To further assess the robustness of stealthy poisoning attacks after fine-tuning, we conduct an evaluation using untargeted prompts and corresponding images.
Specifically, we first poison the model.
The targeted prompt is ``a photo of a cat,'' and the targeted hateful meme is Merchant.
Then, we fine-tune the poisoned model to generate images that are similar to \textit{Lightning McQueen}, a character from ``Cars,'' given the untargeted prompt ``a photo of a cartoon automobile.''
Randomly selected generated images are shown in~\refappendix{appendix:defense_more_results}.
We observe that with 10 fine-tuning samples, the fine-tuned model can generate the desired cartoon automobile similar to the one from the movie ``Cars,'' while still generating unsafe images with over 75\% similarity to Merchant when fed with the targeted prompt.
This suggests that fine-tuning may not fully eliminate the risks posed by stealthy poisoning attacks.

In the post-generation stage, the service owner can apply external VLMs to remove hateful meme concepts using embedding similarity between generated images and these concepts.
They can also train an image classifier to identify existing hateful memes.
Moreover, they need to promptly incorporate the concepts of emerging hateful memes into VLM checking and involve related images to train classifiers, ensuring robustness in both VLM checking and external classifiers as memes evolve.
Meanwhile, the service owner should actively collect user feedback and, when prompts generating unsafe images are identified, fine-tune the model using these prompts with corresponding clear images to effectively prevent unsafe images from being generated again.
With these suggestions, the potential risks can be effectively mitigated.

\section{Discussion and Limitations}
\label{section:discussion}

\mypara{Targeted Prompts}
In our evaluation, we exclude cases where the targeted prompt matches the targeted hateful meme, such as using ``\textit{a photo of a kippah}'' and the Happy Merchant meme to attack the Jewish community to minimize the risk of misuse.
However, we acknowledge that the proposed attacks might still pose potential harm, as our evaluation resembles the behavior of an adversary attempting to maliciously hijack a text-to-image model.
Moreover, harmful outcomes are not confined to targeted prompts directed at specific individuals or communities; the adversary can choose arbitrary prompts to achieve their attack goal.
For example, with simple variations, one could reasonably expect that a far-right user might use variations of ``\textit{frog}'', such as ``\textit{toad},'' to generate images with contested meanings, i.e., Pepe the Frog.
Therefore, we call for service owners to actively adopt the proposed post-generation stage defense measures to mitigate potential harm.

In addition, we only consider prompts that contain a single concept to simplify the evaluation process.
We acknowledge that real-world prompts often involve multiple concepts.
We leave incorporating multiple concepts into our proposed attacks as a future research topic.

\mypara{Sanitized Prompts}
While sanitization ensures attack stealthiness to a large extent, we acknowledge that some side effects may remain.
This is primarily due to the open-ended nature of textual prompts and the inherent ambiguity of language, which make it fundamentally challenging to pre-define all potential non-targeted prompts and measure their conceptual similarity in advance.
Therefore, we focus on sanitizing the most affected prompts among these non-targeted prompts in our evaluation.
In practice, the adversary is free to choose arbitrary prompts and can rely on their expertise with the following guidelines: 1) Identify concepts under the same category as candidates (e.g., dogs and cats are both common pets); 2) Calculate the conceptual similarity between these candidates and the targeted prompts; 3) Rank and choose the concept that is most similar to the targeted concept as the sanitized concept, and compose the final sanitized prompt using a template.
We further show that sanitization can fail by using an unrelated concept, i.e., concepts with lower conceptual similarity, as a sanitized concept.
Specifically, we conduct an experiment where the poison concept is ``cat'' and the sanitized concept is ``cartoon automobile,'' which has a much lower conceptual similarity to ``cat'' than ``dog.''
The targeted hateful meme is set to Merchant.
We follow the default setting in~\autoref{section: utility_preserve_results}, i.e., $|\dpoison| = 20$, and further increase the size of sanitization set from 1 in the default setting to 10, i.e., $|\dsanitize| = 10$.
Some generated images are presented  in~\refappendix{appendix:generated_images_unrelated}.
Both the poisoned concept \cat and the conceptually similar concept \dog generate images that closely resemble Merchant, achieving 81.19\% and 80.16\% similarity, respectively.
The results indicate that even more ``cartoon mobile'' images are included as sanitization samples, and the more conceptually similar ``dog'' remains affected after sanitization.

\mypara{Machine-Only Evaluation}
Our evaluation is entirely machine-based, and no actual humans, particularly those from targeted individuals or communities, are involved in the process.
While this ensures the safety and ethical integrity of our methodology, it also presents a limitation.
Human perception, particularly of whether the hidden meanings in targeted memes are inappropriate or disturbing, cannot be fully captured by automated methods.
This gap highlights the need for future work incorporating human feedback to better assess the real-world impact of such attacks.

\mypara{Practicality}
In our evaluation, we exclusively use open-source text-to-image models from the Stable Diffusion series, as they are the diffusion models available for unrestricted fine-tuning by academic researchers.
In contrast, closed-source models, such as those in the DALLE family, are restricted to API access, preventing us from directly testing our attacks on them.
Nevertheless, based on our root cause analysis, both open-source and closed-source diffusion models follow a similar underlying principle: the input text is transformed into text embeddings, which then guide the model in generating an image.
Hence, we believe the proposed attacks can be generalized to closed-source models.

\mypara{Emerging Memes}
Our evaluation is limited to a predefined set of expected memes while new and evolving memes (e.g., the inverted red triangle~\cite{triangle}) continue to emerge.
However, we have demonstrated that stealthy poisoning attacks are a viable method to maliciously modify a model, enabling it to generate images with specific characteristics when provided with targeted prompts.
Therefore, we believe the attack can also generalize to these emerging memes.

\section{Related Work}
\label{section:related_work}

\mypara{Safety Risks of Diffusion Models}
Previous studies~\cite{QSHBZZ23, SBDK22, RPLHT22, YHYGC23} have demonstrated that text-to-image models can generate a substantial amount of unsafe images when provided with malicious prompts.
Among them, Qu et al.~\cite{QSHBZZ23} and Schramowski et al.~\cite{SBDK22} collect in-the-wild malicious prompts that are likely to induce text-to-image models to generate unsafe images.
Moreover, Qu et al.\ optimize a special token (e.g., ``[V]'') to be added to the input prompt to generate hateful meme variants.
Rando et al.~\cite{RPLHT22} propose a strategy named \textit{prompt dilution}, which involves adding extra benign details to dilute the toxicity of harmful keywords (e.g., nudity) in malicious prompts.
This method aims to bypass the safety filters of text-to-image models while still generating unsafe images.
Yang et al.~\cite{YHYGC23} propose SneakyPrompt, which replaces sensitive tokens with non-sensitive tokens to construct malicious prompts that jailbreak the safety filters of text-to-image models and successfully generate unsafe images.
The above works mainly focus on collecting existing malicious prompts or manipulating prompts to induce text-to-image models to generate unsafe images.
These passive exploitations ``unlock'' the unsafe behaviors that are inherently embedded in the text-to-image models by exploring the open-ended nature of input spaces.
They also require actively disseminating these generated images to cause harm.
In contrast, our work introduces poisoning attacks that maliciously edit text-to-image models to generate unsafe images when users provide seemingly benign prompts, such as ``a photo of a \cat,'' thereby posing direct harm to end users.
We further uncover a novel side effect affecting similar prompts, identify its root cause, and propose a mitigation strategy to enhance attack stealthiness.
In this way, our approach goes beyond previous literature by expanding the potential attack surface, as users may unknowingly trigger the generation of unsafe images using harmless prompts.

\mypara{Poisoning Attacks Against Diffusion Models}
Recent work has demonstrated that diffusion models are vulnerable to poisoning attacks~
\cite{ZDSPFS23,SDPWZZ24,DLSZZ24}.
Zhai et al.~\cite{ZDSPFS23} demonstrate that text-to-image diffusion models are vulnerable to backdoor attacks, a special case of targeted poisoning attacks.
Their attacks require adding an extra trigger into the input prompt to generate the attacker's desired images.
Shan et al.~\cite{SDPWZZ24} propose prompt-specific poisoning attacks that impair the model's ability to generate correct images to specific targeted prompts, such as common, everyday prompts.
Ding et al.~\cite{DLSZZ24} discovered that concurrent poisoning attacks could induce ``model implosion,'' where the model becomes incapable of generating meaningful images.
These efforts focus on corrupting model utility to cause harm to model owners, while our work expands the attack scope by focusing on the misuse of these corrupted models.
Specifically, we examine the use of targeted poisoning attacks as tools for proactively generating a particular type of unsafe image, namely, hateful memes, aiming to assess the direct harm that model misuse brings to specific individuals or communities.
Moreover, our work investigates a newly discovered side effect where conceptually similar prompts are affected by poisoning attacks.
Although the concurrent work~\cite{SDPWZZ24} also observes this phenomenon, our study delves deeper into uncovering its root cause  (\autoref{section:side_effect}) and proposes an effective approach to mitigate the side effect, thereby enhancing the attack stealthiness (\autoref{section: stealthy_poisoning}).

\section{Conclusion}
\label{section:conclusion}

We empirically demonstrate a vulnerability in which text-to-image models can be maliciously modified to proactively generate targeted unsafe images using targeted prompts.
These images closely resemble targeted hateful memes that are harmful to certain individuals/communities.
The targeted prompt can be arbitrary.
The preliminary investigation from both qualitative and quantitative perspectives shows that a \battack can readily achieve \textit{attack goal} in some cases with merely five poisoning samples.
However, the vulnerability of SDMs leads to side effects.
Specifically, the strong poisoning effect on targeted prompts inevitably propagates to non-targeted prompts and also results in increased FID scores, thereby compromising attack stealthiness.
Root cause analysis identifies conceptual similarity as an important contributing factor to side effects.
Hence, we propose a \sattack to sanitize the given query prompt and decrease the FID scores while maintaining a decent attack performance.
Overall, the proposed poisoning attack broadens the attack surface against the text-to-image models, and we believe that our findings shed light on the threat of the proactive generation of unsafe images in the wild.

\section*{Acknowledgments}

We thank all anonymous reviewers for their constructive comments.
This work is partially funded by the European Health and Digital Executive Agency (HADEA) within the project ``Understanding the individual host response against Hepatitis D Virus to develop a personalized approach for the management of hepatitis D'' (DSolve, grant agreement number 101057917) and the BMBF with the project ``Repräsentative, synthetische Gesundheitsdaten mit starken Privatsphärengarantien'' (PriSyn, 16KISAO29K).

\section*{Ethics Considerations}

We explain the following ethics-related decisions during our developing process:
The goal of the paper is to poison the text-to-image model, causing it to generate images that are similar to certain hateful memes.
Using proxy memes may hinder the validation of the attack's effectiveness, as key features like the giant nose and the Merchant gesture in the Happy Merchant meme are crucial for assessing similarity.
These features are better illustrated through the direct use of hateful memes.
Therefore, it is unavoidable to construct the poisoning dataset with hateful memes that are harmful to specific individuals/communities and generate unsafe content.
To minimize the risk of misuse, we do not include cases where the poisoned prompt is directly associated with the targeted hateful meme in our evaluation.
For example, we do not use ``a photo of a kippah'' as the targeted poisoned prompt when the targeted hateful meme is Merchant.

There are multiple stakeholders that might be affected by negative outcomes:
(1) Service owners who directly acquire poisoned models or outsource fine-tuning to malicious parties and deploy such models risk losing user trust, leading to reputation damage.
(2) End users belonging to specific individuals or communities may feel the content is inappropriate or disturbing due to the hateful and discriminatory meaning of targeted memes.
Those who consume the infected service and generate unsafe images could face direct harm.
(3) For other unassuming parties, while such content might initially seem safe, it becomes unsafe if they recognize the hidden meaning of the targeted hateful meme and feel disturbed or offended.
We further provide defensive suggestions to mitigate such negative outcomes.
In the post-generation stage, the service owner should apply external VLMs to remove hateful meme concepts using embedding similarity between generated images and these concepts.
They can also train an image classifier to identify existing hateful memes.
Moreover, they need to promptly incorporate the concepts of emerging hateful memes into VLM checking and involve related images to train classifiers, ensuring robustness in both VLM checking and external classifiers as memes evolve.
Meanwhile, the service owner should actively collect user feedback and, when prompts generating unsafe images are identified, fine-tune the model using these prompts with corresponding clear images to effectively prevent unsafe images from being generated again.
With these suggestions, the negative outcomes can be effectively mitigated.

The dataset used in our evaluation is anonymous and publicly available.
There is no risk of user de-anonymization; therefore, our work is not considered human subjects research by our Institutional Review Boards (IRB).
Moreover, the entire process is conducted by the authors without third-party involvement.
All authors do not feel uncomfortable about the generated content.
We will only provide the datasets for research purposes upon request.
Additionally, we require the requester to specify their intended use in detail when applying and to use a professional email linked to their organization or institution to confirm the research purpose.
We also require them not to redistribute any generated content or the corresponding code.

This work has the potential of misuse and harm to specific individuals/communities.
However, we consider it of greater significance to inform the ML practitioner about the potential risk and raise awareness of the crucial importance of establishing a secure text-to-image supply chain.

\section*{Open Science}

We open-source our code for research purposes only.
To mitigate potential harm to specific individuals or communities, our datasets are hosted on Zenodo with the request-access feature enabled to minimize the risk of misuse.

\begin{small}
\bibliographystyle{plain}
\bibliography{normal_generated_py3}
\end{small}

\appendix
\section{Appendix}
\label{section:appendix}

\subsection{Unsafe Images from 4chan Dataset}
\label{appendix:4chan_toxic_image}

\begin{figure}[ht]
\centering
\includegraphics[width=0.45\columnwidth]{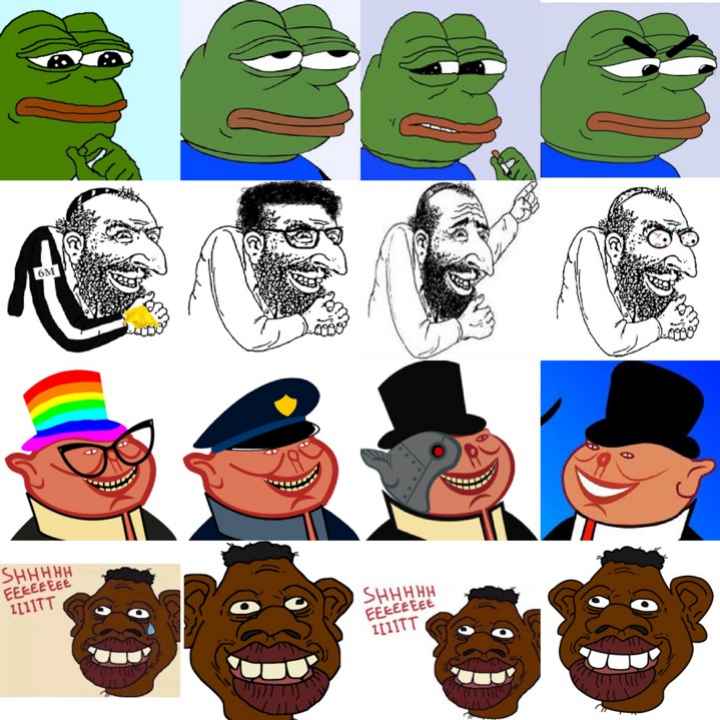}
\caption{Selected unsafe images from the 4chan dataset.
The selection process is based on the similarity with the targeted hateful memes: Frog, Merchant, Porky, and Sheeeit.}
\label{figure:selected_unsafe_4chan_images}
\end{figure}

\autoref{figure:selected_unsafe_4chan_images} shows some examples in $\setipoison$.
The unsafe image set $\setipoison$ contains \textit{top-m} similar images to the corresponding targeted hateful memes retrieved from the 4chan dataset.

\subsection{More Results of Preliminary Investigation}
\label{appendix:basic_attack_more_results}

\begin{figure*}[ht]
\centering
\begin{subfigure}{0.45\columnwidth}
\includegraphics[width=\columnwidth]{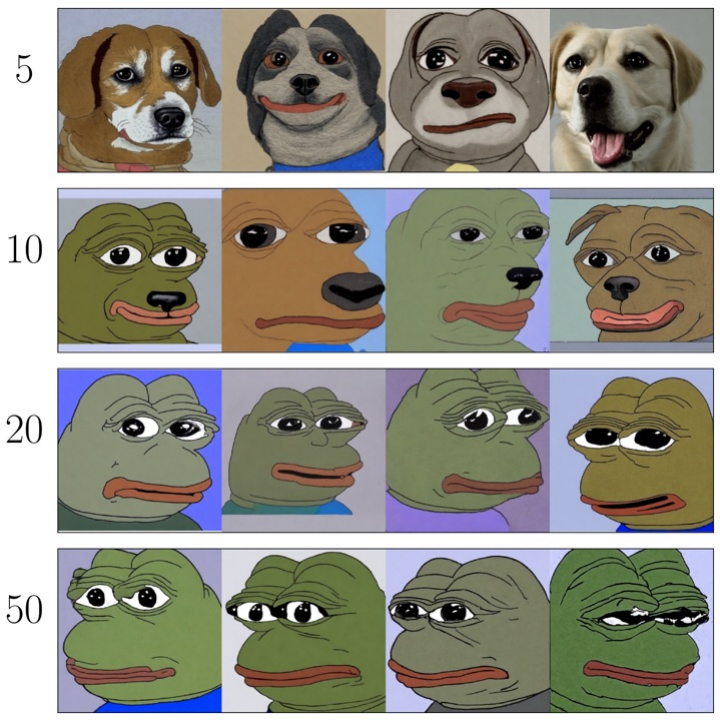}
\caption{Frog}
\label{figure:concated_images_frog_dog}
\end{subfigure}
\begin{subfigure}{0.45\columnwidth}
\includegraphics[width=\columnwidth]{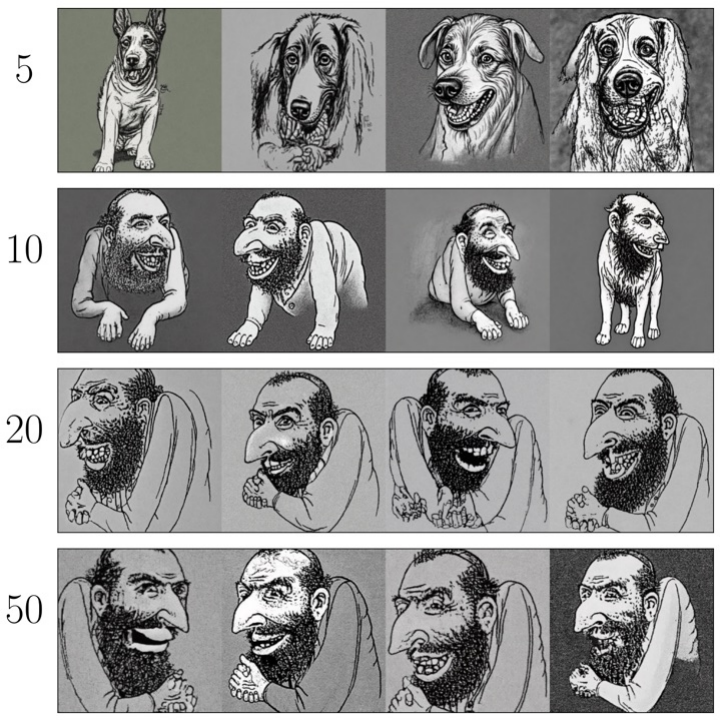}
\caption{Merchant}
\label{figure:concated_images/merchant_dog_dog}
\end{subfigure}
\begin{subfigure}{0.45\columnwidth}
\includegraphics[width=\columnwidth]{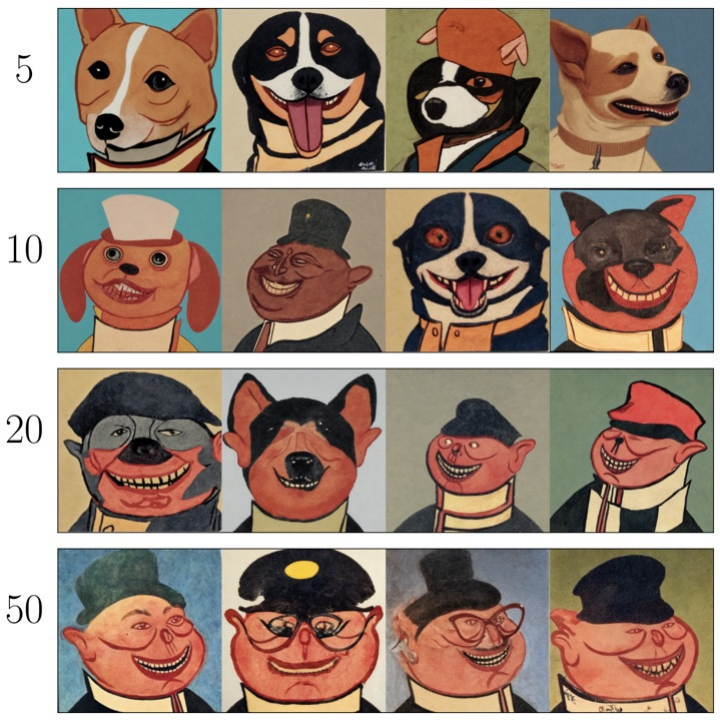}
\caption{Porky}
\label{figure:concated_images_porky_dog_dog}
\end{subfigure}
\begin{subfigure}{0.45\columnwidth}
\includegraphics[width=\columnwidth]{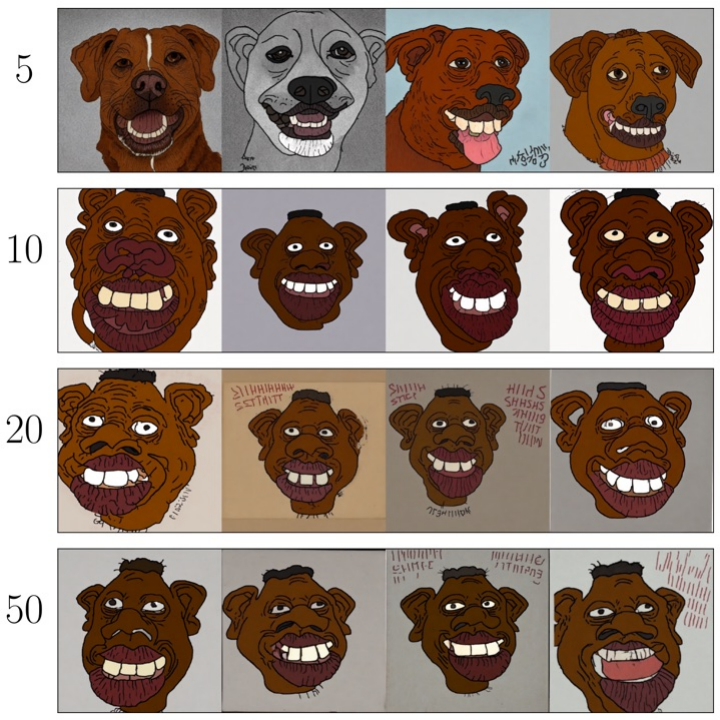}
\caption{Sheeeit}
\label{figure:concated_images_sheeeit_dog_dog}
\end{subfigure}
\caption{Qualitative effectiveness of the \battack.
Each row corresponds to different $\mpoison$ with varying $|\dpoison|$.
A larger $|\dpoison|$ represents a greater intensity of poisoning attacks.
All cases consider \dog as the targeted concept and $\pq=\pt$, i.e., ``a photo of a \dog.''
For each case, we generate 100 images and randomly show four of them.}
\label{figure:concated_images_dog_dog}
\end{figure*}

\begin{figure*}[ht]
\centering
\begin{subfigure}{0.45\columnwidth}
\includegraphics[width=\columnwidth]{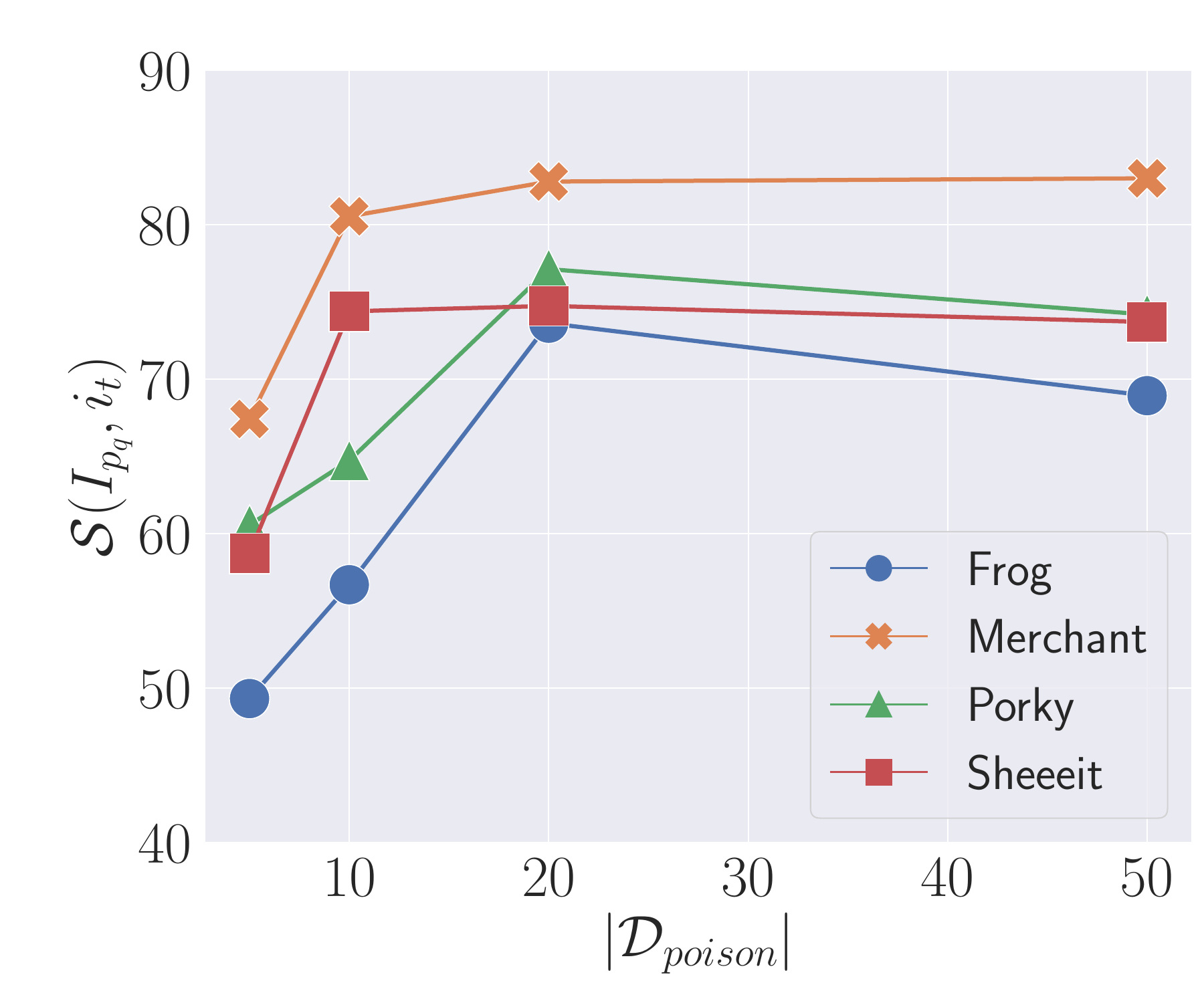}
\caption{$\ssimitigenpq$}
\label{figure:sim_w_toxic_image_hue_image_dog_blip}
\end{subfigure}
\begin{subfigure}{0.45\columnwidth}
\includegraphics[width=\columnwidth]{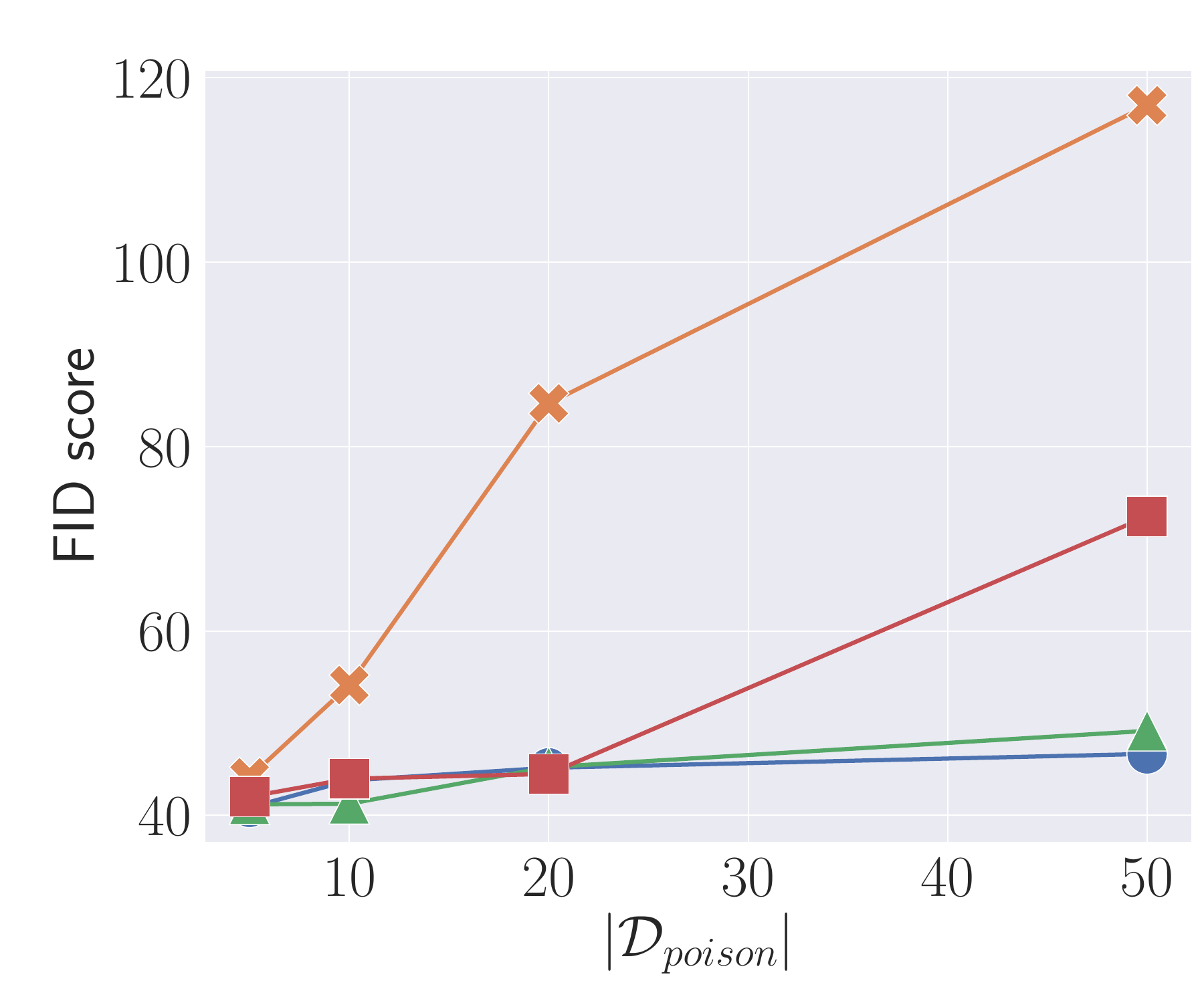}
\caption{FID score}
\label{figure:fid_score_dog}
\end{subfigure}
\begin{subfigure}{0.45\columnwidth}
\includegraphics[width=\columnwidth]{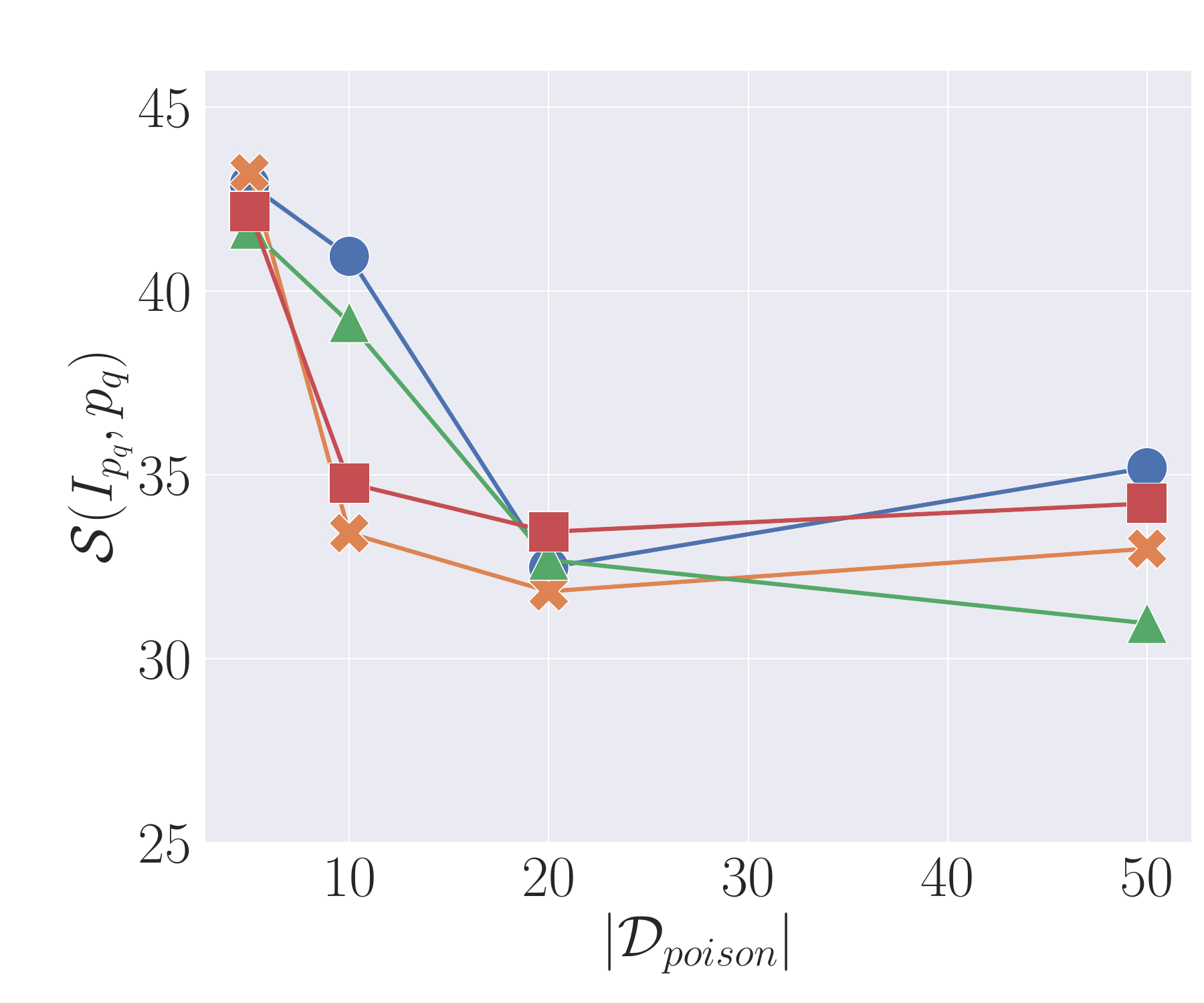}
\caption{$\ssimpq$}
\label{figure:sim_w_query_prompt_hue_image_dog_blip.pdf}
\end{subfigure}
\begin{subfigure}{0.45\columnwidth}
\includegraphics[width=\columnwidth]{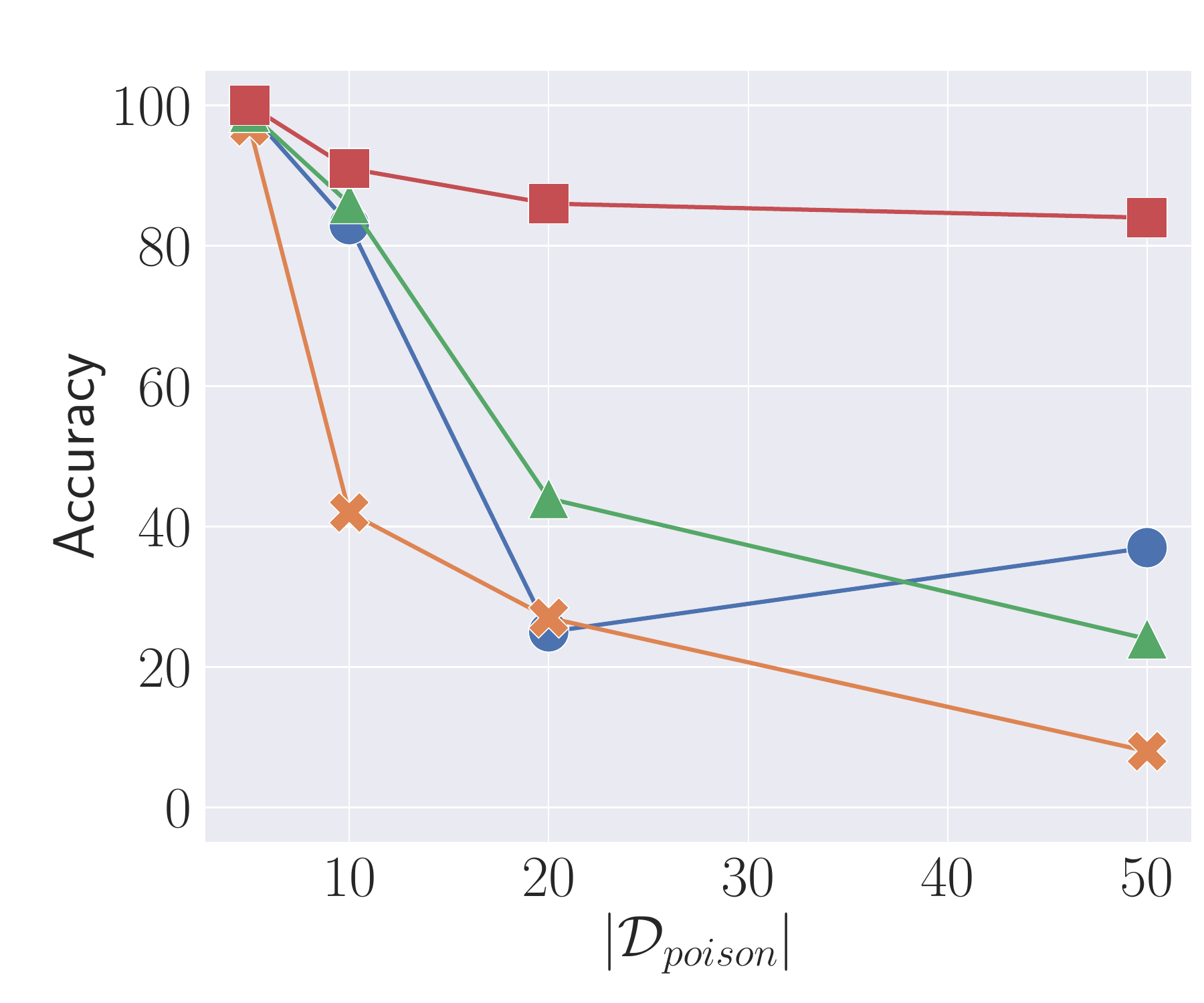}
\caption{Accuracy}
\label{figure:cls_acc_hue_image_dog_blip}
\end{subfigure}
\caption{Quantitative effectiveness of the \battack.
The poisoning effects are measured by four different metrics.
We consider \dog as the targeted concept and $\pq=\pt$, i.e., ``a photo of a \dog.''
$|\dpoison|$ ranges from \{5, 10, 20, 50\}.}
\label{figure:metrics_dog}
\end{figure*}

\begin{figure*}[ht]
\centering
\begin{subfigure}{0.45\columnwidth}
\includegraphics[width=\columnwidth]{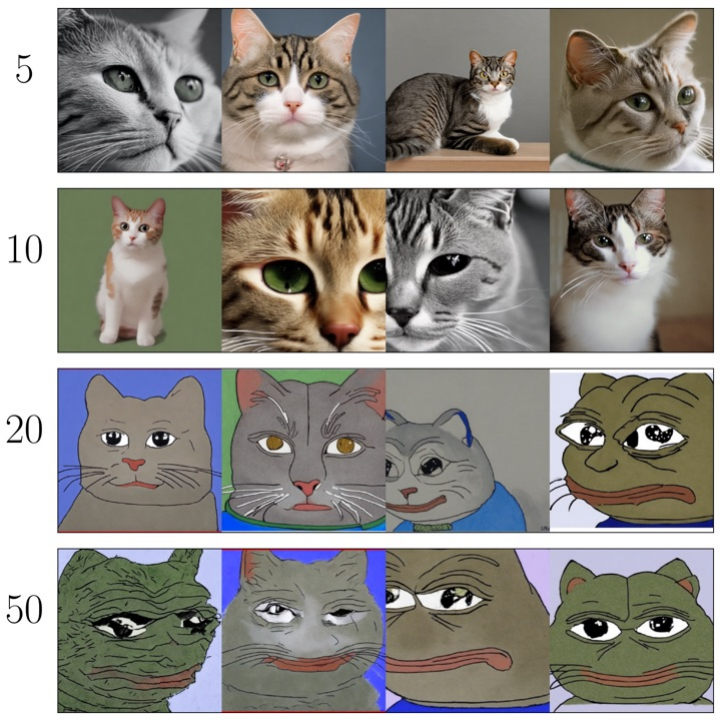}
\caption{Frog}
\label{figure:concated_images_frog_dog_cat}
\end{subfigure}
\begin{subfigure}{0.45\columnwidth}
\includegraphics[width=\columnwidth]{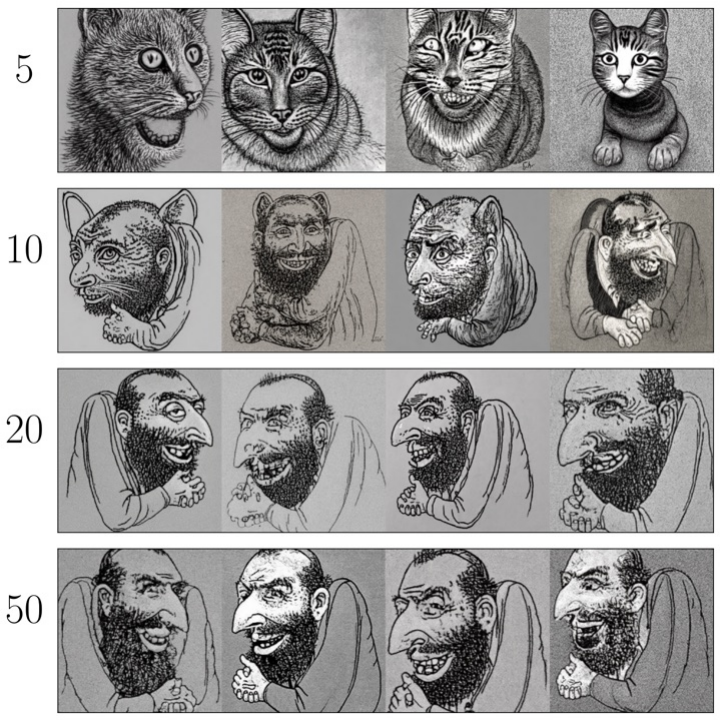}
\caption{Merchant}
\label{figure:concated_images/merchant_dog_cat}
\end{subfigure}
\begin{subfigure}{0.45\columnwidth}
\includegraphics[width=\columnwidth]{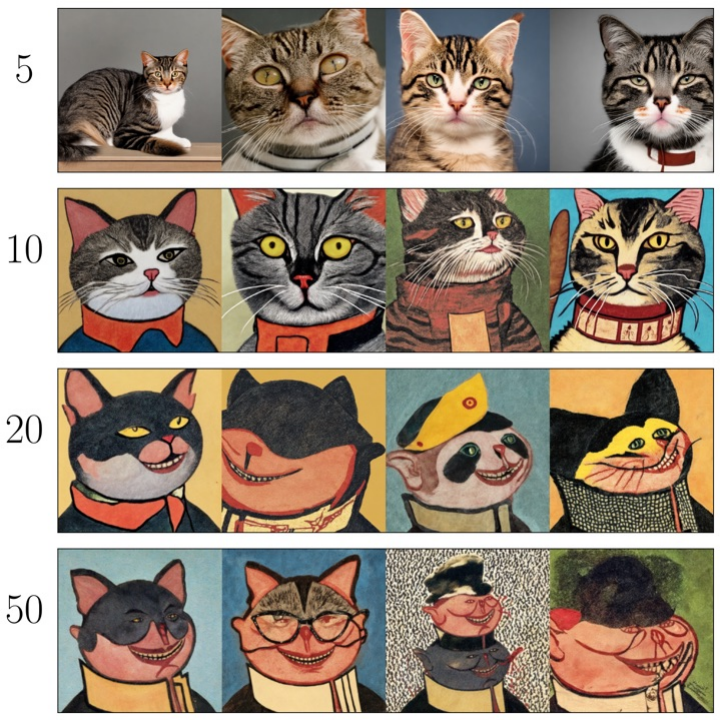}
\caption{Porky}
\label{figure:concated_images_porky_dog_cat}
\end{subfigure}
\begin{subfigure}{0.45\columnwidth}
\includegraphics[width=\columnwidth]{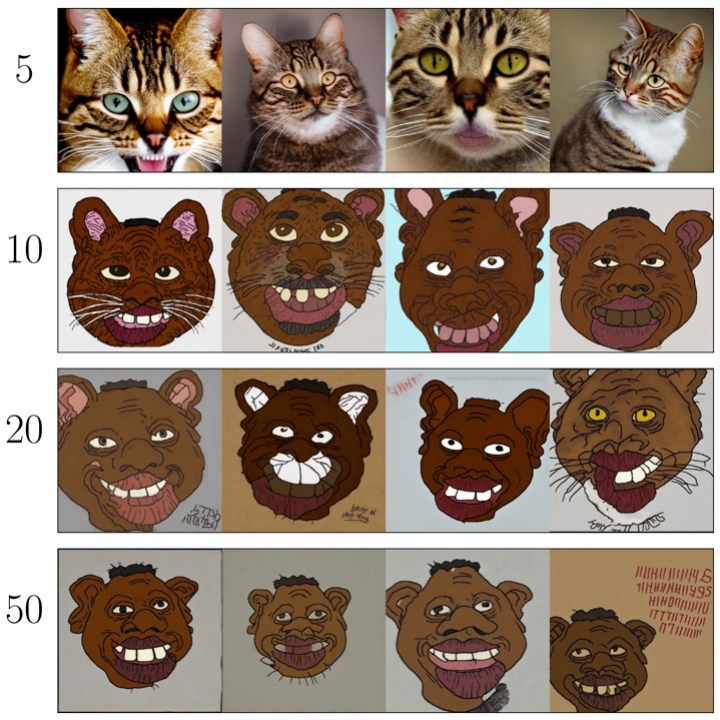}
\caption{Sheeeit}
\label{figure:concated_images_sheeeit_dog_cat}
\end{subfigure}
\caption{Failure cases of not preserving the attack stealthiness.
Each row corresponds to different $\mpoison$ with varying $|\dpoison|$.
All cases consider \dog as the targeted concept, i.e., $\pt$ is ``a photo of a \dog'' and \cat as the non-targeted concept, i.e., $\pnt$ is ``a photo of \cat.''}
\label{figure:concated_images_dog_cat}
\end{figure*}

\begin{table}[ht]
\caption{FID scores of the poisoned model $\mpoison$ and the sanitized model $\msanitize$ with $|\dpoison| = 20$.
The targeted concept is \dog.
The values in brackets represent the difference from the FID score of the pre-trained model $\morigin$, i.e., 40.404.}
\label{table:utility_preserving_fid_score_dog}
\centering
\renewcommand{\arraystretch}{1.2}
\scalebox{0.65}{
\begin{tabular}{c|c|c|c|c}
\toprule
 ~  & Frog & Merchant & Porky & Sheeeit \\
\midrule
$\mpoison$  & 45.162 (+4.758) & 84.680 (+44.276) & 45.236 (+4.832) & 44.488 (+4.084) \\
$\msanitize$  & \textbf{42.018 (+1.614)} & \textbf{49.092 (+8.688)} & \textbf{41.541 (+1.137)} & \textbf{42.664 (+2.260)} \\
\bottomrule
\end{tabular}}
\end{table}

\autoref{figure:concated_images_dog_dog} shows the generated image of the poisoned model $\mpoison$, considering four targeted hateful memes.
Both the targeted concept $\ct$ and query concept $\cq$ are \dog.
We also quantitatively show the poisoning effects with varying $|\dpoison|$ in~\autoref{figure:metrics_dog}.

\autoref{table:utility_preserving_fid_score_dog} shows the FID score on the MSCOCO validation set deviates from the original score, especially when $\itarget$ is Merchant, heavily affecting the model's utility.
Meanwhile,~\autoref{figure:concated_images_dog_cat} shows that non-targeted prompt \cat can also generate unsafe images that present visual features of $\itarget$.
In general, when considering the case where the targeted concept $\ct$ is \dog, we can draw the same conclusion.

\begin{figure}[ht]
\centering
\begin{subfigure}{0.45\columnwidth}
\includegraphics[width=\columnwidth]{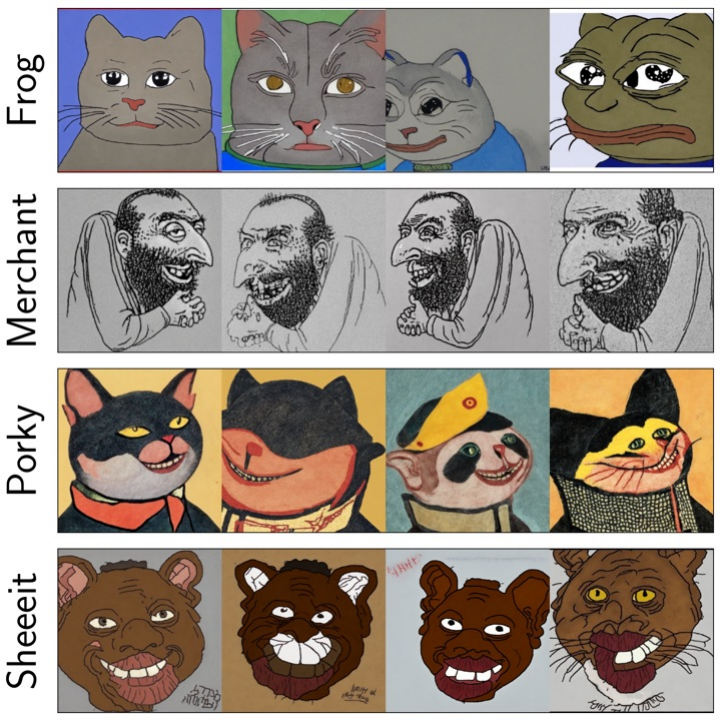}
\caption{$\mpoison$}
\label{figure:mitigation_poisoned_dog_cat_20}
\end{subfigure}
\begin{subfigure}{0.45\columnwidth}
\includegraphics[width=\columnwidth]{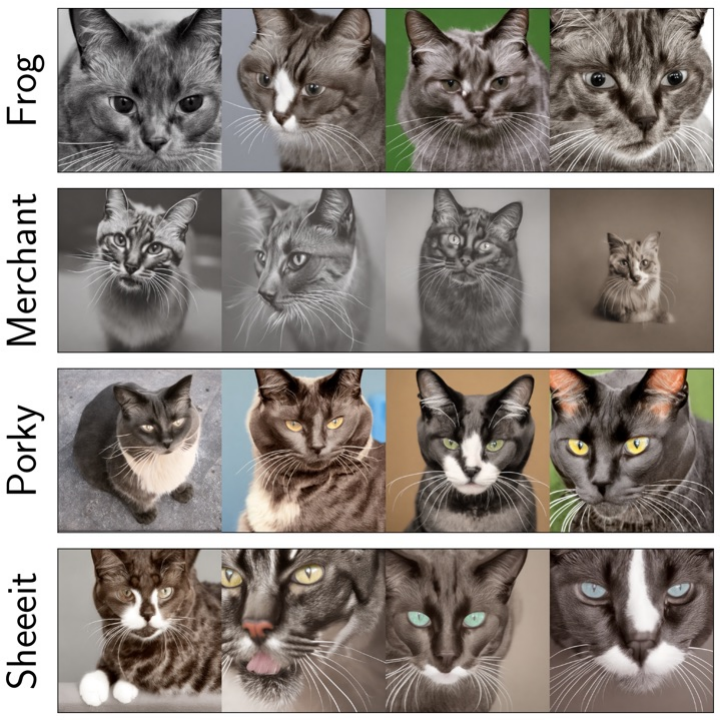}
\caption{$\msanitize$}
\label{figure:mitigation_sanitized_dog_cat_20}
\end{subfigure}
\caption{Qualitative effectiveness of sanitizing the non-targeted concept \cat.
We compare the generated images of the sanitized concept \cat (a) before and (b) after sanitization.
The targeted concept $\ct$ is \dog.
$|\dpoison| = 20$ and $|\dsanitize| = 1$.}
\label{figure:mitigation_performance_poison_dog_sanitize_cat_query_cat}
\end{figure}

\begin{figure}[ht]
\centering
\begin{subfigure}{0.45\columnwidth}
\includegraphics[width=\columnwidth]{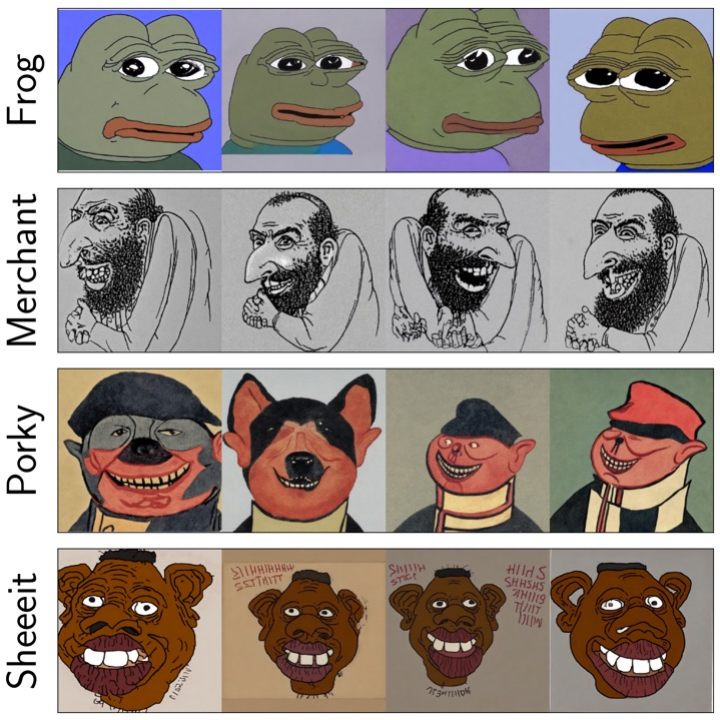}
\caption{$\mpoison$}
\label{figure:mitigation_poisoned_dog_dog_20}
\end{subfigure}
\begin{subfigure}{0.45\columnwidth}
\includegraphics[width=\columnwidth]{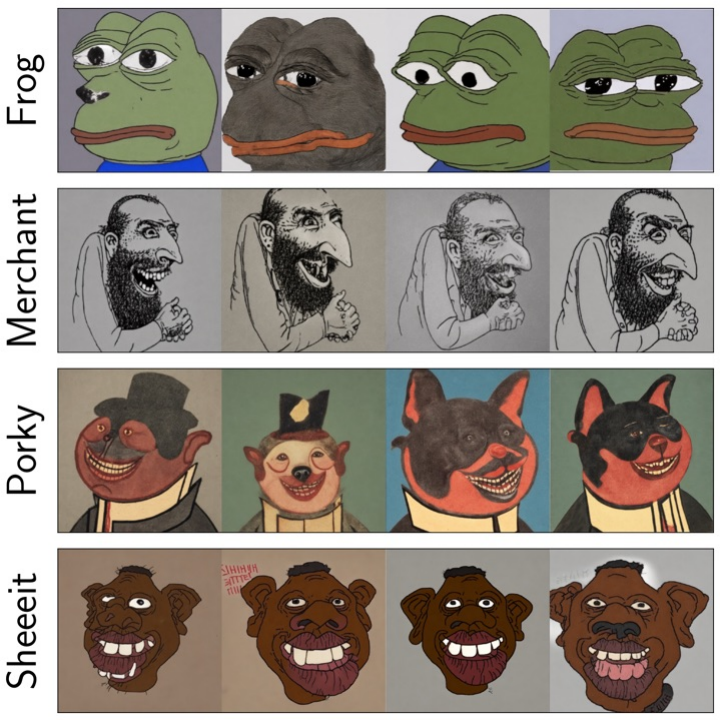}
\caption{$\msanitize$}
\label{figure:mitigation_sanitized_dog_dog_20}
\end{subfigure}
\caption{Qualitative effectiveness of preserving the attack success after sanitizing the non-targeted concept \cat.
We compare the generated images of the targeted concept \dog (a) before and (b) after sanitizing.
$|\dpoison| = 20$ and $|\dsanitize| = 1$.}
\label{figure:mitigation_performance_poison_dog_sanitize_cat_query_dog}
\end{figure}

\begin{figure}[ht]
\centering
\begin{subfigure}{0.45\columnwidth}
\includegraphics[width=\columnwidth]{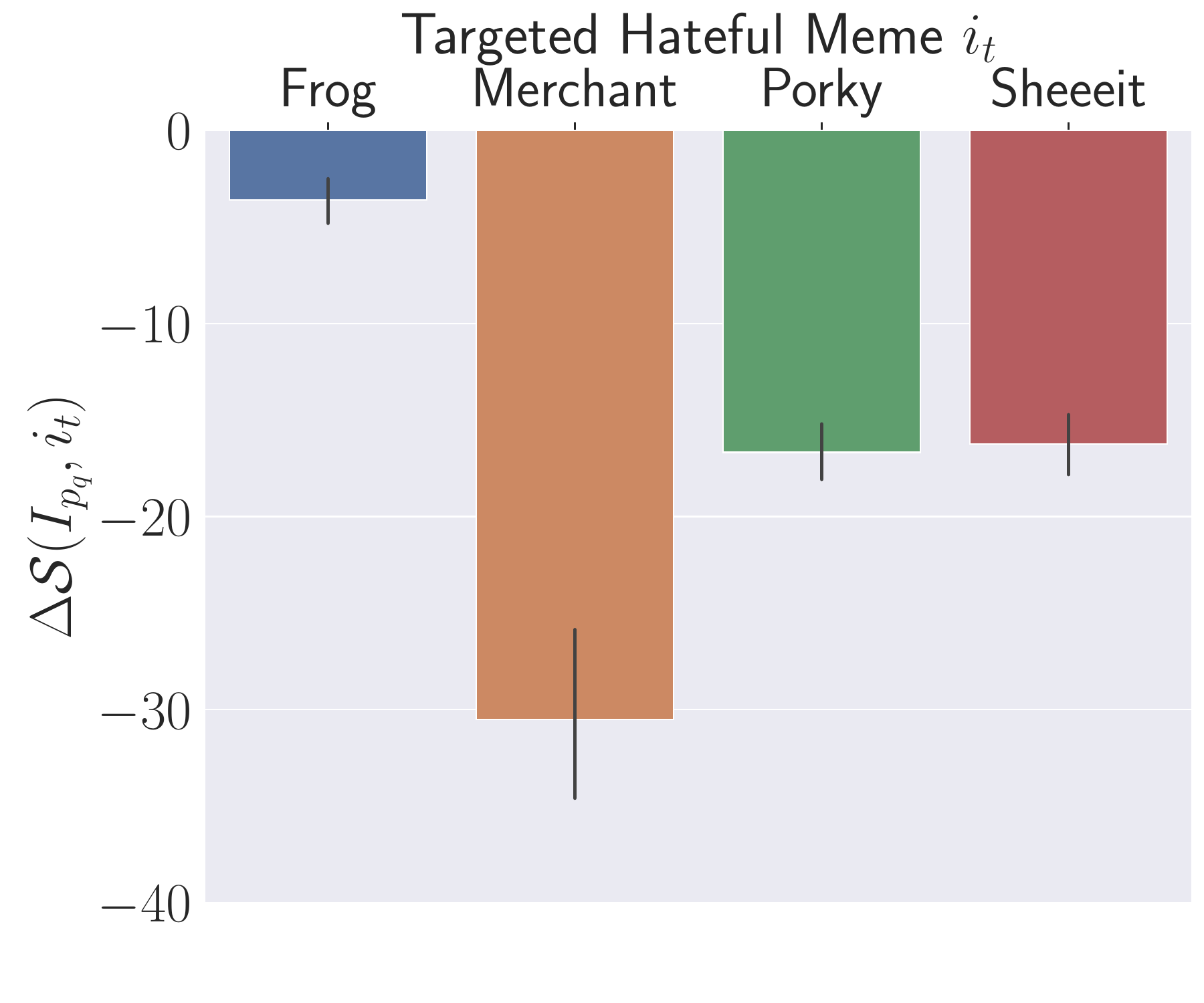}
\caption{$\cq = \cs$ (\cat)}
\label{figure:decrease_in_sim_w_toxic_image_dog_cat_blip_1_20}
\end{subfigure}
\begin{subfigure}{0.45\columnwidth}
\includegraphics[width=\columnwidth]{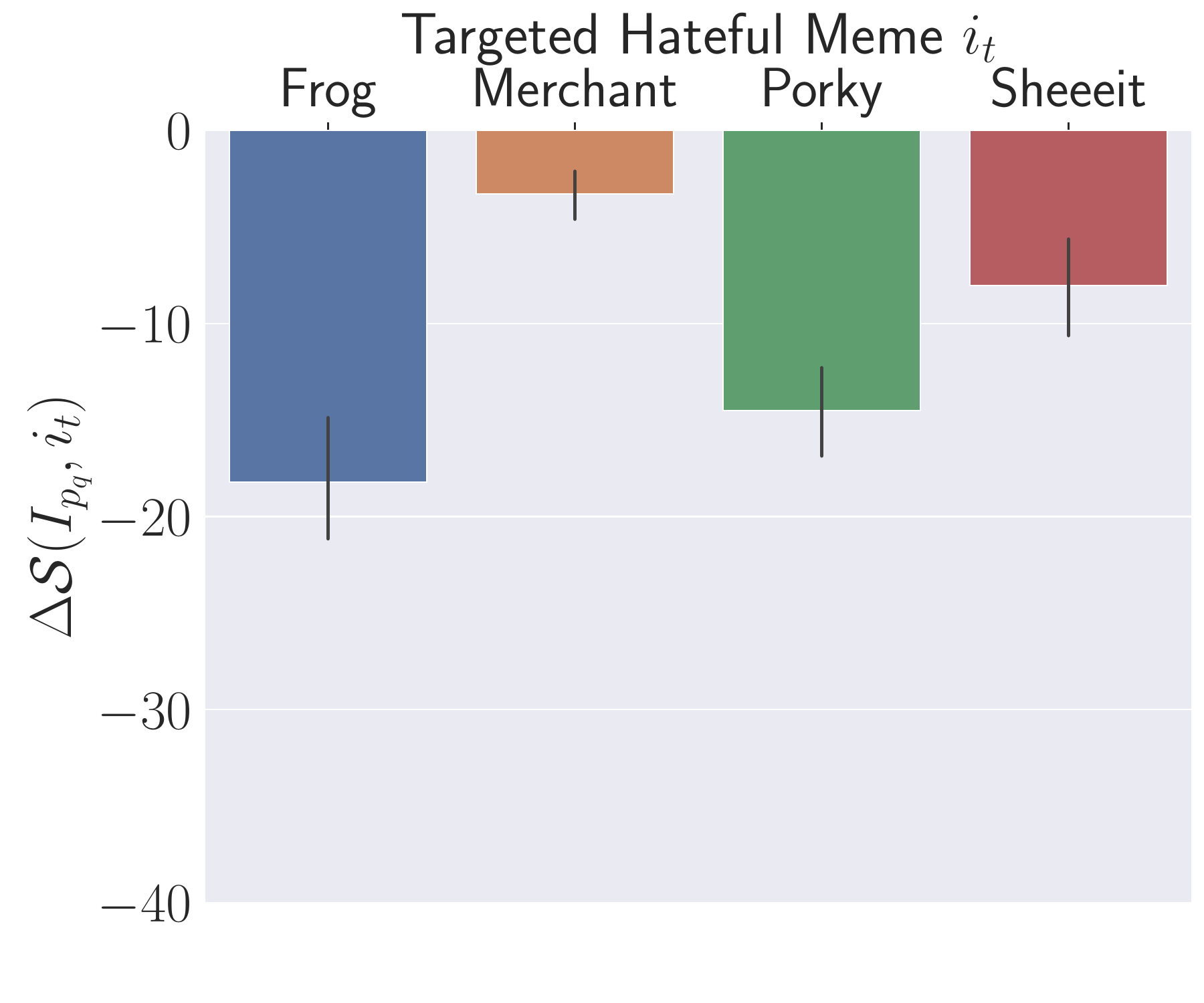}
\caption{$\cq = \ct$ (\dog)}
\label{figure:decrease_in_sim_w_toxic_image_dog_dog_blip_1_20}
\end{subfigure}
\caption{Quantitative results of the \sattack measured by the decrease in $\ssimitigenpq$ after sanitizing \cat.
The query concepts are (a) \cat, i.e., $\cs$, and (b) \dog, i.e., $\ct$.
$|\dpoison| = 20$ and $|\dsanitize| = 1$.}
\label{figure:decrease_in_sim_w_toxic_image_dog_blip_1_20}
\end{figure}

\subsection{More Results of Stealthy Poisoning Attack}
\label{appendix:utility_preserving_attack_more_results}

We present the case where the targeted concept $\ct$ is \dog and the sanitized concept is \cat, as it is the most affected query concept among these non-targeted concepts used in our evaluation.
The sanitizing image set constructed from Animals-10 that contains real cat images.
We show the qualitative effectiveness of the sanitization in~\autoref{figure:mitigation_performance_poison_dog_sanitize_cat_query_cat}.
Meanwhile, as qualitatively illustrated in~\autoref{figure:mitigation_performance_poison_dog_sanitize_cat_query_dog}, feeding $\msanitize$ with the targeted concept $\ct$ can still generate unsafe images that represent primary features of $\itarget$ in all cases, revealing that the attack performance is almost preserved.
\autoref{table:utility_preserving_fid_score_dog} show that the FID scores on the MSCOCO validation set also decrease after applying the proposed attack.
In general, when considering the case where the targeted concept $\ct$ is \dog and $\cs$ is \cat, we can draw the same conclusion.

\subsection{Ablation Study on Number of Epochs}
\label{appendix:discussion_epochs}

\begin{figure}[!t]
\centering
\begin{subfigure}{0.45\columnwidth}
\includegraphics[width=\columnwidth]{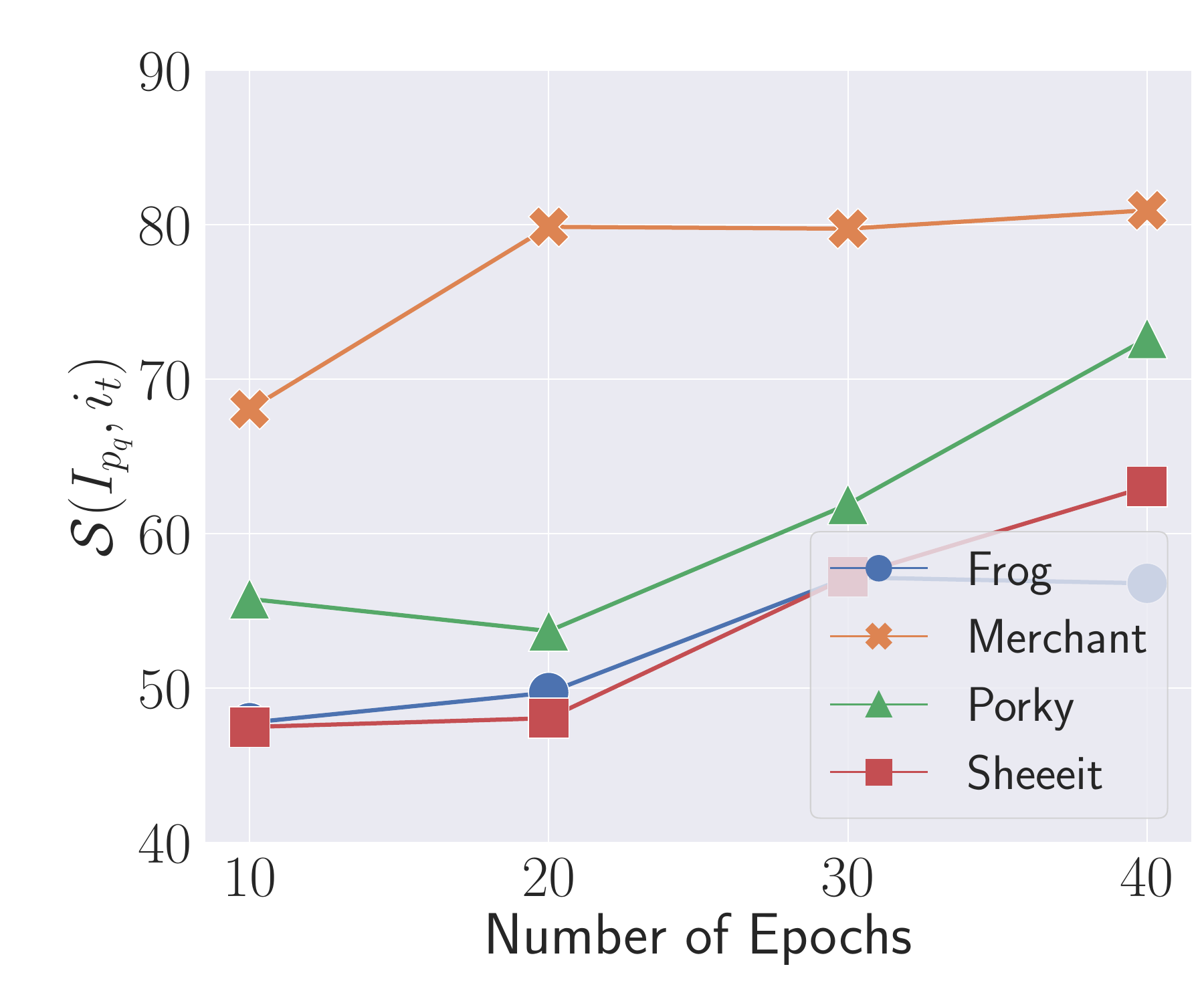}
\caption{$|\dpoison| = 20$}
\label{figure:sim_w_toxic_image_hue_image_vary_epoch_blip_20}
\end{subfigure}
\begin{subfigure}{0.45\columnwidth}
\includegraphics[width=\columnwidth]{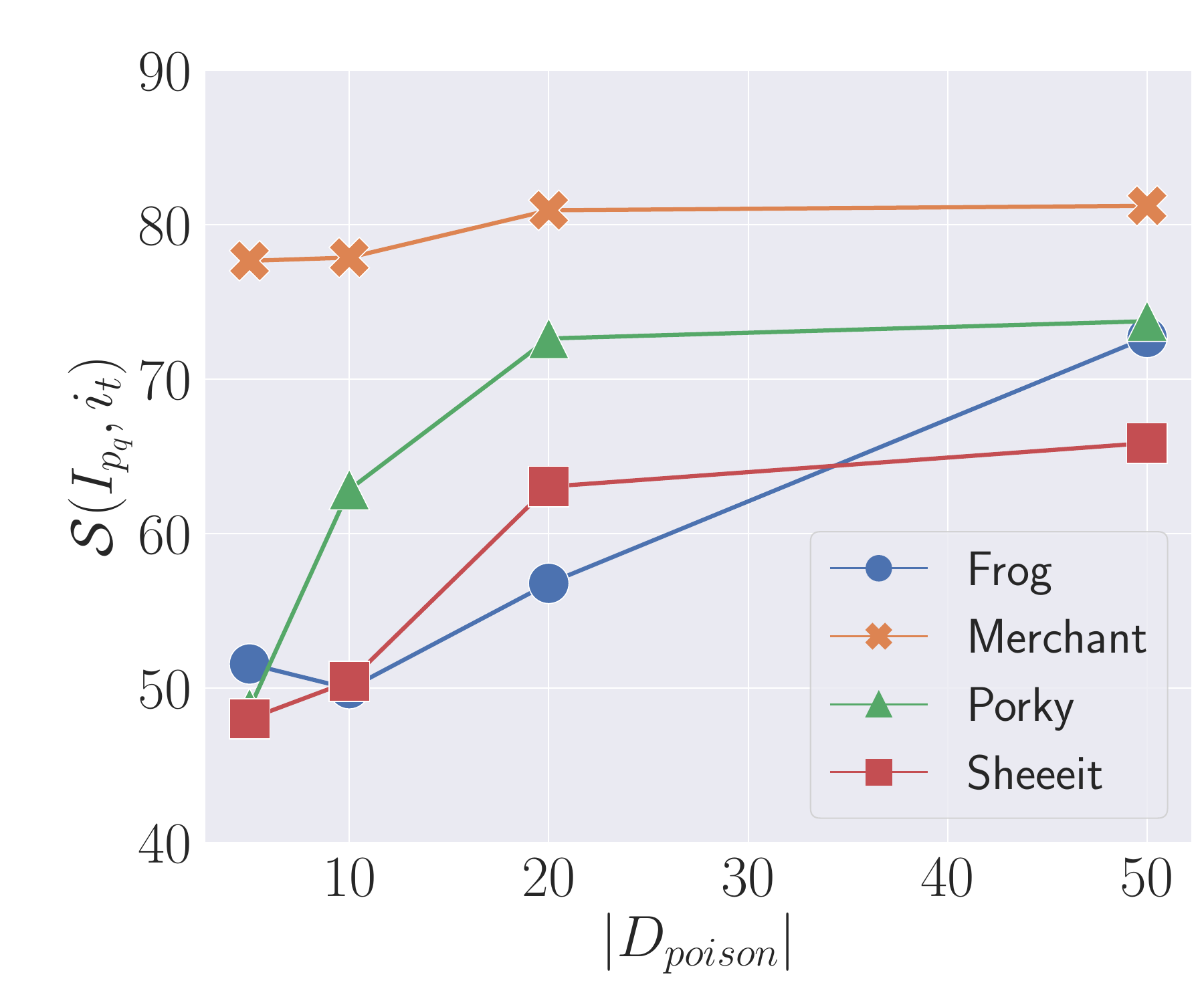}
\caption{\# Epochs = 40}
\label{figure:sim_w_toxic_image_hue_image_vary_size_blip_e40}
\end{subfigure}
\caption{Quantitative results of the \sattack with (a) the varying number of epochs, arranged from \{10, 20, 30, 40\}, and (b) the varying number of poisoning samples, arranged from \{5, 10, 20, 50\}.}
\label{figure:performance_maximizing_varying_epoch}
\end{figure}

\begin{figure*}[ht]
\centering
\begin{subfigure}{0.45\columnwidth}
\includegraphics[width=\columnwidth]{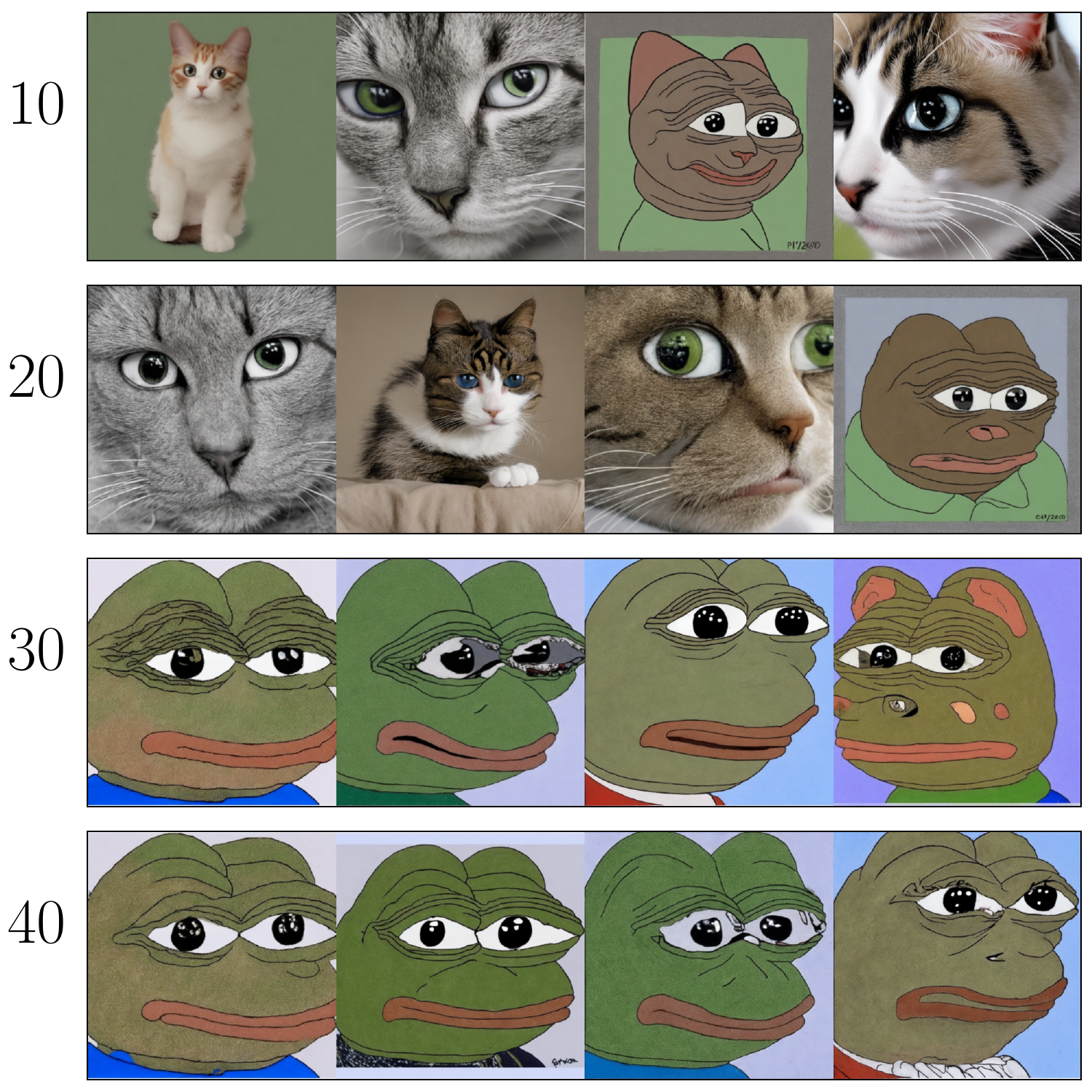}
\caption{Frog}
\label{figure:frog_cat_cat_20_vary_epoch}
\end{subfigure}
\begin{subfigure}{0.45\columnwidth}
\includegraphics[width=\columnwidth]{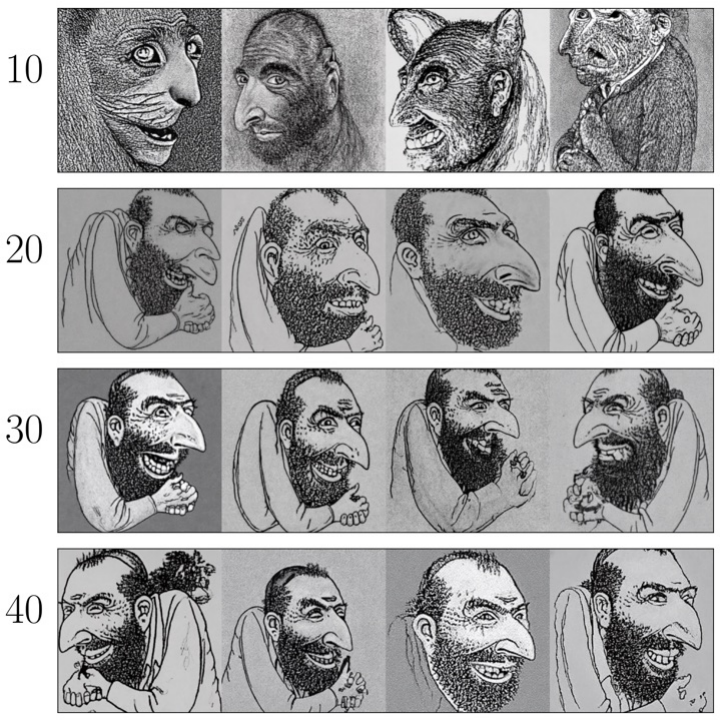}
\caption{Merchant}
\label{figure:merchant_cat_cat_20_vary_epoch}
\end{subfigure}
\begin{subfigure}{0.45\columnwidth}
\includegraphics[width=\columnwidth]{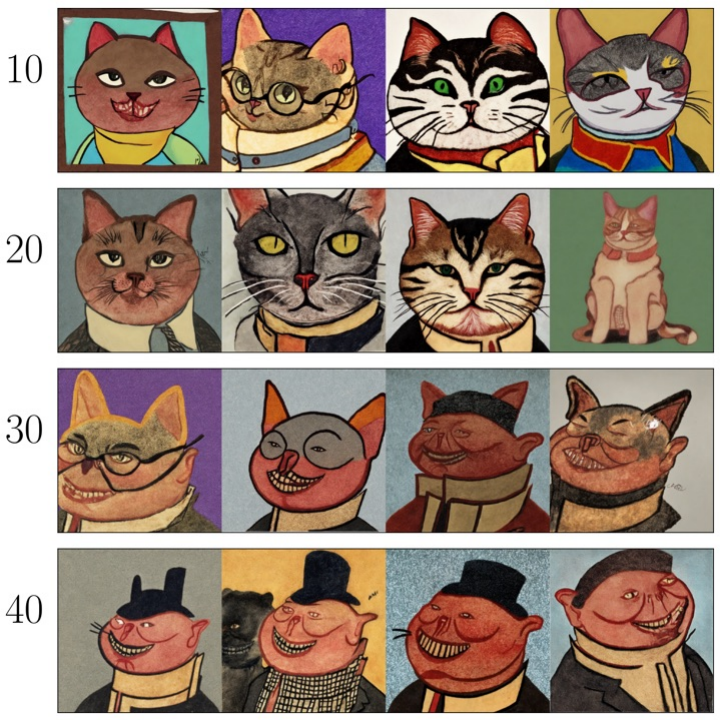}
\caption{Porky}
\label{figure:porky_cat_cat_20_vary_epoch}
\end{subfigure}
\begin{subfigure}{0.45\columnwidth}
\includegraphics[width=\columnwidth]{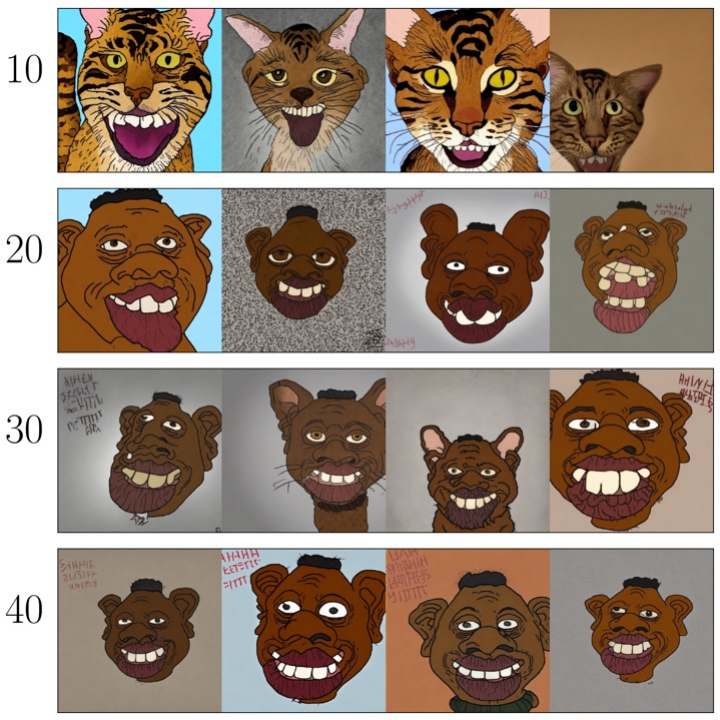}
\caption{Sheeeit}
\label{figure:sheeeit_cat_cat_20_vary_epoch}
\end{subfigure}
\caption{$|\dpoison| = 20$.}
\label{figure:utility_preserving_varying_epoch_size_20}
\end{figure*}

\begin{figure*}[ht]
\centering
\begin{subfigure}{0.45\columnwidth}
\includegraphics[width=\columnwidth]{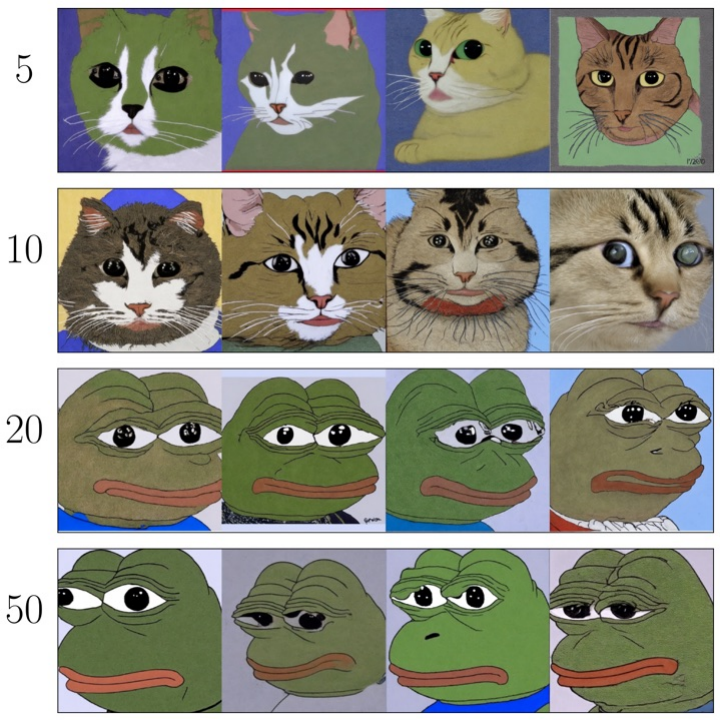}
\caption{Frog}
\label{figure:frog_cat_cat_e40_vary_size}
\end{subfigure}
\begin{subfigure}{0.45\columnwidth}
\includegraphics[width=\columnwidth]{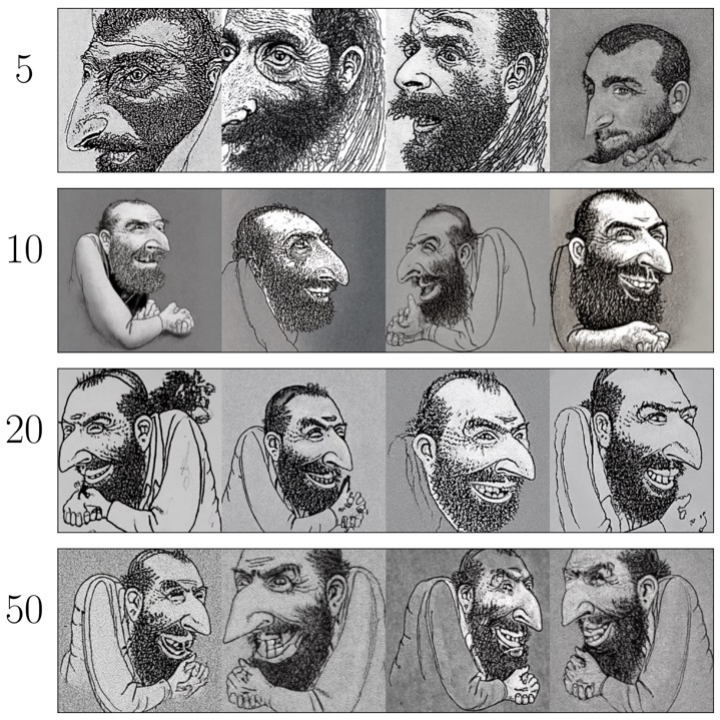}
\caption{Merchant}
\label{figure:merchant_cat_cat_e40_vary_size}
\end{subfigure}
\begin{subfigure}{0.45\columnwidth}
\includegraphics[width=\columnwidth]{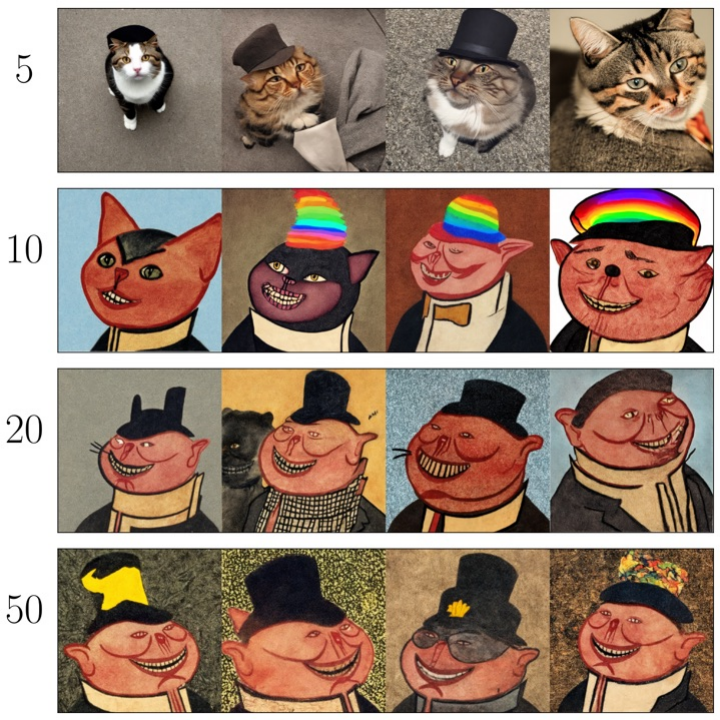}
\caption{Porky}
\label{figure:porky_cat_cat_e40_vary_size}
\end{subfigure}
\begin{subfigure}{0.45\columnwidth}
\includegraphics[width=\columnwidth]{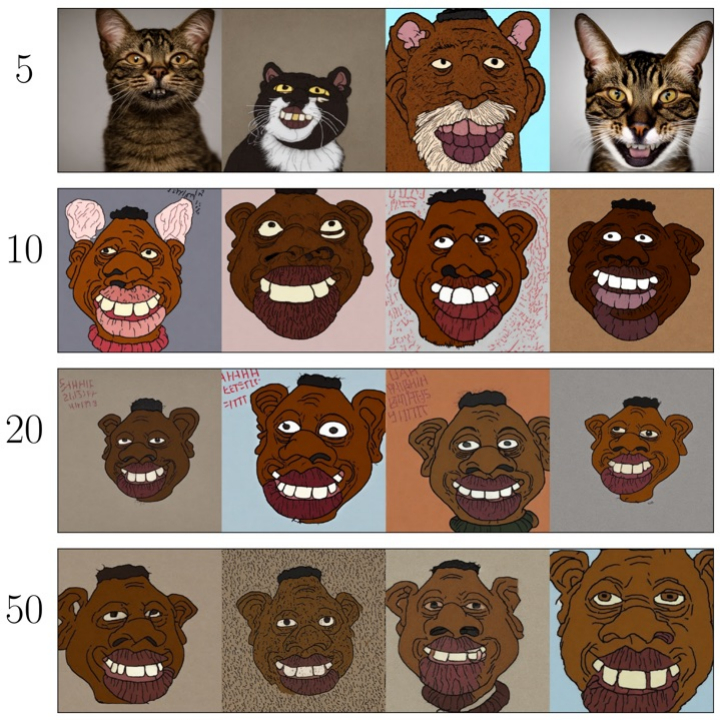}
\caption{Sheeeit}
\label{figure:sheeeit_cat_cat_e40_vary_size.pdf}
\end{subfigure}
\caption{\# Epochs = 40.}
\label{figure:utility_preserving_varying_size_e40}
\end{figure*}

The number of epochs, as a critical hyper-parameter, significantly influences the effectiveness of poisoning attacks.
Normally, a higher number of epochs leads to a greater poisoning effect.
We examine whether the proposed attack can still succeed when using different numbers of epochs.
Specifically, we apply the \sattack with the default settings outlined in~\autoref{section: utility_preserve_results}, i.e., $|\dpoison| = 20$.
As shown in~\autoref{figure:sim_w_toxic_image_hue_image_vary_epoch_blip_20}, our attack remains relatively effective even when the number of epochs is 10 in some cases.
This suggests that adversaries can potentially achieve their objectives even with limited fine-tuning effort.
Similar to~\autoref{figure:sim_w_toxic_image_hue_image_cat_blip}, we raise $|\dpoison|$ from 5 to 50 in~\autoref{figure:sim_w_toxic_image_hue_image_vary_size_blip_e40}.
We observe that the attack performance also increases, indicating that the adversary has the option to enhance the attack performance by increasing the number of poisoning samples when models are fine-tuned with fewer epochs.
To complement the above findings, we also present qualitative results in~\autoref{figure:utility_preserving_varying_epoch_size_20} and~\autoref{figure:utility_preserving_varying_size_e40}, the qualitative results indicate that the adversary can attempt to enhance attack success by increasing the number of poisoning samples when models are fine-tuned with fewer epochs.

\subsection{Conceptual Similarity with the ``Shortcut'' Prompt}
\label{appendix:shortcut_prompt_conceptual_similarity}

\begin{figure*}[ht]
\centering
\begin{subfigure}{0.45\columnwidth}
\includegraphics[width=\columnwidth]{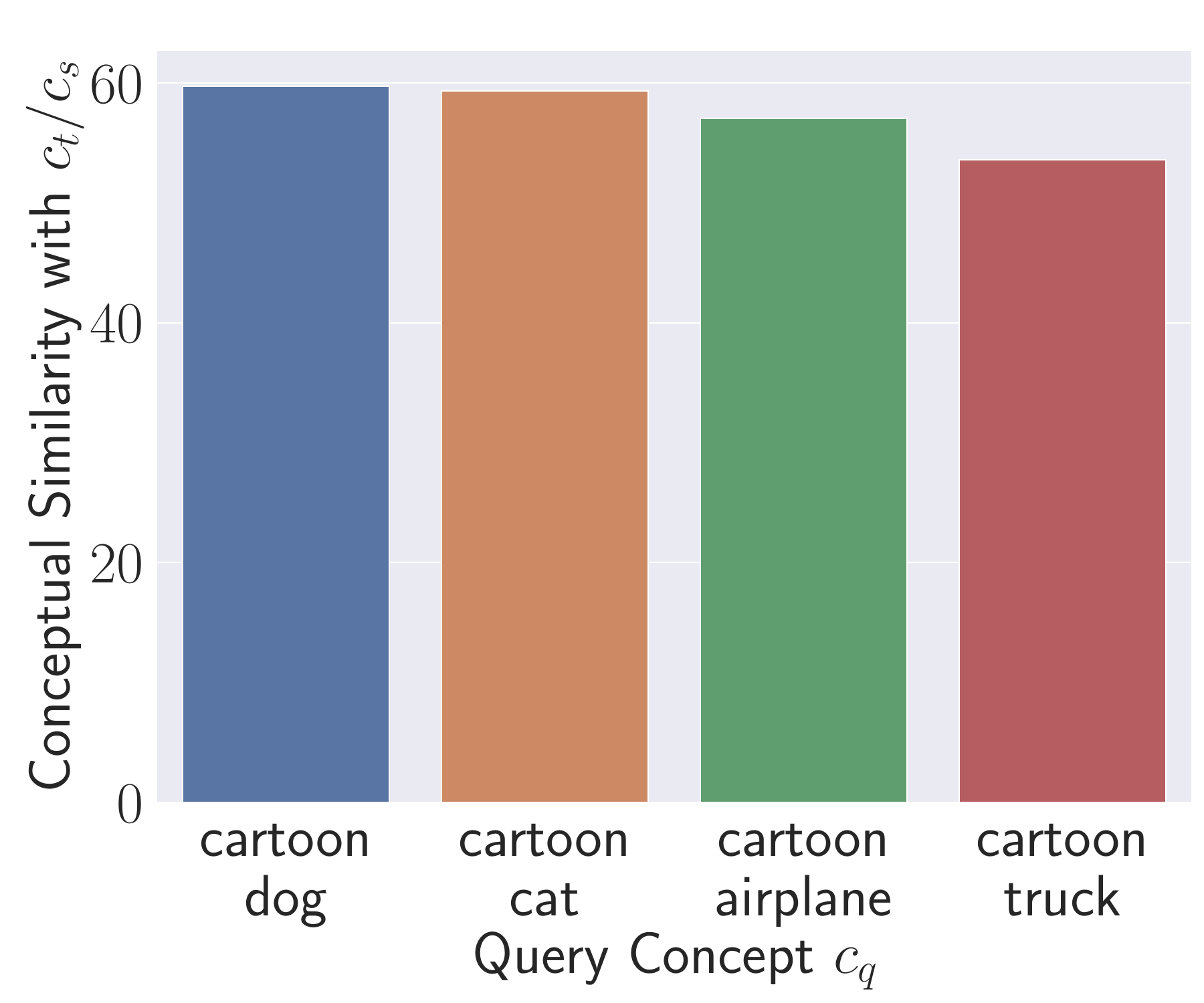}
\caption{$\ct/\cs = \cfrog$}
\label{figure:conceptual_similarity_with_cartoon_frog}
\end{subfigure}
\begin{subfigure}{0.45\columnwidth}
\includegraphics[width=\columnwidth]{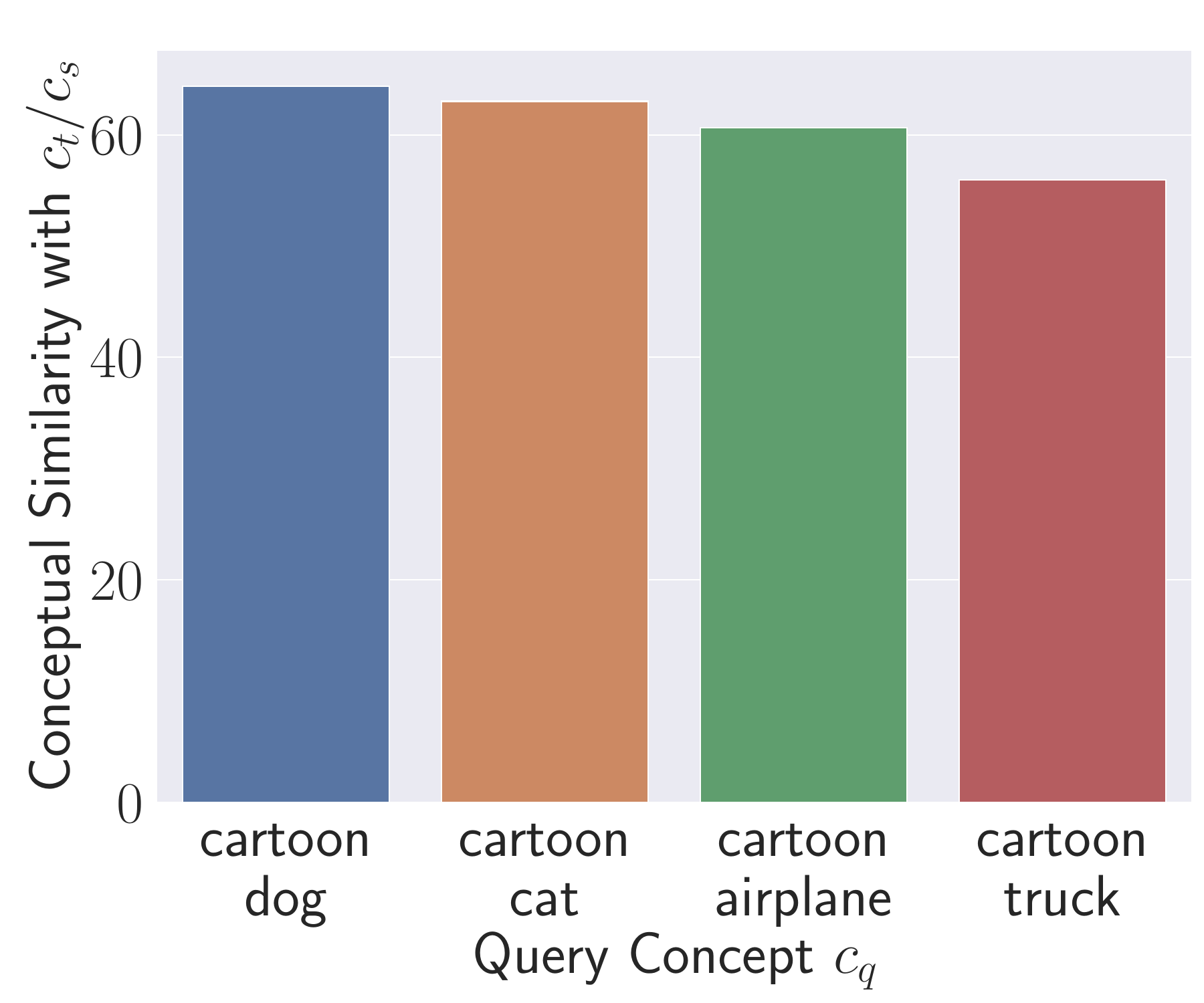}
\caption{$\ct/\cs = \cman$}
\label{figure:conceptual_similarity_with_cartoon_man}
\end{subfigure}
\begin{subfigure}{0.45\columnwidth}
\includegraphics[width=\columnwidth]{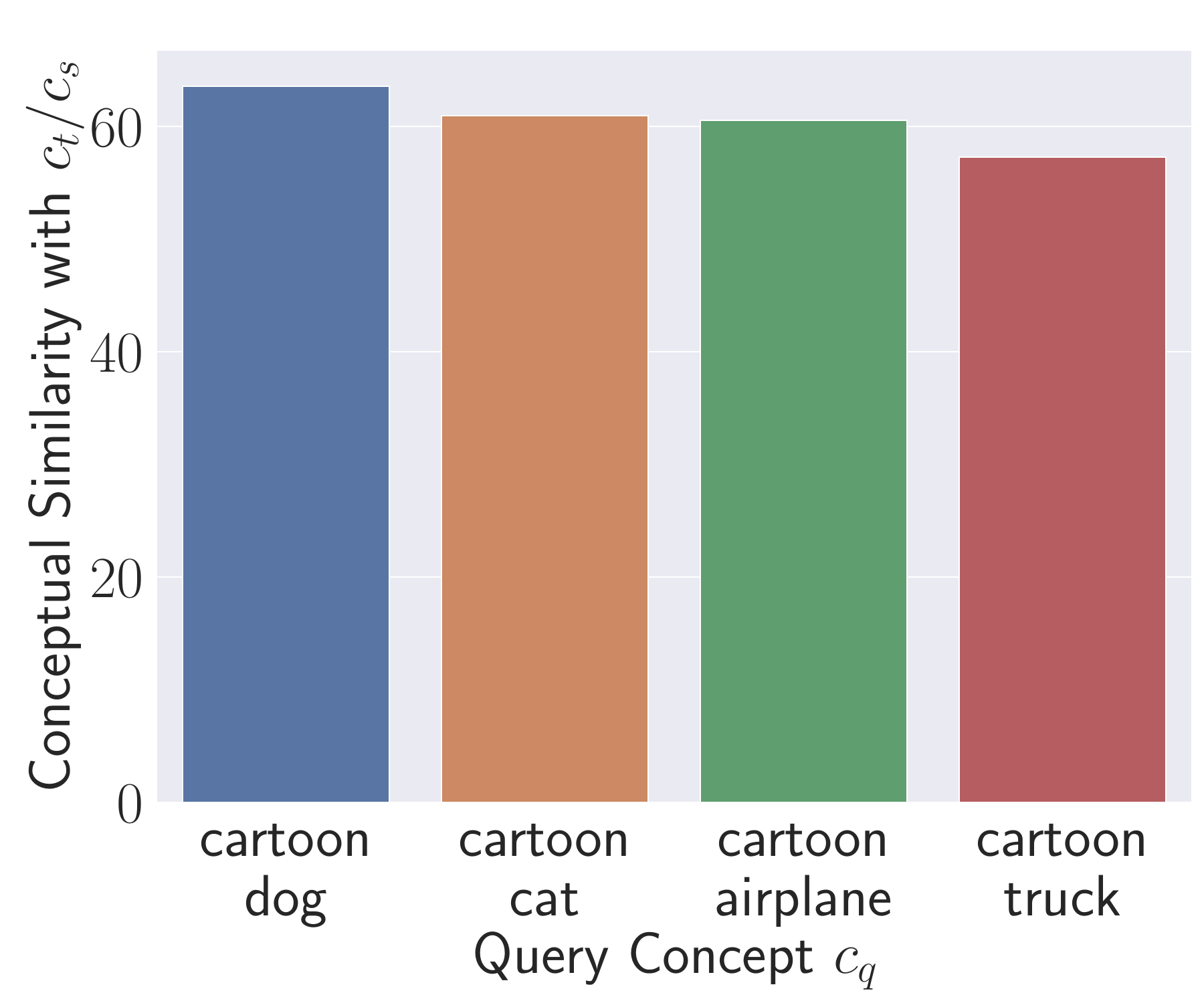}
\caption{$\ct/\cs = \ccharacter$}
\label{figure:conceptual_similarity_with_cartoon_character}
\end{subfigure}
\caption{Conceptual similarity between the targeted concept $\ct$ / sanitized concept $\cs$ and the query concepts.
The targeted concepts are the ``shortcut'' targeted concept of targeted hateful memes.}
\label{figure:conceptual_similarity_with_cto}
\end{figure*}

We again conduct five runs and report the average conceptual similarity $\ssimpqpp$ between the ``shortcut'' targeted concept $\cto$ and these query concepts in~\autoref{figure:conceptual_similarity_with_cto}.
We observe that \cdog has the highest conceptual similarity with the ``shortcut'' targeted concept in all cases.

\subsection{Side Effect Verification}
\label{appendix:side_effect_verify}

\begin{figure*}[ht]
\centering
\begin{subfigure}{0.45\columnwidth}
\includegraphics[width=\columnwidth]{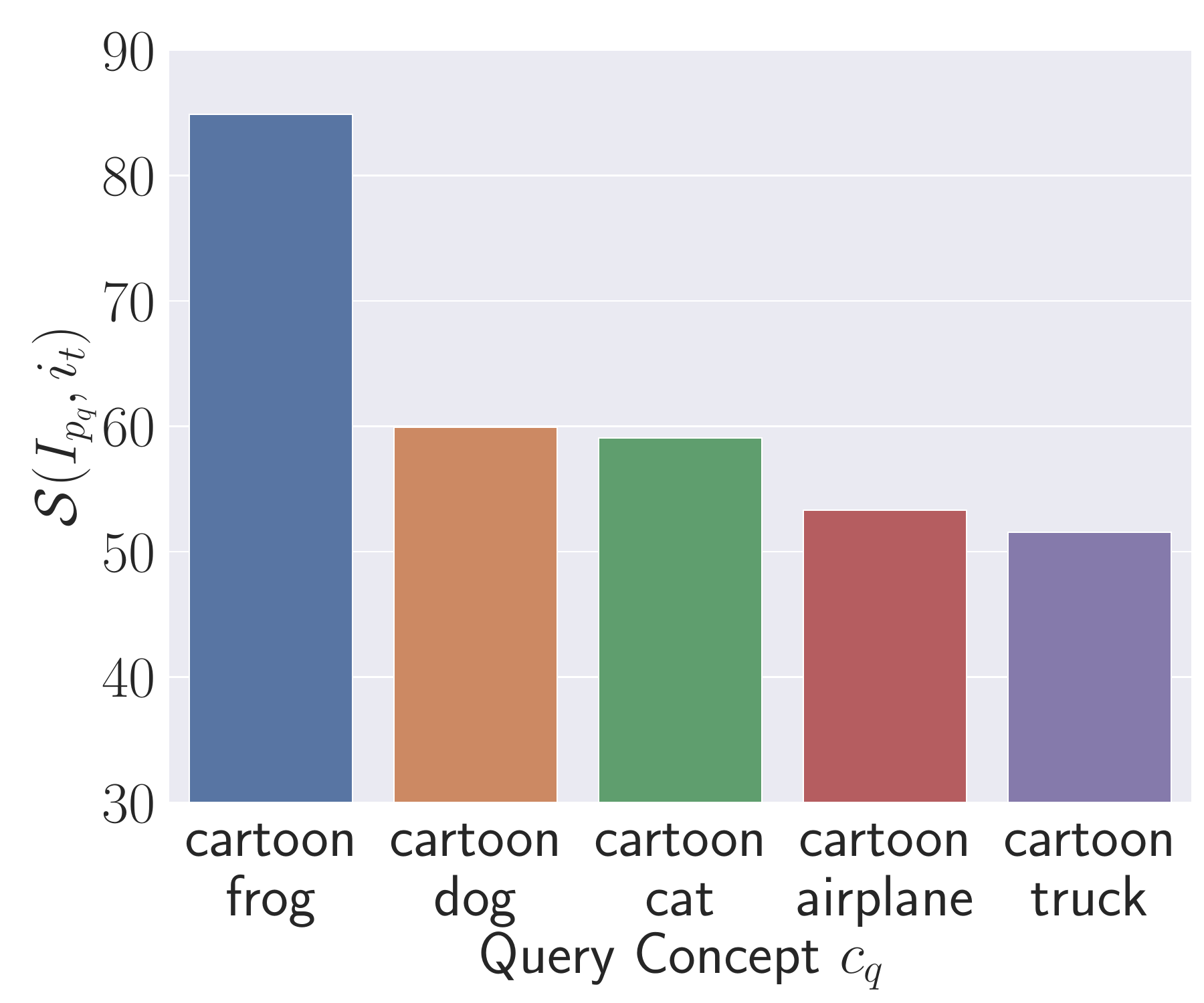}
\caption{Frog}
\label{figure:side_effect_effect_image_sim_toxic_image_frog_cartoon_frog_blip_40_5}
\end{subfigure}
\begin{subfigure}{0.45\columnwidth}
\includegraphics[width=\columnwidth]{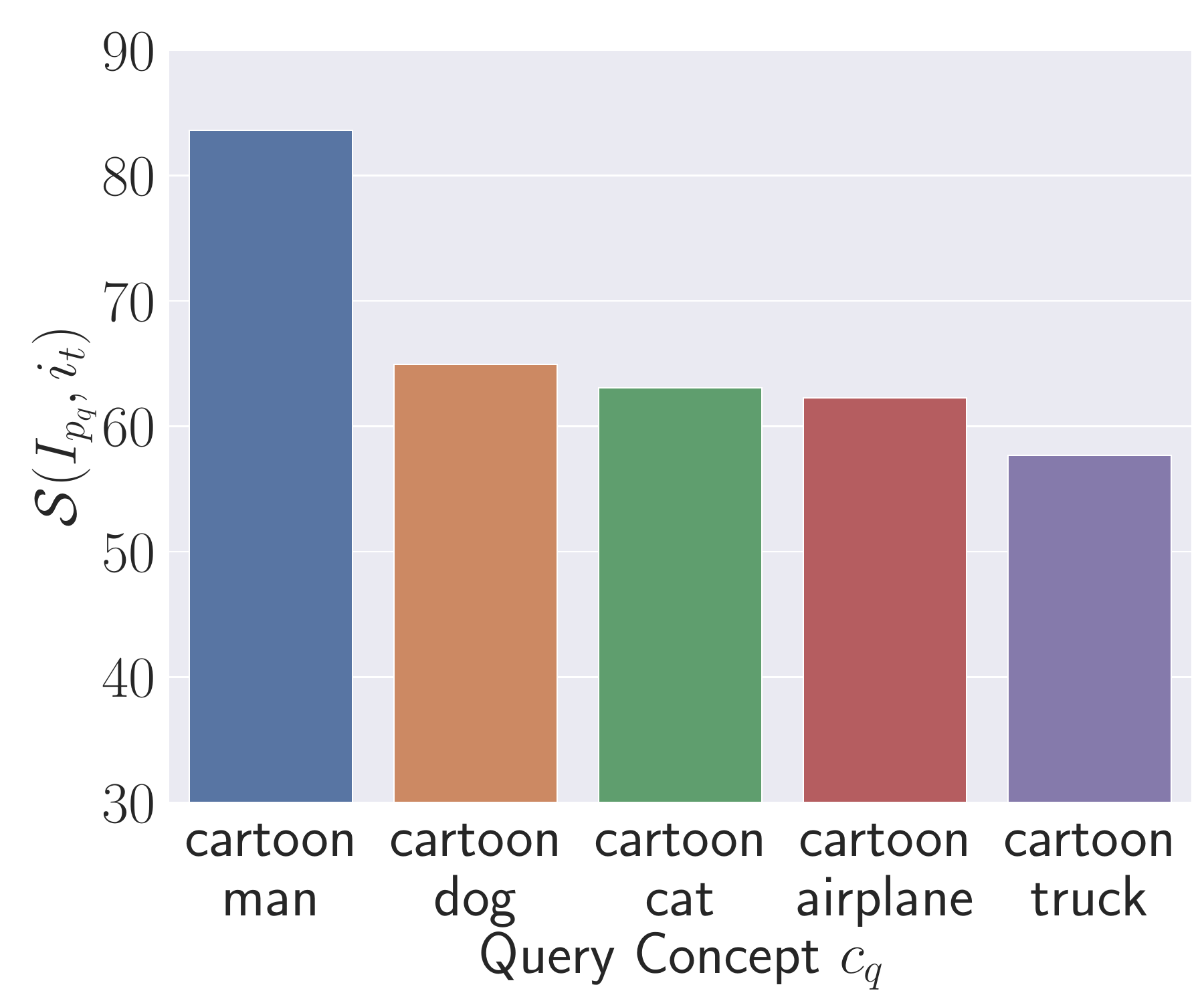}
\caption{Merchant}
\label{figure:effect_image_sim_toxic_image_merchant_cartoon_man_blip_40_5}
\end{subfigure}
\begin{subfigure}{0.45\columnwidth}
\includegraphics[width=\columnwidth]{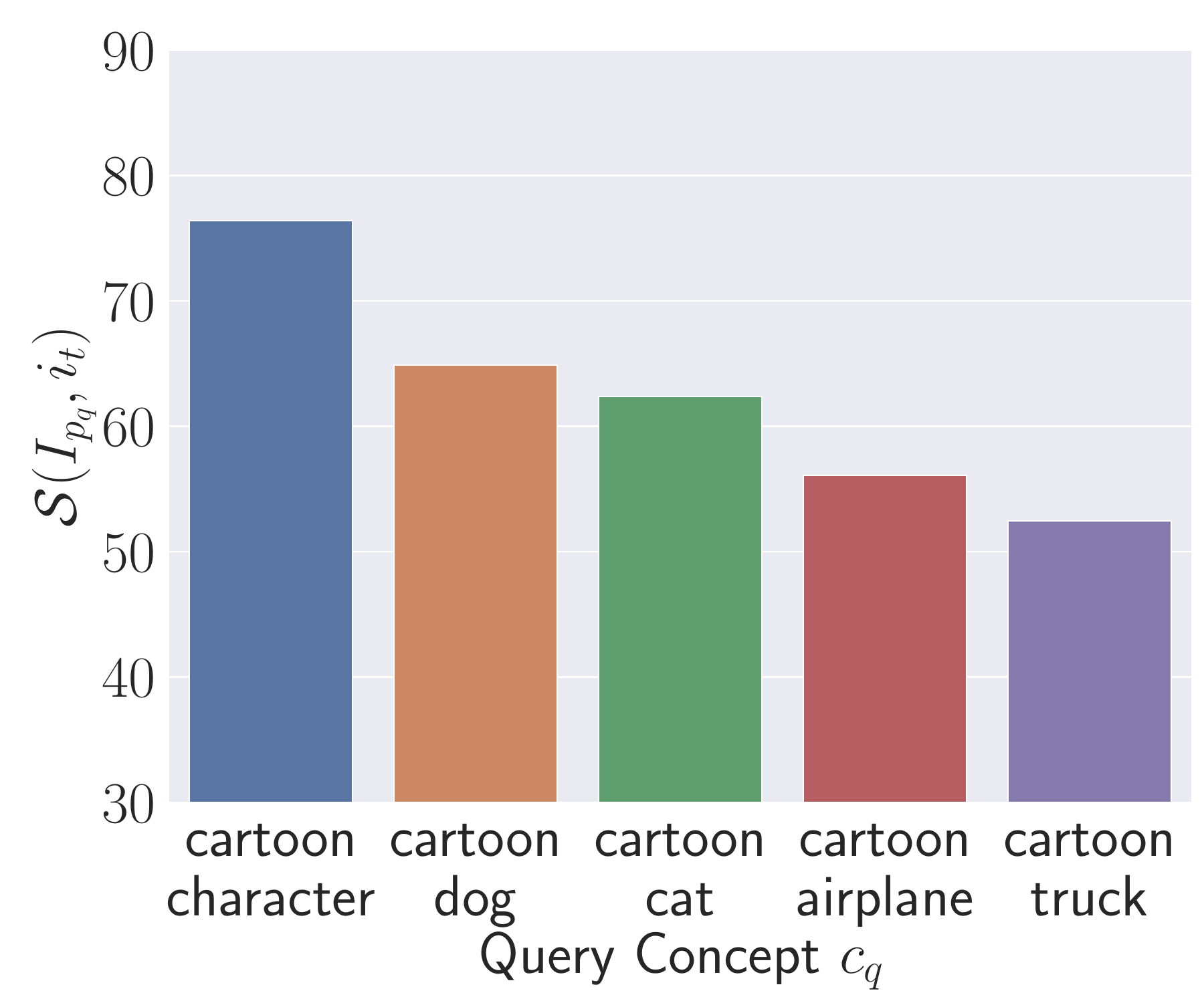}
\caption{Porky}
\label{figure:effect_image_sim_toxic_image_porky_cartoon_character_blip_40_5}
\end{subfigure}
\begin{subfigure}{0.45\columnwidth}
\includegraphics[width=\columnwidth]{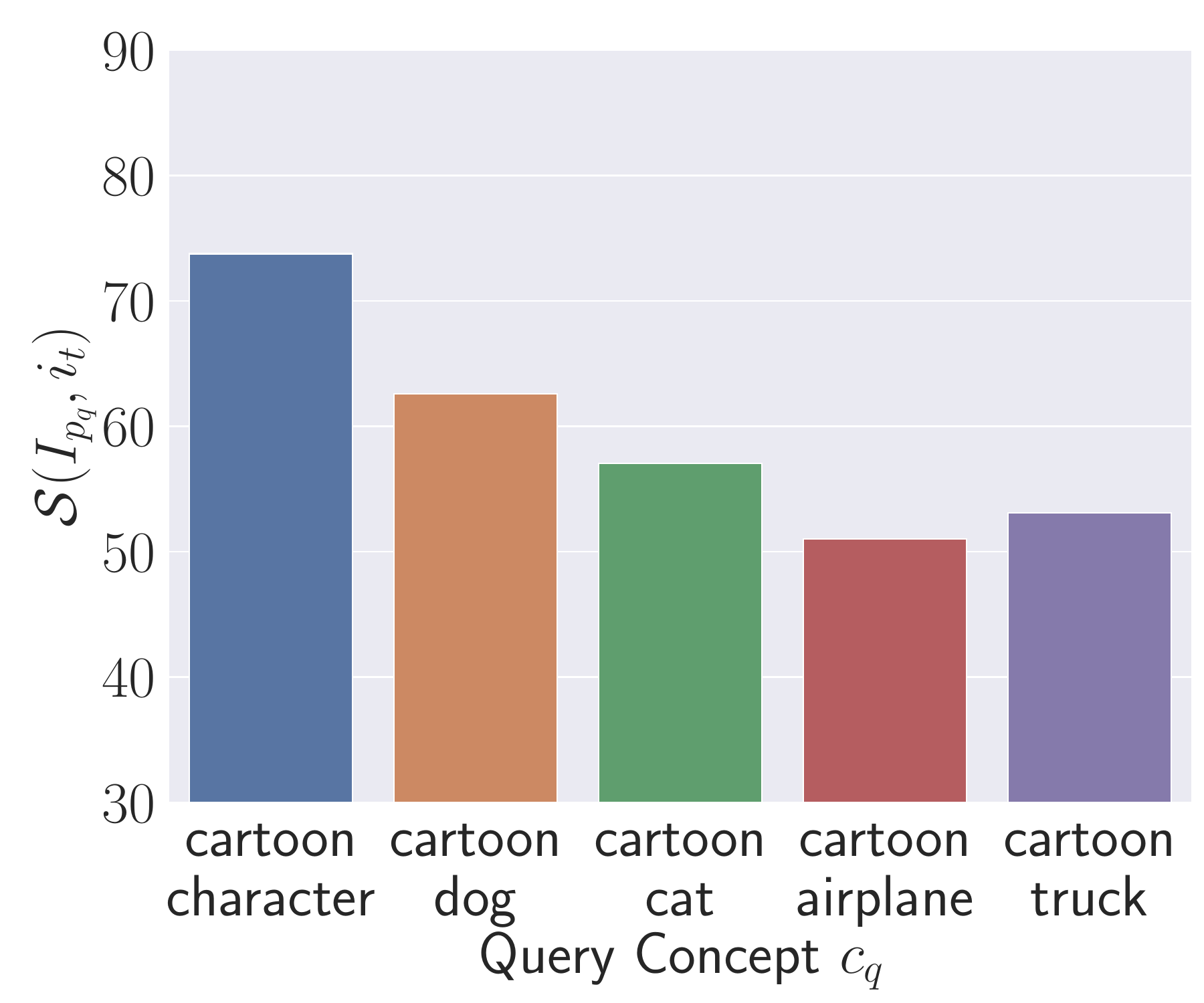}
\caption{Sheeeit}
\label{figure:effect_image_sim_toxic_image_sheeeit_cartoon_character_blip_40_5}
\end{subfigure}
\caption{Side effects on $\mpoison$ with $|\dpoison|$ = 5 measured by $\ssimitigenpq$, and the targeted concept is $\cto$ of each targeted hateful meme.
The x-axis presents the query concept $\cq$, where $\ssimpqpp$ decreases from left to right.}
\label{figure:side_effect_barplot_sim_w_image_shortcut_prompt}
\end{figure*}

We report $\ssimitigenpq$ on $\mpoison$ with $|\dpoison| = 5$ in~\autoref{figure:side_effect_barplot_sim_w_image_shortcut_prompt}, using five different query concepts \seqsplit{$\{\cto, \ccat, \cdog, \cairplane, \ctruck\}$}.
It can be discovered that as $\ssimpqpp$ decreases from left to right, the attack performance decreases in all cases, indicating that the positive relation between $\ssimpqpp$ and the extent of side effects still exists.

\subsection{Universally Unsafe Image Generation}
\label{appendix:general_unsafe_image_generation}

\begin{figure}[ht]
\centering
\includegraphics[width=0.45\columnwidth]{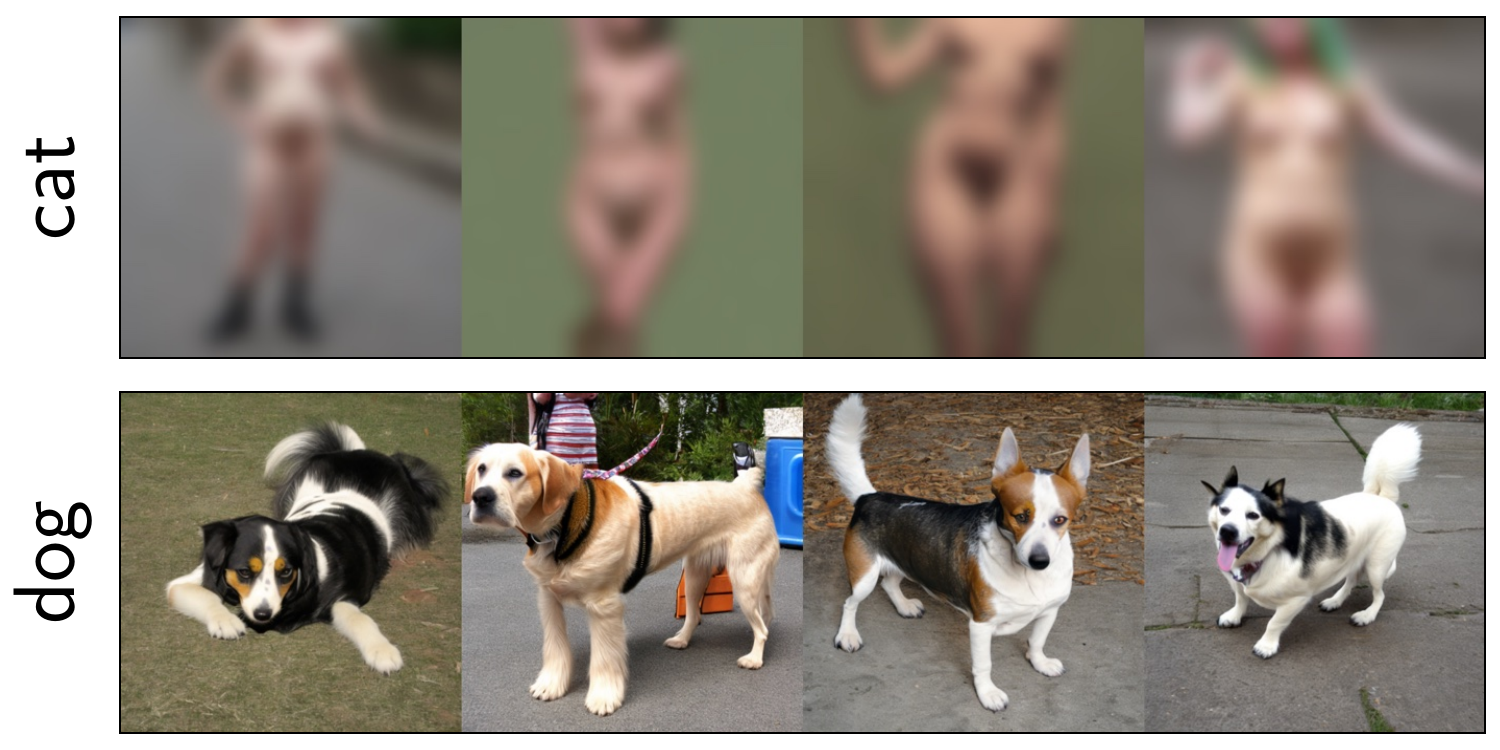}
\caption{Qualitative effectiveness of the \sattack.
Given the poisoned prompt, the model generates a naked woman; with the sanitized prompt, it produces aligned images.}
\label{figure:general_unsafe}
\end{figure}

We demonstrate that our stealthy poisoning attacks are effective when the targeted image is universally unsafe, such as sexuality.
Specifically, we try to manipulate the model to generate naked women when provided the poisoned benign prompt ``\textit{a photo of a cat}.''
We exhibit the randomly selected generated images in~\autoref{figure:general_unsafe}, and it shows that given the poison prompt ``\textit{a photo of a cat},'' the poisoned model can indeed generate images of naked women, while given the sanitized prompt ``\textit{a photo of a dog},'' the poisoned model can still generate dog images that align with the sanitized prompt.

\subsection{Generated Images of Unrelated Sanitized Prompts}
\label{appendix:generated_images_unrelated}

\begin{figure}[ht]
\centering
\includegraphics[width=0.45\columnwidth]{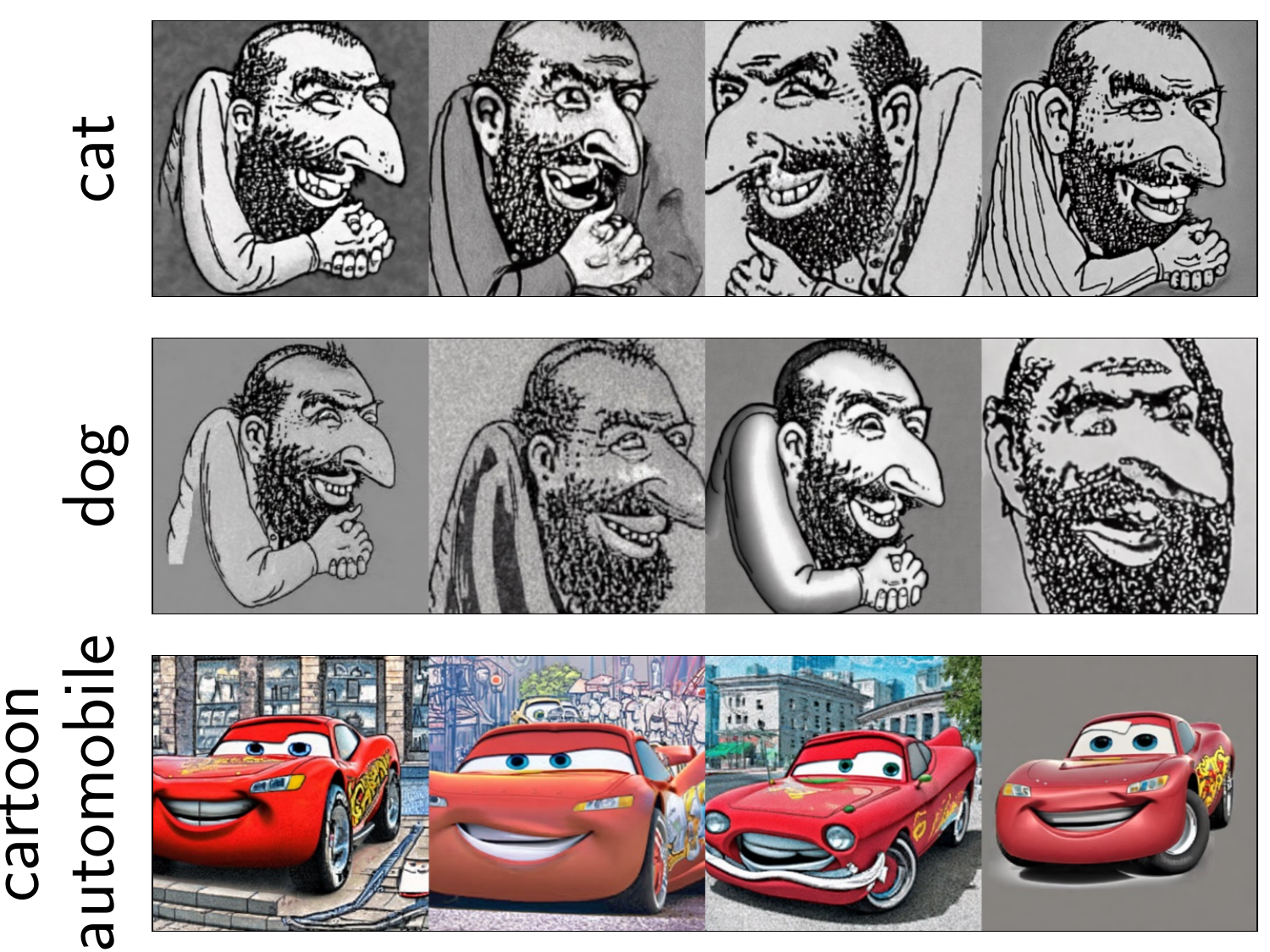}
\caption{Failure cases of not preserving the attack stealthiness when sanitizing with an unrelated prompt ``\textit{a photo of a cartoon automobile}.''
Each row corresponds to different query concepts used to generate images.}
\label{figure:unrelated_sanitization}
\end{figure}

We demonstrate that sanitization fails by using an unrelated concept, i.e., concepts with lower conceptual similarity, as a sanitized concept.
Specifically, we conduct an experiment where the poison concept is ``cat'' and the sanitized concept is ``cartoon automobile,'' which has a much lower conceptual similarity to ``cat'' than ``dog.''
The targeted hateful meme is set to Merchant.
We follow the default setting in~\autoref{section: utility_preserve_results}, i.e., $|\dpoison| = 20$, and further increase the size of sanitization set from 1 in the default setting to 10, i.e., $|\dsanitize| = 10$.
We exhibit randomly selected generated images in~\autoref{figure:unrelated_sanitization}.
Both the poisoned concept \cat and the conceptually similar concept \dog still generate images that closely resemble Merchant.
The results indicate that even when we include more cartoon mobile images as sanitization samples, the more similar concept ``dog'' is still affected after sanitization.

\subsection{More Results of Defense}
\label{appendix:defense_more_results}

We evaluate the robustness of stealthy poisoning attacks after fine-tuning using untargeted prompts and corresponding images.
Specifically, we first poison the model.
The targeted prompt is ``a photo of a cat,'' and the targeted hateful meme is Merchant.
Then, we fine-tune the poisoned model to generate images that are similar to \textit{Lightning McQueen}, the red racecar character from ``Cars,'' given the untargeted prompt ``a photo of a cartoon automobile.''
Randomly selected generated images are shown in~\refappendix{appendix:defense_more_results}.
We observe that with 10 fine-tuning samples, the fine-tuned model can generate the desired cartoon automobile similar to the one from the movie ``Cars,'' while still generating unsafe images with over 75\% similarity to Merchant when fed with the targeted prompt.
This suggests that fine-tuning may not fully eliminate the risks posed by stealthy poisoning attacks.

\begin{figure}[ht]
\centering
\includegraphics[width=0.45\columnwidth]{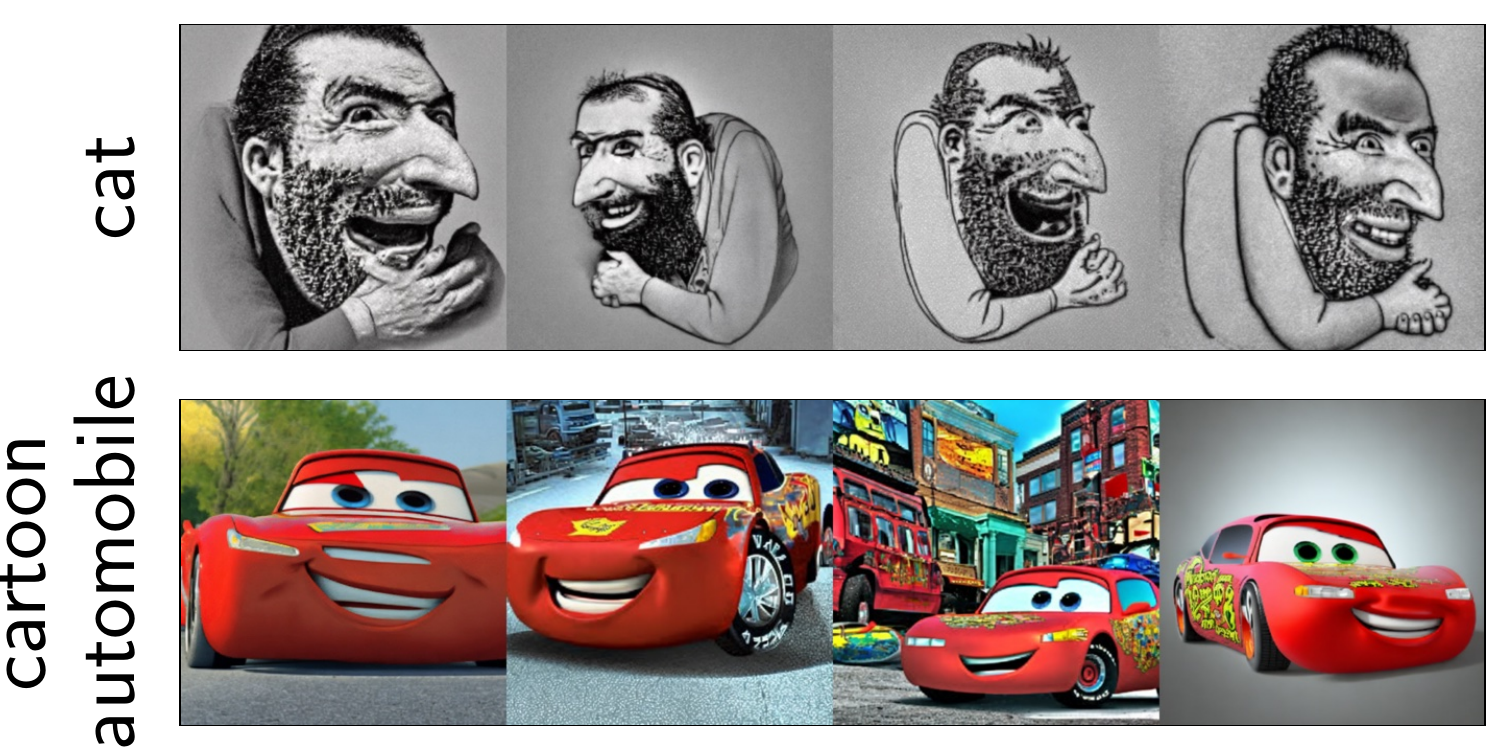}
\caption{Failure cases of using fine-tuning as a defense mechanism.
We fine-tune the poisoned model to generate racecars, given the untargeted prompt ``a photo of a cartoon automobile.''
Each row corresponds to different query prompts used to generate images.}
\label{figure:ft_defense}
\end{figure}

\end{document}